\pdfoutput=1
\documentclass[12pt]{article}
\usepackage{cite}
\usepackage{graphicx}
\usepackage{latexsym}   
\usepackage{mathrsfs}
\usepackage[overload]{textcase}

\font\grande=cmr9.5 scaled \magstep4
\font\medio=cmr9.5 scaled \magstep2
\outer\def\beginsection#1\par{\medbreak\bigskip
      \message{#1}\leftline{\bf#1}\nobreak\medskip
\vskip-\parskip
      \noindent}

\begin{document}
\bibliographystyle{unsrt}

\titlepage

\vspace{1cm}
\begin{center}
{\grande Large-scale gauge spectra and pseudoscalar couplings}\\
\vspace{1.5 cm}
Massimo Giovannini \footnote{e-mail address: massimo.giovannini@cern.ch}\\
\vspace{1cm}
{{\sl Department of Physics, CERN, 1211 Geneva 23, Switzerland }}\\
\vspace{0.5cm}
{{\sl INFN, Section of Milan-Bicocca, 20126 Milan, Italy}}
\vspace*{1cm}
\end{center}
\vskip 0.3cm
\centerline{\medio  Abstract}
\vskip 0.5cm
It is shown that the slopes of the superhorizon hypermagnetic spectra produced by the variation of the gauge couplings are practically unaffected by the relative strength of the parity-breaking terms. A new method is proposed for the estimate of the gauge power spectra in the presence of pseudoscalar interactions during inflation. To corroborate the general results, various concrete examples are explicitly analyzed. Since the large-scale gauge spectra also determine the late-time magnetic fields it turns out that the pseudoscalar contributions have little impact on the magnetogenesis requirement. Conversely the parity-breaking terms crucially affect the gyrotropic spectra that may seed, in certain models, the baryon asymmetry of the Universe. In the most interesting regions of the parameter space the modes reentering prior to symmetry breaking lead to a sufficiently large baryon asymmetry while the magnetic power spectra associated with the modes reentering after symmetry breaking may even be of the order of a few hundredths of a nG over typical length scales comparable with the Mpc prior to the collapse of the protogalaxy.  From the viewpoint of the effective field theory description of magnetogenesis scenarios these considerations hold generically for the whole class of inflationary models where the inflaton is not constrained by any underlying symmetry.
\noindent
\vspace{5mm}
\vfill
\newpage

\renewcommand{\theequation}{1.\arabic{equation}}
\setcounter{equation}{0}
\section{Introduction}
\label{sec1}
The conventional lore for the generation of the temperature and polarization anisotropies of the Cosmic Microwave Background (CMB) relies on the adiabatic paradigm \cite{ONE0a} that has been observationally tested by the various releases of the WMAP collaboration \cite{ONE0b,ONE0c,ONE0d}  and later confirmed by  terrestrial and space-borne observations including the Planck experiment  \cite{ONE0f,ONE0e}. One of the most studied lores for the generation of adiabatic and Gaussian large-scale curvature inhomogeneities is represented by the single-field scenarios that are described by a scalar-tensor action where the inflaton field $\varphi$ is minimally coupled to the four-dimensional metric (see e.g. \cite{ONE0g,ONE0h}).

In the framework of the adiabatic paradigm (possibly complemented by an early stage of inflationary expansion) it has been argued that the large-scale gauge fields could be parametrically amplified and eventually behave as vector random fields that do not break the spatial isotropy. In this context the problem is however shifted to the invariance under Weyl rescaling that forbids any efficient amplification of gauge fields in conformally flat (and four-dimensional) background geometries \cite{ONE0m}. One of the first suggestions along this direction has been the introduction of a pseudoscalar coupling \cite{TWOa,TWOb,TWOc} not necessarily coinciding with the Peccei-Quinn axion \cite{ONE0n,ONE0o,ONE0p}. It has been later suggested that the resulting action could be complemented by a direct coupling of the inflaton with the kinetic term of the gauge fields both in the case of inflationary and contracting Universes \cite{FOUR,FIVE,SIX,SEVEN,EIGHT}. The direct interaction with the inflaton plays  the role of an effective gauge coupling and a similar interpretation follows when the internal dimensions are dynamical \cite{SEVEN,EIGHT}. The origin of the scalar and of the pseudoscalar couplings may involve not only the inflaton but also some other  spectator field with specific physical properties \cite{NINE}. In the last two decades the problem dubbed magnetogenesis in Ref. \cite{SEVEN} gained some attention (see \cite{TEN, ELEVEN, TWELVEa, TWELVEb, TWELVEc, TWELVEd, TWELVEdd, TWELVEe, TWELVEf, TWELVEg, TWELVEh, TWELVEi, TWELVEl, TWELVEm, TWELVEo, TWELVEp, TWELVEq, TWELVEr, TWELVEs, TWELVEt, TWELVEu, TWELVEv} for an incomplete list of references dealing with different aspects of the problem). 

Even if the pseudoscalar coupling only leads to weak breaking of Weyl invariance, it efficiently modifies the topological properties of the hypermagnetic flux lines in the electroweak plasma \cite{FIVEa0} as also discussed in Ref. \cite{FIVEa1} by taking into account the chemical potentials associated with the finite density effects. If the gyrotropy is sufficiently large the produced Chern-Simons condensates may decay and eventually produce the baryon asymmetry \cite{FIVEb1a,FIVEa1,FIVEb1,FIVEb2,FIVEb3,FIVEb4,FIVEc}. These gyrotropic and helical fields play a key role in various aspects of anomalous magnetohydrodynamics \cite{SIXa}. In the collisions of heavy ions this phenomenon is often dubbed chiral magnetic effect \cite{SIXa,SIXb} (see also \cite{SIXe,SIXf}). There some differences between the formulation of anomalous magnetohydrodynamics \cite{SIXc,SIXd} and the chiral magnetic effects: while Ohmic and chiral currents are concurrently present in the former, only chiral currents are typically considered in the latter.

In conventional inflationary scenarios a large class of magnetogenesis models considered so far in the literature can be summarized in terms of the following schematic action\footnote{The Greek indices run from $0$ to $3$; $G$ denotes the determinant of the four-dimensional metric $G_{\alpha\beta}$; $R = G^{\mu\nu} R_{\mu\nu}$ is the Ricci scalar defined from the contraction of the Ricci tensor }:
\begin{eqnarray}
S &=& \int d^{4}x \, \,\sqrt{- G} \biggl[ - \frac{R}{2 \ell_{P}^2} + \frac{1}{2} G^{\alpha\beta} \partial_{\alpha} \varphi \partial_{\beta} \varphi 
+ \frac{1}{2} G^{\alpha\beta} \partial_{\alpha} \psi \partial_{\beta} \psi - {\mathcal V}(\varphi,\psi) 
\nonumber\\
&-& \frac{\lambda(\varphi,\psi)}{16\pi} Y_{\alpha\beta} \, Y^{\alpha\beta} - \frac{\overline{\lambda}(\varphi,\psi)}{16 \pi } Y_{\alpha\beta} \, \widetilde{\,Y\,}^{\alpha\beta} \biggr],
\label{action1}
\end{eqnarray}
where $\ell_{P}^2 = 8 \pi G$; $Y_{\alpha\beta}$ and $\widetilde{\,Y\,}^{\alpha\beta}$ denote, respectively, the gauge field strength and its dual; throughout the paper we shall employ both  the reduced Planck mass $\overline{M}_{P} = \ell_{P}^{-1}$ and the standard Planck mass 
$\overline{M}_{P}$ with $M_{P} = \sqrt{8 \pi} \,\,\overline{M}_{P}$.  In  Eq. (\ref{action1}) $\phi$ and $\psi$ are, respectively, the inflaton field and a generic spectator field.  There are no compelling reasons why $\lambda(\varphi, \psi)$ and $\overline{\lambda}(\varphi,\psi)$ must coincide or just scale with the same law. To appreciate this statement it suffices to consider the simplest situation where the potential is given by ${\mathcal V}(\varphi,\psi)= V(\varphi) + W(\psi)$ while $ \lambda= \lambda(\varphi)$ and $\overline{\lambda} = \overline{\lambda}(\psi)$. In this case while the evolution of the gauge 
coupling $e = \sqrt{4\pi/\lambda}$ is controlled by the inflaton,  $\overline{\lambda}$ 
only depends on the spectator field $\psi$. If $W(\psi) =0$ and $\overline{\lambda} = \overline{\lambda}(\varphi)$ we could argue that the action for the hypercharge fields only depends on the gauge coupling but this is actually incorrect unless $\overline{\lambda}$ coincides with $\lambda$. This statement gets more clear if the gauge part of the action (\ref{action1}) is rewritten as:
\begin{equation}
S_{gauge} = - \frac{1}{4} \int \, d^{4} x \frac{\sqrt{ -G}}{e^2} \biggl[ Y_{\alpha\beta} Y^{\alpha\beta} + \biggl(\frac{\overline{\lambda}}{\lambda}\biggr) Y_{\alpha\beta} \widetilde{Y}^{\alpha\beta} \biggr],
\label{action2}
\end{equation}
where $e^{2} = 4 \pi/\lambda$.  In spite of possible tunings,  the generic situation would imply that $\overline{\lambda}$ and $\lambda$ are neither equal nor proportional.

The purpose of this paper is to demonstrate that the slopes of the superhorizon hypermagnetic and hyperelectric spectra produced by the variation of the gauge coupling do not depend, in practice, on the relative weight of $\overline{\lambda}$ and $\lambda$.
This conclusion suggests that the magnetogenesis scenarios are not affected by the parity-breaking terms that instead 
determine the gyrotropic contributions of the gauge power spectra. Provided the pseudoscalar interactions 
have certain scaling properties, they can provide, at least in principle, a mechanism for the generation of the baryon asymmetry. For the estimate of the gauge power spectra in the presence of pseudoscalar interactions it will not be assumed, as always done so far, that $\overline{\lambda}$ and $\lambda$ either scale in the same way in time or even coincide. We rather use an approximate method that is corroborated by explicit examples. A similar approximate method has been recently suggested for the estimate of the polarized backgrounds of relic gravitons \cite{POLWKB}.  The obtained results have been employed to analyze a phenomenological scenario where 
the modes reentering prior to symmetry breaking lead to the production of hypermagnetic gyrotropy. The 
obtained Chern-Simons condensate affects the Abelian anomaly and ultimately leads to the baryon asymmetry of the Universe (BAU in what follows). The modes reentering after electroweak symmetry breaking lead 
instead to ordinary magnetic fields that can seed the large-scale magnetic fields and 
therefore provide a magnetogenesis mechanism. We argue that there exist 
regions of the parameter space where the BAU is sufficiently large and the magnetic power spectra may even be of the order of a few hundredths of a nG over typical length scales comparable with the Mpc prior to the collapse of the protogalaxy.  As we see this 
analysis holds generically for the whole class of inflationary models where the inflaton is not constrained by any underlying symmetry.

The layout of this paper is the following. After a preliminary discussion of the problem and of its motivations (see section \ref{sec2}),  an approximate method for the estimate of the gauge spectra is described in section \ref{sec3}. This strategy is based on the Wentzel-Kramers-Brillouin (WKB) approach where, however, the turning points are fixed by the structure of the polarized mode functions.  In section \ref{sec4} the obtained results are corroborated by explicit examples. In section \ref{sec5}  the obtained 
results are examined in the light of the late-time gauge spectra with particular attention to the requirements associated with the BAU and with the magnetogenesis constraints. Section \ref{sec6} contains our concluding remarks. In appendix \ref{APPA} we reported the exact form of the mode functions for different expressions of 
the pseudoscalar couplings; in appendix \ref{APPB} the results of the WKB approach have been 
explicitly compared with the exact solutions of the mode functions; finally in appendix \ref{APPC} we reported some useful 
results holding in the case of decreasing gauge coupling. 

\renewcommand{\theequation}{2.\arabic{equation}}
\setcounter{equation}{0}
\section{Preliminary considerations}
\label{sec2}
\subsection{The general lore}
While we consider, for the sake of simplicity, the evolution in a conventional inflationary background, this 
choice is not exclusive since most of the present considerations could also be applied to different models (e.g. 
contracting scenarios). In a conformally flat Friedmann-Robertson-Walker metric 
$G_{\mu\nu} = a^2(\tau) \eta_{\mu\nu}$ (where $\eta_{\mu\nu}$ is the four-dimensional Minkowski metric) the Hamiltonian constraint stemming from the equations derived from the total action (\ref{action1}) is given by:
\begin{equation}
3 \overline{M}_{P}^2 {\mathcal H}^2 = \biggl[\frac{{\varphi'}^2}{2} + V a^2\biggr] 
+ \biggl[\frac{{\psi'}^2}{2} + W a^2\biggr]  + \frac{1}{8\pi a^2} (B^2 + E^2).
\label{EqP1}
\end{equation}
In Eq. (\ref{EqP1}) the prime denotes the derivation with respect to the conformal time coordinate 
$\tau$ and ${\mathcal H} = a^{\prime}/a$ is related to the Hubble rate $H$ as 
$ a H = {\mathcal H}$; finally the fields $\vec{E}$ and $\vec{B}$ are 
the comoving hyperelectric and hypermagnetic fields. The dominant 
source of the background geometry is the field $\varphi$. During slow-roll, as usual,  
the kinetic energy of $\varphi$ can be neglected and the inflaton potential is generally dominant against 
the potential of the spectator field:
\begin{equation}
V \gg \frac{\varphi^{\prime\, 2}}{2 a^2 } \gg \biggl( W + \frac{\psi^{\prime\, 2}}{2 a^2 }\biggr) \gg \frac{E^2 +B^2}{8 \pi a^4}.
\label{EqP2}
\end{equation} 
The last inequality in Eq. (\ref{EqP2}) guarantees that the gauge fields
will not affect the evolution of the geometry.  The comoving fields appearing in Eqs. (\ref{EqP1}) and (\ref{EqP2}) are related to their physical counterpart as $\vec{\,E\,}  =  \, a^2 \, \sqrt{\lambda}\, \vec{\,E\,}^{(phys)}$ and as $\vec{\,B\,} = a^2 \, \sqrt{\lambda}\,\vec{\,B\,}^{(phys)}$. Furthermore the components of the field strengths are directly expressible as $Y_{i\, 0} = - a^2 E_{i}^{(phys)}$,  $Y^{i\, j} = - \epsilon^{i\, j\, k} B_{k}^{(phys)}/a^2$ and similarly for the dual strength. The evolution of 
 the comoving fields follows from:
\begin{eqnarray}
&& \vec{\nabla} \times \biggl(\sqrt{ \lambda} \, \vec{\,\,B\,\,}\biggr) = \partial_{\tau} \biggl( \sqrt{\lambda}  \vec{\,\,E\,\,}\biggr) + \biggl(\frac{\overline{\lambda}^{\,\prime}}{\sqrt{\lambda}}\biggr) \vec{B} +  \frac{\vec{\nabla} \, \overline{\lambda} \times \vec{\,\,E\,\,}}{\sqrt{\lambda}},
\label{EqP3}\\
&&\vec{\nabla}\cdot\biggl( \sqrt{\lambda}\, \vec{\,\,E\,\,} \biggr) = 
\frac{ \vec{\,\,B\,\,}\cdot \vec{\nabla} \lambda}{\sqrt{\lambda}},\qquad \vec{\nabla} \cdot\biggl(\frac{\vec{B}}{\sqrt{\lambda}}\biggr) =0,
\label{EqP4}\\
&& \vec{\nabla} \times \biggl(\frac{\vec{E}}{\sqrt{\lambda}}\biggr) + \partial_{\tau} \biggl( \frac{\vec{B}}{\sqrt{\lambda}}\biggr) =0,
\label{EqP5}
\end{eqnarray}
where, to avoid potential confusions, the derivations with respect to $\tau$ have been made explicit. While both $\lambda$ and $\overline{\lambda}$ enter Eqs. (\ref{EqP3}), (\ref{EqP4}) and (\ref{EqP5}) their evolution is unrelated; no one orders that the two couplings 
are either proportional or even equal. This aspect can be better appreciated by considering the case where 
$\lambda$ and $\overline{\lambda}$ are both homogeneous  so that Eq. (\ref{EqP3}) becomes:
\begin{equation}
\vec{\nabla} \times \vec{B} = \frac{1}{\sqrt{\lambda}} \partial_{\tau} \biggl( \sqrt{\lambda}\,\, \vec{E} \biggr) + 
\frac{\overline{\lambda}^{\prime}}{\lambda} \, \vec{B}.
\label{EqP6}
\end{equation}
With standard manipulations Eqs. (\ref{EqP5}) and (\ref{EqP6}) 
can be directly combined so that the evolution of the comoving hypermagnetic field is ultimately given by: 
\begin{equation}
\frac{1}{\sqrt{\lambda}} \partial_{\tau} \biggl[ \lambda \, \partial_{\tau} \biggl(\frac{\vec{B}}{\sqrt{\lambda}}\biggr) \biggr] - \nabla^2\, \vec{B} 
- \frac{\overline{\lambda}^{\,\prime}}{\lambda} \vec{B} =0.
\label{EqP7}
\end{equation}
Equation (\ref{EqP7}) shows that $\overline{\lambda}$ and $\lambda$ are not bound to scale in the same manner unless they 
are proportional or even coincide.

\subsection{Few examples}
Let us now consider, in this respect, the parametrization $ \overline{\lambda} = \psi/M_{1}$ where $M_{1}$ is a typical scale and $\psi$ is constrained by Eq. (\ref{EqP2}). Since we want $\psi$ to be  light during inflation, we require $\psi_{*} \ll \overline{M}_{\mathrm{P}}$ and $m < H$ where $H$ is the typical curvature scale during inflation. The governing equation for $\psi$ 
\begin{equation}
 \psi'' + 2  {\mathcal H} \psi'  +  \frac{\partial W}{\partial\psi} a^2 =0,\qquad \qquad W(\psi) = m^2 (\psi -\psi_{*})^2/2,
\label{EqP8a}
\end{equation}
can be rephrased in terms of $\mu=m/H$ and $\epsilon= - \dot{H}/H^2$:
\begin{equation}
(a \psi)'' +[ \mu^2 - (2 - \epsilon)] a^2 H^2 (a \psi)  =0,\qquad\qquad \epsilon= - \dot{H}/H^2 = \frac{\overline{M}_{\mathrm{P}}^2}{2} (V_{,\varphi}/V)^2.
\label{EqP9}
\end{equation}
For $\mu\ll 1$ the evolution of $\psi$ is simply given by $\psi \simeq \psi_{*} + \psi_{i} (- \tau/\tau_{i})^{\zeta}$ 
where $\zeta = (3 - 2 \epsilon)/(1 -\epsilon)$.

Consider next a  generic example for the evolution of $\lambda= \lambda(\varphi)$. In the slow--roll approximation it can be argued that the dependence of $\lambda$   upon $\varphi$ follows from:
\begin{equation}
\lambda(\varphi) = e^{ - c \, {\mathcal I}(\varphi_{*}, \varphi)}, \qquad {\mathcal I}(\varphi_{*}, \varphi) = \frac{1}{\overline{M}_{P}^2} \int_{\varphi_{*}}^{\varphi} \frac{V}{V_{\, , \varphi}} \, d\varphi,
\label{EqP8}
\end{equation}
where $c$ is an arbitrary constant which may be either positive or negative. Equation (\ref{EqP8}) typically arises in various classes of models where $\lambda$ is proportional to $(a/a_{*})^{c}$ and this parametrization is plausible as long as the inflaton slowly rolls; depending on the model the relation between $(a/a_{*})$ and $\varphi$ might be  different but still the general parametrization of $\lambda= \lambda(a)$ must hold. For monomial potentials (e.g. $W = M_{2}^4 \phi^{p}$) we would have, for instance, $  \lambda(\varphi) \propto e^{- c\,\varphi^2/( 2 p \overline{M}_{P}^2)}$; note that, in general terms, $M_{2} \neq M_{1}$. Within the same parametrization other models can be analyzed  like  the case of small field 
and hybrid models where the potential is approximately given by $V(\varphi) = M_{2}^4 ( 1 \pm \kappa \varphi^{p})$ (where the plus corresponds  to the hybrid models while the minus to the small field models).
In the case of plateau-like potentials, we have\footnote{In the slow-roll approximation
(and in terms of the cosmic time coordinate $t$) we obtain from Eq. (\ref{EqP10}) that $\Phi = \ln{[ 2 M_{2} (t_{*} - t)/3]}$ (valid for $t \ll - t_{*}$); 
this also means that, within the same approximation, $H \simeq M_{2}(1 - e^{-\Phi})/2$. } 
\begin{equation}
V(\varphi) = \frac{3 M_{2}^2 \overline{M}_{P}^2}{4} \biggl( 1 - e^{-\Phi}\biggr)^2,\qquad\qquad \Phi= \sqrt{\frac{2}{3}} \biggl(\frac{\varphi}{\overline{M}_{P}}\biggr).
\label{EqP10}
\end{equation}
From the above examples we have therefore to acknowledge that, in Eq. (\ref{EqP6}), there are three different quantities determining the evolution of the hypermagnetic fields; two of them have well defined scaling properties in $\tau$ while the third one strongly depends on the model:  
\begin{equation}
\frac{\lambda^{\prime}}{\lambda} = {\mathcal O}(\tau^{-1}), \qquad \frac{\lambda^{\prime\prime}}{\lambda} = {\mathcal O}(\tau^{-2}), \qquad \frac{\overline{\lambda}^{\prime}}{\lambda} = \biggl(\frac{\psi^{\prime}}{M_{1}}\biggr) e^{c \,{\mathcal I}(\varphi_{*}, \varphi)}.
\label{EqP11}
\end{equation} 
Equation (\ref{EqP11}) shows that there are no obvious reasons to assume that $\overline{\lambda}^{\prime}$ 
must be  proportional to $\lambda^{\prime}$. Even assuming that $\psi$ coincides with the inflaton, from Eq. (\ref{EqP11}) 
we would have $(\varphi^{\prime}/M_{1}) e^{c {\mathcal I}(\varphi_{*}, \varphi)}$ which depends on the particular model
and does not necessarily demand  $\lambda= \overline{\lambda}$. 

\subsection{Evolutions of the gauge coupling}
In what follows, we therefore assume that $\lambda$ and $\overline{\lambda}$ scale differently, and 
the gauge power spectra are estimated as a function of the possible dynamical evolutions. For the sake of concreteness, during the inflationary stage we posit that:
\begin{eqnarray}
\lambda(\tau) &=& \lambda_{1} \biggl(- \frac{\tau}{\tau_{1}}\biggr)^{2 \gamma}, \qquad \tau \leq - \tau_{1},
\label{EqP12a}\\
\overline{\lambda}(\tau) &=& \overline{\lambda}_{2} \biggl(- \frac{\tau}{\tau_{2}}\biggr)^{2 \beta}, \qquad \tau \leq - \tau_{1},
\label{EqP12b}
\end{eqnarray}
where $\tau_{1}$ marks the end of inflation while $|\tau_{2}| \geq \tau_{1}$. While this choice is purely illustrative 
we stress that other complementary situations could be discussed with the same techniques developed here\footnote{
Without specific fine-tunings in the simplest situation it is however plausible to think that $\tau_{2} = {\mathcal O}(\tau_1)$.}.
For $\tau \geq - \tau_{1}$ we also posit that the evolution of $\lambda$ is continuously matched 
to the radiation-dominated phase. Since in the evolution equations of $\vec{E}$ and $\vec{B}$ there are terms going as $\lambda^{\prime\prime}/\lambda$, the continuity of $\lambda$ and $\lambda^{\prime}$ is essential. Conversely it is sufficient to demand that only $\overline{\lambda}$ is continuous 
since, in the corresponding equations, only terms going as $\overline{\lambda}^{\prime}$ may arise. With these precisions we have that for $\tau \geq - \tau_{1}$  the evolution of $\lambda$ and $\overline{\lambda}$ 
is  parametrized as:
\begin{eqnarray}
\lambda(\tau) &=& \lambda_{1} \biggl[\frac{\gamma}{\delta}\biggl(\frac{\tau}{\tau_{1}} +1\biggr) +1\biggr]^{ 2 \delta}, \qquad \qquad \tau \geq - \tau_{1},
\label{EqP13a}\\
\overline{\lambda}(\tau) &=& \overline{\lambda}_{2} \biggl(\frac{\tau_{1}}{\tau_{2}}\biggr)^{2 \beta}, \qquad \qquad \tau \geq - \tau_{1}.
\label{EqP13b}
\end{eqnarray}
Equation (\ref{EqP13a}) describes the situation where the gauge coupling (introduced in Eq. (\ref{action2}) and related to the inverse of $\lambda$) increases during the inflationary phase and then flattens out later on. 
Equation (\ref{EqP13a}) could be complemented with the dual evolution where the gauge coupling decreases and then flattens out; in this case we have 
\begin{eqnarray}
\lambda(\tau) &=& \lambda_{1} \biggl(- \frac{\tau}{\tau_{1}}\biggr)^{- 2 \widetilde{\,\gamma\,}}, \qquad \tau \leq - \tau_{1},
\label{EqP14a}\\
\lambda(\tau) &=& \lambda_{1} \biggl[\frac{\widetilde{\gamma}}{\widetilde{\,\delta\,}}\biggl(\frac{\tau}{\tau_{1}} +1\biggr) +1\biggr]^{ - 2 \widetilde{\,\delta\,}}, \qquad \qquad \tau \geq - \tau_{1}.
\label{EqP14b}
\end{eqnarray}
We shall often critically compare the physical situations 
implied by Eqs. (\ref{EqP13a})--(\ref{EqP13b}) 
and by Eqs. (\ref{EqP14a})--(\ref{EqP14b}). While we shall 
consider as more physical the case described by Eqs.  (\ref{EqP13a})--(\ref{EqP13b}) the methods discussed below 
can also be applied to the case where the gauge coupling 
is initially very large and then decreases. 

\renewcommand{\theequation}{3.\arabic{equation}}
\setcounter{equation}{0}
\section{The general argument and the power spectra}
\label{sec3}
\subsection{Quantum fields and their evolution}
In what follows the right (i.e. $R$) and left (i.e. $L$) polarizations shall be defined, respectively, by the subscripts $\pm$:
\begin{equation}
\hat{\varepsilon}^{(\pm)}(\hat{k}) = \frac{\hat{e}^{\oplus}(\hat{k}) \pm i \, \hat{e}^{\otimes}(\hat{k})}{\sqrt{2}}, 
\qquad  \hat{\varepsilon}^{(+)}(\hat{k}) \equiv \hat{\varepsilon}_{R}(\hat{k}), \qquad 
\hat{\varepsilon}^{(-)}(\hat{k}) \equiv \hat{\varepsilon}_{L}(\hat{k}),
 \label{WKB1}
 \end{equation}
where $\hat{k}$, $\hat{e}_{\oplus}$ and $\hat{e}_{\otimes}$ denote a triplet of mutually orthogonal 
unit vectors defining, respectively, the direction of propagation and  the two linear (vector) polarizations.  From Eq. (\ref{WKB1}) the vector product of $\hat{k}$ with the circular polarizations will be given by $\hat{k} \times \hat{\varepsilon}^{(\pm)} = \mp \, i\,  \hat{\varepsilon}^{(\pm)}$.  If both $\lambda$ and $\overline{\lambda}$ are homogeneous 
(as we shall assume hereunder), the quantum Hamiltonian 
associated with the gauge action (\ref{action2}) becomes:
\begin{equation}
\widehat{H}_{Y}(\tau) = \frac{1}{2} \int d^3 x \biggl[ \widehat{\pi}_{i}^{2}  + 
{\mathcal F}\biggl( \widehat{\pi}_{i}\,\widehat{\mathcal Y}_{i} + \widehat{\mathcal Y}_{i}\,\widehat{\pi}_{i}\biggr)+
\partial_{i}\widehat{{\mathcal Y}}_{k} \,\,\partial^{i} \widehat{{\mathcal Y}}_{k} - \biggl(\frac{\overline{\lambda}^{\, \prime}}{ \lambda} \biggr)\, \widehat{{\mathcal Y}}_{i} \,\,\partial_{j} \widehat{{\mathcal Y}}_{k} \,\, \epsilon^{i \, j\, k}\biggr],
\label{WKB2}
\end{equation}
where $ \widehat{{\mathcal Y}}_{i}$ is the quantum field operator corresponding to the (rescaled) 
vector potential ${\mathcal Y}_{i} = \sqrt{\lambda/(4\pi)} \,\, Y_{i}$ defined in the Coulomb gauge \cite{VP} which is probably the most convenient for this problem since it is invariant 
under Weyl rescaling. In Eq. (\ref{WKB2}) $\widehat{\pi}_{i} = \widehat{{\mathcal Y}}_{i}^{\,\,\prime} - {\mathcal F} \, \widehat{{\mathcal Y}}_{i}$ denotes the canonical momentum operator; to make the notation more concise, the rate of variation of the gauge coupling  ${\mathcal F} = \sqrt{\lambda}^{\,\prime}/\sqrt{\lambda}$ has been introduced throughout. The evolution equations of the field operators following form the Hamiltonian (\ref{WKB2}) are (units $\hbar = c =1$ will be adopted):
\begin{eqnarray}
\widehat{\pi}_{i}^{\,\,\prime} &=& i\, \biggl[ \widehat{H}_{Y}, \widehat{\pi}_{i} \biggr] = - {\mathcal F} \, \widehat{\pi}_{i} + \nabla^2 \widehat{{\mathcal Y}}_{i} + \frac{\overline{\lambda}^{\,\prime}}{\lambda} \, \epsilon_{i\, j\, k} \partial^{j} \, \widehat{{\mathcal Y}}^{k},
\nonumber\\
\widehat{{\mathcal Y}}_{i}^{\,\,\prime} &=&   i\, \biggl[ \widehat{H}_{Y}, \widehat{{\mathcal Y}}_{i} \biggr] = \widehat{\pi}_{i} + {\mathcal F} \, \widehat{{\mathcal Y}}_{i}.
\label{WKB3}
\end{eqnarray}
The initial data of the field operators fore Eqs. (\ref{WKB2})--(\ref{WKB3}) must obey the canonical commutation relations at equal times:
\begin{equation}
\biggl[\widehat{{\mathcal Y}}_{i}(\vec{x}_{1}, \tau),\widehat{\pi}_{j}(\vec{x}_{2}, \tau)\biggr] = i \Delta_{ij}(\vec{x}_{1} - \vec{x}_{2}),\qquad 
\Delta_{ij}(\vec{x}_{1} - \vec{x}_{2}) = \int \frac{d^{3}k}{(2\pi)^3} e^{i \vec{k} \cdot (\vec{x}_{1} - \vec{x}_2)} p_{ij}(\hat{k}), 
\label{WKB4}
\end{equation}
where $p_{ij}(\hat{k}) = (\delta_{ij} - \hat{k}_{i} \hat{k}_{j})$. The function $\Delta_{ij}(\vec{x}_{1} - \vec{x}_{2})$ is the transverse generalization of the Dirac delta function ensuring that  both the field operators and the canonical momenta are divergenceless. The mode expansion for the hyperelectric and 
hypermagnetic fields can be easily written in the circular basis of Eq. (\ref{WKB1}) as\footnote{ We note that the creation and annihilation operators $\widehat{a}_{k,\,\alpha}$ and $\widehat{a}_{k,\,\alpha}^{\dagger}$ 
are directly defined in the circular basis and they obey the standard 
commutation relation $[\widehat{a}_{\vec{k}, \, \alpha}, \, \widehat{a}_{\vec{p}, \, \beta}] = \delta^{(3)}(\vec{k}- \vec{p})\, \delta_{\alpha\beta}$. In Eqs. (\ref{WKB5})--(\ref{WKB6}) ``h.c.'' denotes the Hermitian conjugate; note, in this 
respect, that, unlike the linear polarizations, the circular polarizations are complex vectors.}:
\begin{eqnarray}
\widehat{E}_{i}(\vec{x},\tau) &=&  -   \sum_{\alpha= \pm} \, \,\int\frac{d^{3} k}{(2\pi)^{3/2}}\,\,
\biggl[ g_{k,\,\alpha}(\tau) \, \widehat{a}_{k,\alpha} \,\,  \varepsilon^{(\alpha)}_{i}(\hat{k})\,\,e^{- i \vec{k} \cdot\vec{x}} + \mathrm{h.c.}\biggr],
\label{WKB5}\\
\widehat{B}_{k}(\vec{x}, \tau)  &=&  - i \, \,\epsilon_{i\,j\,k} \,   \sum_{\alpha= \pm}\,\,\int\, \frac{d^{3} k}{(2\pi)^{3/2}}\,\, k_{j} \,\,
\biggl[ f_{k,\, \alpha}(\tau) \, \widehat{a}_{k,\,\alpha}\, \,\,  \varepsilon^{(\alpha)}_{i}(\hat{k})\, e^{- i \vec{k} \cdot\vec{x}} - \mathrm{h.c.} \biggr],
\label{WKB6}
\end{eqnarray}
where, incidentally, the hyperelectric field operator coincides (up to a sign) with the canonical momentum 
(i.e. $\widehat{E}_{i} =  - \widehat{\pi}_{i} = -  \sqrt{\lambda} (\widehat{{\mathcal Y}}_{i}/\sqrt{\lambda})^{\, \prime}$) while the hypermagnetic operator is simply $\widehat{B}_{k}=\epsilon_{i\,j \, k} \,\partial_{i}\, \widehat{{\mathcal Y}}_{j}$. The  hypermagnetic and hyperelectric mode functions (i.e. $f_{k,\, \alpha}(\tau)$ and $g_{k,\, \alpha}(\tau)$ respectively) must preserve the commutation relations (\ref{WKB4}) 
and this is why their Wronskian $W_{\alpha} = f_{k,\, \alpha} \, g^{\ast}_{k,\, \alpha} - f_{k,\, \alpha}^{\ast} \, g_{k,\, \alpha}$ must be normalized as $W_{\alpha} = i$ for 
$\alpha = \pm$; in other words the Wronskian normalization must be independently enforced  for each of the two circular polarizations. 

The actual evolution of the mode functions follows by inserting the expansions (\ref{WKB5})--(\ref{WKB6}) into Eq. (\ref{WKB3}) and the final result is: 
\begin{eqnarray}
f_{k,\, \pm}^{\,\prime} &=& g_{k,\,\pm} + {\mathcal F} f_{k,\, \pm},
\label{WKB7}\\
g_{k,\,\pm}^{\,\prime} &=& - k^2 \, f_{k,\, \pm} - {\mathcal F} \,g_{k,\,\pm}  \mp \,  \biggl(\frac{\overline{\lambda}^{\,\prime}}{\lambda}\biggr)\, k \, f_{k,\, \pm}.
\label{WKB8}
\end{eqnarray}
Equations (\ref{WKB7})--(\ref{WKB8}) have actually the same content of Eqs. (\ref{EqP5})--(\ref{EqP6}) and their solutions will be thoroughly discussed in the last part of this section and also in sec. \ref{sec4}. 

\subsection{General forms of the gauge power spectra}
 From the Fourier transform of the field operators 
(\ref{WKB5})--(\ref{WKB6}) 
\begin{eqnarray}
\widehat{E}_{i}(\vec{q},\tau) &=& - \,\sum_{\alpha= \pm}\biggl[ \varepsilon_{i}^{(\alpha)}(\hat{q}) \, g_{q,\,\alpha} \widehat{a}_{\vec{q},\,\alpha} +\varepsilon_{i}^{(\alpha)\ast}(-\hat{q}) \, g_{q,\,\alpha}^{\ast} \widehat{a}_{-\vec{q},\,\alpha}^{\dagger}\biggr],
\label{WKB9a}\\
\widehat{B}_{k}(\vec{p},\tau) &=& - \, i\, \epsilon_{i\,j\,k} \,\sum_{\alpha= \pm}\biggl[ p_{i} \, \varepsilon_{j}^{(\alpha)}(\hat{p}) \, f_{p,\,\alpha} \widehat{a}_{\vec{p},\,\alpha} + p_{i} \, \varepsilon_{j}^{(\alpha)\ast}(-\hat{p})\, f_{p,\,\alpha}^{\ast} \widehat{a}_{-\vec{p},\,\alpha}^{\dagger} \biggr],
\label{WKB9b}
\end{eqnarray}
 As a consequence the  two-point functions constructed from Eqs. (\ref{WKB9a}) and (\ref{WKB9b}) will consists of the symmetric contribution and of the corresponding antisymmetric part\footnote{The expectation values are  computed from Eqs. (\ref{WKB9a})--(\ref{WKB9b}), by recalling that $2 \varepsilon_{i}^{(+)}(\hat{k}) \varepsilon_{j}^{(-)}(\hat{k}) = [ p_{ij}(\hat{k}) - i \, \epsilon_{i j \ell} \, \hat{k}^{\ell}]$ where, as in Eq. (\ref{WKB4}), $p_{ij}(\hat{k}) = (\delta_{ij} - \hat{k}_{i}\hat{k}_{j})$ is the traceless projector and $\epsilon_{i j \ell}$ is the Levi-Civita symbol in three spatial dimensions. See, in this respect, the definitions of Eq. (\ref{WKB1}).}
\begin{eqnarray}
&& \langle \widehat{E}_{i}(\vec{k},\tau)\, \widehat{E}_{j}(\vec{p},\tau) \rangle = \frac{ 2 \pi^2 }{k^3} \biggl[\, P_{E}(k,\tau) \, p_{ij}(\hat{k}) 
+ i\,P_{E}^{(G)}(k,\tau)\,\epsilon_{i\, j\, \ell} \, \hat{k}^{\ell}\biggr] \, \delta^{(3)}(\vec{p} + \vec{k}),
\label{WKB10}\\
&& \langle \widehat{B}_{i}(\vec{k},\tau)\, \widehat{B}_{j}(\vec{p},\tau) \rangle = \frac{ 2 \pi^2 }{k^3} \biggl[\, P_{B}(k,\tau) \, p_{ij}(\hat{k}) 
+ i\,P_{B}^{(G)}(k,\tau)\,\epsilon_{i\, j\, \ell} \, \hat{k}^{\ell}\biggr] \, \delta^{(3)}(\vec{p} + \vec{k}).
\label{WKB11}
\end{eqnarray}
 In Eqs. (\ref{WKB10})--(\ref{WKB11}) $P_{E}(k,\tau)$ and $P_{B}(k,\tau)$ denote the hyperelectric and the hypermagnetic power spectra
whose explicit expression is given by:
 \begin{equation}
P_{E}(k,\tau) = \frac{k^{3}}{4 \pi^2} \biggl[ \bigl| g_{k,\,-}\bigr|^2 + \bigl| g_{k,\,+}\bigr|^2 \biggr], \qquad
P_{B}(k,\tau) = \frac{k^{5}}{4 \pi^2} \biggl[ \bigl| f_{k,\,-}\bigr|^2 + \bigl| f_{k,\,+}\bigr|^2 \biggr].
\label{WKB12}
\end{equation}
When either 
 $\overline{\lambda} \to 0$ or $\overline{\lambda}^{\prime} \to 0$  the anomalous coupling disappears from the Hamiltonian (\ref{WKB2}) and Eqs. (\ref{WKB7})--(\ref{WKB8}) imply that the hyperelectric and hypermagnetic mode functions have a common limit: if $f_{k}$ and $g_{k}$ denote the common solutions of Eqs. (\ref{WKB7})--(\ref{WKB8})
for $\overline{\lambda}\to 0$  we have that  $\lim_{\overline{\lambda} \to 0} f_{k,\,\pm}  = e^{-i \pi/4} f_{k}$ and  
$\lim_{\overline{\lambda} \to 0} g_{k,\,\pm}  = e^{i \pi/4} g_{k}$; the phase factor follows from the definition of the circular modes 
of Eq. (\ref{WKB1}).  In the limit $\overline{\lambda} \to 0$ the gyrotropic contributions appearing in Eqs. (\ref{WKB10}) and (\ref{WKB11}) 
\begin{equation}
P_{E}^{(G)}(k,\tau) =  \frac{k^{3}}{4 \pi^2} \biggl[  \bigl| g_{k,\,-}\bigr|^2 - \bigl| g_{k,\,+}\bigr|^2 \biggr],\qquad
P_{B}^{(G)}(k,\tau) =  \frac{k^{5}}{4 \pi^2} \biggl[ \bigl| f_{k,\,-}\bigr|^2  -  \bigl| f_{k,\,+}\bigr|^2\biggr]
\label{WKB13}
\end{equation}
will vanish. The superscript $(G)$ reminds that power spectra of Eq. (\ref{WKB13}) 
 determine the corresponding gyrotropies defined, respectively, by the 
 expectation values of the two pseudoscalar quantities $\langle \vec{B} \, 
 \cdot \vec{\nabla}\times \vec{B} \rangle$ and $\langle \vec{E} \, 
 \cdot \vec{\nabla}\times \vec{E} \rangle$. While the magnetic gyrotropies are gauge-invariant 
 (and have been originally introduced by Zeldovich in the context of the mean-field 
 dynamo theory \cite{zeld2,kaza,kraich}) the corresponding helicities (e.g. $\vec{{\mathcal Y}}\cdot \vec{E}$ and 
 $\vec{{\mathcal Y}}\cdot \vec{B}$) are not gauge-invariant and this is why we shall refrain 
 from using them\footnote{The  hypermagnetic gyrotropy 
has some advantages in comparison with the case of the Chern-Simons number density (i.e. $n_{CS} \propto \vec{Y}\cdot\vec{B}$). The difference of $n_{CS}$ at different times is always gauge-invariant\cite{CKN1}. However, at a fixed time, $n_{CS}$ (unlike the corresponding gyrotropy) is gauge-dependent. }.

Even if we mainly introduced the comoving power spectra, for the phenomenological applications what matters are not directly 
the comoving spectra of Eqs. (\ref{WKB12})--(\ref{WKB13}) but rather their physical counterpart. From the relations between the physical and the comoving fields introduced prior to Eqs. (\ref{EqP3})--(\ref{EqP5}) the physical power spectra are given by:
\begin{equation}
P_{X}^{(phys)}(k,\tau) = \frac{P_{X}(k,\tau)}{\lambda(\tau) \, a^4(\tau)},
\label{PXphys}
\end{equation}
where $P_{X}(k,\tau)$ generically denotes one of the comoving quantities 
listed in Eqs. (\ref{WKB12})--(\ref{WKB13}). Let us finally remind that the energy density of the gauge fields 
follows from the corresponding energy-momentum tensor 
derived from the action (\ref{action2}). Using Eqs. (\ref{WKB10})--(\ref{WKB11}) we can obtain $\langle \hat{\rho}_{Y} \rangle$.
To compare energy density of the parametrically amplified gauge fields
with the energy density of the background geometry we introduce the spectral energy density in critical units:
\begin{eqnarray}
\Omega_{Y}(k,\tau) &=& \frac{1}{\rho_{crit}} \frac{ d \langle \hat{\rho} \rangle}{d \ln{k}} = \frac{2}{3 H^2 \, M_{P}^2 a^4}  \biggl[ P_{E}(k,\tau) + P_{B}(k,\tau)\biggr]
\nonumber\\
&=& \frac{2}{3 H^2 \, M_{P}^2} \,\lambda\,\biggl[ P^{(phys)}_{E}(k,\tau) + P_{B}^{(phys)}(k,\tau)\biggr],
\label{OMY}
\end{eqnarray}
where we expressed $\Omega_{Y}(k,\tau)$ both in terms of the comoving and of the physical power spectra.
To guarantee the absence of dangerous backreaction effects $\Omega_{Y}(k,\tau)$ must always be subcritical throughout the various stages of the evolution and for all relevant scales; this requirement must be separately verified both during and after inflation.

\subsection{WKB estimates of the mode functions}
 We are now going to solve Eqs. (\ref{WKB7})--(\ref{WKB8})  by using the  WKB approximation for each of the two circular 
modes since the presence of the anomalous contribution 
slightly modifies the structure of the turning points\footnote{A similar technique  has been originally employed in the context of the polarized backgrounds of relic gravitons \cite{POLWKB}.}.  After combining Eqs. (\ref{WKB7}) and (\ref{WKB8}) the evolution of the hypermagnetic mode functions is given by:
\begin{equation}
f_{k,\, \pm}^{\prime\prime} + \biggl[ k^2 \pm k \frac{\overline{\lambda}^{\prime}}{\lambda} - \frac{\sqrt{\lambda}^{\prime\prime}}{\sqrt{\lambda}} \biggr] f_{k,\,\pm} =0,
\label{WKB14}
\end{equation}
which is ultimately analogous to the decoupled equation already discussed in Eq. (\ref{EqP7}). Having determined the solution of Eq. (\ref{WKB14}), the hyperelectric mode functions follow directly from Eq. (\ref{WKB7}) which is in fact a definition of $g_{k,\,\pm}$, i.e. $g_{k,\,\pm} = f_{k,\,\pm}^{\prime} - {\mathcal F} f_{k,\,\pm}$. For the present ends Eq. (\ref{WKB14}) can be viewed as: 
\begin{equation}
f_{k,\,\pm}^{\prime\prime}  + \biggl[ k^2 - \frac{w_{\pm}^{\prime\prime}}{w_{\pm}} \biggr] f_{\pm}=0,
\label{WKB15}
\end{equation}
where now $w_{\pm}(k,\tau)$ are two undetermined functions obeying 
\begin{equation}
\frac{w_{\pm}^{\prime\prime}}{w_{\pm}} = \mp \,k \, \frac{\overline{\lambda}^{\prime}}{\lambda} \, + \, \frac{\sqrt{\lambda}^{\prime\prime}}{\sqrt{\lambda}}.
\label{WKB16}
\end{equation}
Since the  left and right modes become of the order of $|w_{\pm}^{\prime\prime}/w_{\pm}|$  at different times,
 from Eq. (\ref{WKB15}) the hypermagnetic mode functions can be formally expressed as:
\begin{eqnarray}
f_{k,\,\pm}(k, \tau) &=& \frac{1}{\sqrt{2 k}} e^{- i \, k\, \tau} , \qquad\qquad k^2 \gg\biggl| \frac{w_{\pm}^{\prime\prime}}{w_{\pm}} \biggr|,
\label{WKB17}\\ 
f_{k,\,\pm}(k, \tau) &=& {\mathcal A}_{k, \,\pm} w_{\pm}(k,\tau) + {\mathcal B}_{k,\, \pm} w_{\pm}(k,\tau) \int^{\tau} \frac{ d\,\tau^{\prime}}{w^2_{\pm}(k,\tau^{\prime})}, \qquad\qquad k^2 \ll \biggl| \frac{w_{\pm}^{\prime\prime}}{w_{\pm}} \biggr|.
\label{WKB18}
\end{eqnarray}
The values of ${\mathcal A}_{k,\pm}$ and ${\mathcal B}_{k,\pm}$ are determined by matching the solutions (\ref{WKB17})--(\ref{WKB18}) at the turning point $\tau_{ex}$ when a given scale exits the effective horizon associated with $|w_{\pm}^{\prime\prime}/w_{\pm}|$. The
 second turning point (denoted by $\tau_{re}$)
corresponds to the moment at which the given scale will reenter 
the effective horizon\footnote{In what follows we shall be interested in the
general expressions of the gauge power spectra prior to $\tau_{re}$.
The discussion of the late-time power spectra will be postponed to 
section \ref{sec5}.}.  The explicit expressions of $f_{k,\,\pm}(\tau)$ and of $g_{k, \pm}(\tau)$ valid for $\tau < \tau_{re}$ are:
\begin{eqnarray}
f_{\pm}(k, \tau) &=& \frac{w_{\pm}(\tau)}{w_{\pm,\, ex}} \biggl\{ f_{\pm,\, ex} + \biggl[g_{\pm,\, ex} 
+ \biggl({\mathcal F}_{ex} - {\mathcal G}_{\pm,\, ex}\biggr)\, f_{\pm,\, ex} \biggr] \,\,{\mathcal I}_{\pm}(\tau_{ex}, \, \tau) \biggr\},
\nonumber\\
g_{\pm}(k,\tau) &=& \frac{w_{\pm,\, ex}}{w_{\pm}(\tau)} \biggl[g_{\pm,\, ex} + \biggl({\mathcal F}_{ex} - {\mathcal G}_{\pm,\, ex}\biggr)\, f_{\pm,\, ex} \biggr]
\label{WKB19}\\
&+& \frac{w_{\pm}(\tau)}{w_{\pm,\, ex}} \bigl({\mathcal G}_{\pm} - {\mathcal F} \bigr) \biggl\{f_{\pm,\, ex} + \biggl[g_{\pm,\, ex} + \biggl({\mathcal F}_{ex} - {\mathcal G}_{\pm,\, ex}\biggr)f_{\pm,\, ex}  \biggr]
{\mathcal I}_{\pm}(\tau_{ex}, \, \tau)\biggr\},
\label{WKB20}
\end{eqnarray}
where ${\mathcal G}_{\pm}= w_{\pm}^{\prime}/w_{\pm}$. For the sake of conciseness  we wrote $f_{\pm,\, ex} = f_{k,\,\pm}(\tau_{ex})$, $g_{\pm,\, ex} = g_{k,\,\pm}(\tau_{ex})$ with the caveat 
that $\tau_{ex}$ is actually different for the left and right modes; the same 
notation has been also adopted for ${\mathcal G}_{\pm, \, ex}$ and for ${\mathcal F}_{ex}$. Finally the integrals ${\mathcal I}_{\pm}(\tau_{ex}, \, \tau)$ appearing in Eq. (\ref{WKB20}) are defined as:
\begin{equation}
{\mathcal I}_{\pm}(\tau_{ex}, \, \tau) = w_{\pm,\, ex}^2 \, \, \int_{\tau_{ex}}^{\tau} \frac{d \, \tau^{\prime}}{w_{\pm}^2(\tau^{\prime})}, \qquad 
\qquad {\mathcal G}_{\pm}= \frac{w_{\pm}^{\prime}}{w_{\pm}}.
\label{WKB21}
\end{equation}
Since the left and right polarization have  different turning points, the two polarizations will hit the effective horizon at slightly 
different times $\tau_{\pm} = - ( 1 + \epsilon_{\pm})/k$ with $| \epsilon_{\pm}(k) | \ll 1$
provided $\sqrt{\lambda}^{\,\prime\prime}/\sqrt{\lambda} = {\mathcal O}(\tau^{-2})$ 
and $\overline{\lambda}^{\prime}/\lambda = {\mathcal O}(\tau^{-1-\alpha})$ with $\alpha > 0$. In the case of Eq. (\ref{EqP12a})
the WKB estimates of the mode functions for the left and right polarizations can be expressed as:
\begin{eqnarray}
f_{k,\,\pm}(\tau) &\simeq& \frac{e^{i k \tau_{\pm}}}{\sqrt{2 k}} \biggl[ \frac{ ( - k \tau)^{\gamma}}{( 1 + \epsilon_{\pm})^{\gamma}} + \frac{( 1 + \epsilon_{\pm})^{\gamma}}{(1 - 2 \gamma)}  ( - k \tau)^{ 1- \gamma}\biggr],
\label{WKB22}\\
g_{k,\,\pm}(\tau) &\simeq& i \, \sqrt{\frac{k}{2}} \biggl[ ( - k \tau)^{\gamma} \, ( 1 + \epsilon_{\pm})^{\gamma} - \frac{i}{2 \gamma +1} \frac{(- k \, \tau)^{\gamma+1}}{( 1 + \epsilon_{\pm})^{\gamma}}\biggr],
\label{WKB23}
\end{eqnarray}
where $\epsilon_{\pm} = \epsilon_{\pm}(k,\beta)$. The same analysis (with different results) can be 
easily applied to different situations, such as the one of Eq. (\ref{EqP14a}).

\subsection{WKB estimates of the power spectra}
The hypermagnetic power spectrum obtained from Eqs. (\ref{WKB12}) and (\ref{WKB22})  is 
different depending upon the value of $\gamma$: if $\gamma >1/2$  and $|\epsilon_{\pm}(k,\beta) |<1$ the second term at the right 
hand side of Eq. (\ref{WKB22}) dominates while the first term gives the dominant contribution for $\gamma < 1/2$. The hypermagnetic power 
spectrum is therefore given by
\begin{equation}
P_{B}(k,\tau) \simeq \frac{a^4\, H^4}{2 \pi^2} |k \,\tau|^{5 - | 2 \gamma -1|} \biggl[ 1 + {\mathcal O}( \epsilon_{+} + \epsilon_{-}) \biggr].
\label{WKB24}
\end{equation} 
The WKB estimates leading to Eq. (\ref{WKB24}) (and to the other 
results of this section) are accurate for the slopes of the power 
spectra while the amplitudes are determined up to
${\mathcal O}(1)$ numerical factors, as we shall see in the following section; 
this is why in Eq. (\ref{WKB24}) we used a sign of approximate equality.
The same analysis leading to Eq. (\ref{WKB24}) can be repeated in the case of the hyperelectric power spectrum:
\begin{equation}
P_{E}(k,\tau) \simeq \frac{a^4\, H^4}{2 \pi^2} | k\tau|^{4 - 2 \gamma} \biggl[ 1 + {\mathcal O}( \epsilon_{+} + \epsilon_{-})\biggr].
\label{WKB25}
\end{equation} 
Note, in this case, the absence of absolute values in the exponent. Inserting the mode 
functions (\ref{WKB23}) in Eq. (\ref{WKB13}) 
the order of magnitude of the gyrotropic components can be easily determined:
\begin{eqnarray}
P_{B}^{(G)}(k,\tau) &\simeq& a^4 H^4 {\mathcal O}(\epsilon_{+} - \epsilon_{-}) |k \,\tau|^{5 - | 2 \gamma -1|},
\label{WKB26a}\\
P_{E}^{(G)}(k,\tau) &\simeq& a^4 H^4 {\mathcal O}(\epsilon_{+} - \epsilon_{-}) |k \,\tau|^{4 - 2\gamma}.
\label{WKB26b}
\end{eqnarray}
The gauge power spectra of Eqs. (\ref{WKB24})--(\ref{WKB25}) and 
(\ref{WKB26a})--(\ref{WKB26b}) have been obtained in the case when the 
gauge coupling increases during inflation (see Eq. (\ref{EqP12a})). The same 
analysis can be repeated  when the gauge coupling decreases 
during inflation, as suggested by Eq. (\ref{EqP14a}). The results 
for the hypermagnetic and for the hyperelectric power spectra will be, this time:
\begin{eqnarray}
\widetilde{\,P\,}_{B}(k,\tau) &\simeq& \frac{a^4\, H^4}{2 \pi^2} | k\tau|^{4 - 2 \widetilde{\gamma}} \biggl[ 1 + {\mathcal O}( \widetilde{\epsilon}_{+} + \widetilde{\epsilon}_{-})\biggr].
\label{WKB27a}\\
\widetilde{\,P\,}_{E}(k,\tau) &\simeq& \frac{a^4\, H^4}{2 \pi^2} |k \,\tau|^{5 - | 2 \widetilde{\gamma} -1|} \biggl[ 1 + {\mathcal O}( \widetilde{\epsilon}_{+} + \widetilde{\epsilon}_{-}) \biggr].
\label{WKB27b}
\end{eqnarray}
In Eqs. (\ref{WKB27a})--(\ref{WKB27b})  we used the tilde to distinguish 
the power spectra obtained in the case of decreasing coupling from the 
ones associated with the increasin coupling. With the same notation 
the gyrotropic spectra are given by: 
\begin{eqnarray}
\widetilde{\,P\,}_{B}^{(G)}(k,\tau) &\simeq& a^4 H^4 {\mathcal O}(\widetilde{\epsilon}_{+} - \widetilde{\epsilon}_{-}) |k \,\tau|^{4 - 2\widetilde{\gamma}},
\label{WKB28a}\\
\widetilde{\,P\,}_{E}^{(G)}(k,\tau) &\simeq& a^4 H^4 {\mathcal O}(\widetilde{\epsilon}_{+} - \widetilde{\epsilon}_{-}) |k \,\tau|^{5 - | 2 \widetilde{\gamma} -1|}.
\label{WKB28b}
\end{eqnarray}
The results of the WKB approximation will be corroborated by a number of examples in the following section. Even if we shall 
preferentially treat the case of increasing coupling, we note that 
the case of decreasing coupling can be formally recovered from 
the one where the gauge coupling increases. For instance 
if $\gamma \to \widetilde{\gamma}$ we have that the gauge 
spectra of Eqs. (\ref{WKB26a})-(\ref{WKB26b}) turn into the ones 
of Eqs. (\ref{WKB27a})-(\ref{WKB27b}) with the caveat 
that $P_{B}(k,\tau) \to \widetilde{\, P\,}_{E}(k,\tau)$ and 
that $P_{E}(k,\tau) \to \widetilde{\,P\,}_{B}(k,\tau)$. This is, after all, a direct consequence of the duality symmetry \cite{NINEaa,TENaa,MGJ}.

\renewcommand{\theequation}{4.\arabic{equation}}
\setcounter{equation}{0}
\section{Explicit examples}
\label{sec4}
The auxiliary equation (\ref{WKB16}) will be solved in a number of explicit cases. 
The obtained solutions will be analyzed in 
the large-scale limit and, in this way, the WKB
power spectra deduced at the end of the previous section will be recovered. 
For a direct solution  it is practical to introduce a new time coordinate
 (conventionally referred to as the $\eta$-time) by positing that $d \tau = N \, d\eta$; in the $\eta$-parametrization  Eq. (\ref{WKB16}) becomes:  
\begin{equation}
\ddot{Z}_{\pm} = \mp k \, \frac{\overline{\lambda}^{\prime}}{\lambda}  \, N^2 \, Z_{\pm} + \frac{\ddot{b}}{b} \, Z_{\pm}, \qquad\qquad Z_{\pm} = \frac{w_{\pm}}{\sqrt{N}}, \qquad \qquad b = \sqrt{\frac{\lambda}{N}},
\label{EX1}
\end{equation}
where the overdot\footnote{It is also common to employ the overdot to denote a derivation with respect to the cosmic time coordinate. To avoid confusions the two notations will never be used in the same context.} now denotes a derivation with respect to $\eta$; $Z_{\pm}$ and $b$ are the rescaled versions of $w_{\pm}$ and $\sqrt{\lambda}$ 
respectively. The form Eq. (\ref{EX1}) can be further simplified by choosing an appropriate form for $N(\tau)$. Since $N$ is, by definition, a real quantity we must have  $N^2>0$;
this means, in particular, that if $\overline{\lambda}^{\prime}/\lambda <0$ it is natural to  posit $(d\eta/d\tau)^2 = N^{-2} \propto - \overline{\lambda}^{\prime}/\lambda$. Conversely if $\overline{\lambda}^{\prime}/\lambda >0$ we would instead choose $(d\eta/d\tau)^2 = N^{-2} \propto \overline{\lambda}^{\prime}/\lambda$. 

\subsection{Solutions of the auxiliary equation}
In Eqs. (\ref{EqP12a})--(\ref{EqP12b})  $\lambda$ and $\overline{\lambda}$ evolve during an inflationary stage of expansion without the constraint of being equal so that the combinations appearing in Eq. (\ref{WKB16}) turn out to be:
\begin{equation}
\frac{\overline{\lambda}^{\prime}}{\lambda} = - \frac{b_{0}}{\tau_{1}} \biggl(- \frac{\tau}{\tau_{1}}\biggr)^{-1-\alpha}, \qquad\qquad \frac{\sqrt{\lambda}^{\prime\prime}}{\sqrt{\lambda}} = \frac{\gamma (\gamma -1)}{\tau^2},
\label{EX2a}
\end{equation}
where $b_{0}$ and $\alpha$ have been introduced and they are:
\begin{equation}
b_{0} = 2 \beta \biggl(\frac{\overline{\lambda}_{2}}{\lambda_{1}}\biggr) \biggl(\frac{\tau_{1}}{\tau_{2}}\biggr)^{2 \beta},\qquad \alpha = 2 (\gamma- \beta).
\label{EX2b}
\end{equation}
The explicit form of  Eqs. (\ref{EX2a})--(\ref{EX2b}) determines the mutual 
relation between the $\eta$-parametrization and the conformal time\footnote{Note that $b_{0}$ has been introduced in Eq. (\ref{EX2b}) while now we also defined $\overline{b}_{0} =(2\, b_{0})/(1-\alpha)$.}  so that, after simple algebra, 
Eq. (\ref{EX1}) becomes:
\begin{equation}
\ddot{Z}_{\pm} \mp q^2 Z_{\pm} - \frac{\nu^2 - 1/4}{\eta^2} Z_{\pm} =0, \qquad \nu = \biggl| \frac{2 \gamma -1}{1 - \alpha} \biggr|,
\qquad q^2 = \frac{b_{0} k }{\tau_{1}} = \frac{\overline{b}_{0}\,k}{\eta_{1}}.
\label{EX3}
\end{equation}
From the relation between $\eta$ and the $\tau$ 
 \begin{equation}
d \eta = \frac{d\tau}{N(\tau)} \qquad \Rightarrow \qquad \biggl(- \frac{\eta}{\eta_{1}}\biggr) = \biggl( - \frac{\tau}{\tau_{1}}\biggr)^{(1-\alpha)/2}, \qquad \eta_{1} = \frac{2 \tau_{1}}{(1-\alpha)},
\label{EX4}
\end{equation}
and depending on the convenience, Eqs. (\ref{EX2a})--(\ref{EX4}) give the explicit form of $N(\tau)$ or $N(\eta)$:
\begin{equation}
N(\tau) = (-\tau/\tau_{1})^{(1+\alpha)/2} \qquad \Rightarrow \qquad N(\eta) = (-\eta/\eta_{1})^{(1+\alpha)/(1-\alpha)}, \qquad \alpha \neq 1.
\label{EX4a}
\end{equation}
From Eqs. (\ref{EX1}) and (\ref{EX4a}) we can deduce $b(\eta)$ and, ultimately, the explicit form of Eq. (\ref{EX3}). All 
in all the solutions of Eq. (\ref{EX3}) in the $\eta$-parametrization are:
\begin{eqnarray}
Z_{+}(q, \eta) &=& \sqrt{ - q \eta} \biggl[ {\mathcal C}_{q,\,+} \, I_{\nu}( - q \eta) + {\mathcal D}_{q,\,+} \, K_{\nu}( - q \eta) \biggr],
\label{EX5}\\
Z_{-}(q,\eta) &=& \sqrt{ - q \eta} \biggl[ {\mathcal C}_{q,\,-} \, J_{\nu}( - q \eta) + {\mathcal D}_{q,\,-} \, Y_{\nu}( - q \eta) \biggr].
\label{EX6}
\end{eqnarray}
In Eq. (\ref{EX5})  $I_{\nu}(- q\eta)$ and $K_{\nu}(-q\eta)$ are the modified Bessel functions while in Eq. (\ref{EX6})  $J_{\nu}(- q\eta)$ and $Y_{\nu}(- q \eta)$ denote the ordinary Bessel functions (see e. g. \cite{MM1,MM2}). It is interesting to remark, at this point, that in the dual case (see Eq. (\ref{EqP14a}) and discussion therein)
the explicit expression of the auxiliary equation (\ref{EX3}) 
has a similar form: 
\begin{equation}
\ddot{\widetilde{\,Z\,}}_{\pm}  \pm \widetilde{\,q\,}^2 \, \widetilde{\,Z\,}_{\pm} - \frac{\widetilde{\,\nu\,}^2 - 1/4}{\eta^2} \widetilde{\,Z\,}_{\pm} =0, \qquad \widetilde{\,\nu\,} =  \frac{2 \widetilde{\gamma} + 1}{\bigl|1 - \widetilde{\,\alpha\,}\bigr|},
\qquad \widetilde{\,q\,}^2 = \frac{\widetilde{\,b\,}_{0} k }{\tau_{1}}, 
\label{EX3a}
\end{equation}
where this time $\widetilde{\, b\,}_{0}$ and $\widetilde{\,\alpha\,}$ are:
\begin{equation}
\widetilde{\, b\,}_{0} = 2 \widetilde{\beta} \biggl(\frac{\tau_{2}}{\tau_{1}}\biggr)^{2 \widetilde{\beta}} \, \biggl(\frac{\overline{\lambda}_{2}}{\lambda_{1}}\biggr), \qquad \widetilde{\,\alpha\,} = 2 (\widetilde{\,\beta\,} - \widetilde{\,\gamma\,}).
\label{EX3b}
\end{equation}
By comparing Eqs. (\ref{EX2a}) and (\ref{EX3a}) it is clear that $\nu \neq \widetilde{\nu}$.
Owing to the different form of Eq. (\ref{EX3a}) the solutions (\ref{EX5})--(\ref{EX6}) are:
\begin{eqnarray}
\widetilde{\,Z\,}_{+}(\widetilde{q}, \eta) &=& \sqrt{ - \widetilde{q} \eta} \biggl[ \widetilde{\,{\mathcal C}\,}_{\widetilde{q},\,+} \, J_{\widetilde{\nu}}( - \widetilde{q} \eta) + \widetilde{{\mathcal D}}_{\widetilde{q},\,+} \, Y_{\widetilde{\nu}}( - \widetilde{q} \eta) \biggr],
\label{EX5a}\\
\widetilde{\,Z\,}_{-}(\widetilde{q},\eta) &=& \sqrt{ - \widetilde{q} \eta} \biggl[ \widetilde{\,{\mathcal C}\,}_{\widetilde{q},\,-}\, I_{\widetilde{\nu}}( - \widetilde{q} \eta) + \widetilde{\,{\mathcal D}\,}_{\widetilde{q},\,-} \, K_{\widetilde{\nu}}( - \widetilde{q} \eta) \biggr], 
\label{EX6a}
\end{eqnarray}
where, as in Eqs. (\ref{EX5})--(\ref{EX6}) we introduced the appropriate Bessel functions.
To avoid digressions the full expressions of the hypermagnetic and hyperelectric mode functions 
can be found in Eqs. (\ref{APPA1})--(\ref{APPA2}) and (\ref{APPA3})--(\ref{APPA4}).
For the explicit evaluations of the power spectra the expressions of the hypermagnetic and hyperelectric mode functions  should be computed
for typical wavelengths larger than the effective horizon and this discussion may be found in appendix 
\ref{APPB}. In what follows we shall concentrate on the most relevant physical aspects and 
encourage the reader to consult the appendices for the technical aspects of the problem. 

\subsection{The exit of the left and right modes}
In view of the large-scale limit of the power spectra it is useful to remind that the exit of a given circular mode is fixed by the equation:
\begin{equation}
k^2 \simeq \mp k \frac{\overline{\lambda}^{\,\prime}}{\lambda} + \frac{\sqrt{\lambda}^{\,\prime\prime}}{\sqrt{\lambda}}.
\label{EX8}
\end{equation}
Equations (\ref{EqP12a})--(\ref{EqP12b}) and Eqs. (\ref{EX2a})--(\ref{EX2b}) imply, in the case of increasing coupling, that the explicit form of Eq. (\ref{EX8}) is:
\begin{equation}
(- k\tau)^2 \simeq \pm ( - k \tau)^{1-\alpha} \, x_{1}^{\alpha} \, b_{0} + \gamma (\gamma -1),
\label{EX9}
\end{equation}
where $x_{1} = k \tau_{1}$ and $\tau_{1}$ approximately denotes, by definition, the end of the inflationary phase.
 For the scales relevant for the present problem $x_{1}$ is so small that the limit $b_{0} x_{1} \ll 1$ is always verified in spite of the value of $b_{0}$.  For a generic wavenumber $k$, assuming the standard post-inflationary thermal history, the actual value of $x_{1}$ is:
\begin{equation}
x_{1} = \frac{k}{a_{1}\, H_{1}} = 10^{-23.05}\,\, \biggl(\frac{k}{\mathrm{Mpc}^{-1}}\biggr)\, \biggl(\frac{r_{T}}{0.01}\biggr)^{-1/4}\,\,\biggl(\frac{h_{0}^2 \Omega_{R0}}{4.15\times 10^{-5}}\biggr)^{-1/4} \,\,\biggl(\frac{{\mathcal A}_{{\mathcal R}}}{2.41\times10^{-9}}\biggr)^{-1/4},
\label{EX10}
\end{equation}
where, as usual,  $\Omega_{R0}$ is the critical fraction of radiation in the concordance paradigm, $r_{T}$ is the tensor to  scalar 
ratio and ${\mathcal A}_{{\mathcal R}}$ is the amplitude of curvature inhomogeneities.  It follows 
from Eq. (\ref{EX10}) that for typical wavelengths ${\mathcal O}(\mathrm{Mpc})$ (and even much shorter) $x_{1}$ is as small as $10^{-23}$. From Eq. (\ref{EX2b}) $b_{0}$ cannot be too large even for quite extreme values of $\overline{\lambda}_{2}/\lambda_{1}$. Since $x_{1}^{\alpha} b_{0} \ll 1$ (provided $\alpha >0$)  the solution of Eq. (\ref{EX9}) is:
\begin{equation}
\tau_{\pm}(k) =  - \frac{1}{k} [ c_{0}(\gamma) + \epsilon_{\pm}(k, \beta,\gamma)], \qquad |\epsilon_{\pm}(k, \beta,\gamma)|\ll 1,
\label{EX11}
\end{equation}
where $c_{0}$ and $\epsilon_{\pm}(k, \beta,\gamma)$ follow by consistency with Eq. (\ref{EX7}):
\begin{equation}
c_{0}(\gamma) = \sqrt{\gamma (\gamma -1)} = {\mathcal O}(1), \qquad \epsilon_{\pm}(k,\beta,\gamma) = \pm \frac{b_{0}}{2\, c_{0}(\gamma)} \biggl[\frac{x_{1}}{c_{0}(\gamma)}\biggr]^{\alpha}.
\label{EX12}
\end{equation}
It is always possible to rescale the value of $c_{0}$  since it is just an ${\mathcal O}(1)$ contribution; 
 in practice, the turning points assume the form
\begin{equation}
\tau_{\pm} \simeq - \frac{1}{k} ( 1 + \epsilon_{\pm}), \qquad \epsilon_{\pm}(k, \alpha) = \pm \frac{b_{0}}{2} x_{1}^{\alpha}.
\label{EX13}
\end{equation}
which basically correspond to the WKB estimate of section \ref{sec3} (see discussion after Eq. (\ref{WKB21})). Even if $\epsilon_{\pm} =\epsilon_{\pm}(k, \alpha)$, for the sake of conciseness in the explicit expressions we shall neglect the dependence upon $k$ and $\alpha$ unless strictly necessary. Concerning the result of Eq. (\ref{EX13}) the following 
three comments are in order:
\begin{itemize}
\item{} the explicit expression of the turning points 
holds provided $\alpha \neq 1$ and $\alpha\neq 0$; 
\item{}  Eqs. (\ref{EX2a})--(\ref{EX2b}) imply that 
$\overline{\lambda}^{\prime}/\lambda$ and $\sqrt{\lambda}^{\prime\prime}/\sqrt{\lambda}$ 
both scale as $\tau^{-2}$ when $\alpha \to 1$;
\item{}  in the limit $\alpha \to 0$ Eqs. (\ref{EX12}) and (\ref{EX13}) imply  that $\epsilon_{\pm} \simeq b_{0}/2$
which may be larger than $1$ as long as $b_{0}>1$; for $ \alpha \to 0$ the structure of the turning points might then be altered in comparison with the results Eqs. (\ref{EX12})--(\ref{EX13}).
\end{itemize}
The above remarks suggest that, besides the case $ \alpha >0$, the limits $\alpha \to 1$ and $\alpha \to 0$
must be separately addressed. When $\alpha \to 1$, Eq. (\ref{EX9}) becomes:
\begin{equation}
(- k \tau^2) \simeq \pm x_{1} \, b_{0} + \gamma (\gamma -1).
\label{EX14a}
\end{equation}
Even if Eq. (\ref{EX14a}) implies a modified the structure of the turning points  the 
final results for the power spectra will be fully compatible with the WKB estimates. If 
 $\alpha \to 0$,  Eq. (\ref{EX9}) becomes:
\begin{equation}
( - k \tau)^2 \mp b_{0} (- k \tau) - \gamma (\gamma -1) \simeq 0.
\label{EX14b}
\end{equation}
As long as $b_{0} \ll 1$ the solution of Eq. (\ref{EX14b}) has again the form (\ref{EX13}).
However,  for $b_{0} \gg 1$ the solution of Eq. (\ref{EX14b}) will rather be $ \tau_{\pm} = \pm b_{0}/k$, as it follows 
by neglecting the third term at the right-hand side of Eq. (\ref{EX14b}). In this limit 
$b_{0}$ may affect the overall amplitudes while the slopes of the gauge power spectra do coincide, as we shall see, 
 with the ones deduced in the original WKB approximation.
It follows from the above considerations that  the solutions 
of the auxiliary equations (\ref{EX5})--(\ref{EX6}) and (\ref{EX5a})--(\ref{EX6a}) must always be evaluated in the small argument limit when the relevant modes are larger than the effective horizon.
Since this point might not be immediately obvious we note that from Eqs. (\ref{EX2b})--(\ref{EX3})  it is immediate to express $- q\eta$ in terms of $\tau$:
\begin{equation}
- q \eta = \frac{ 2 \sqrt{ x_{1} \, b_{0}}}{|1 - \alpha|} \biggl( - \frac{\eta}{\eta_{1}}\biggr) = c(z), \qquad\qquad c(z)=   \frac{ 2 \sqrt{ x_{1} \, b_{0}}}{|1 - \alpha|} \, z^{(1-\alpha)/2},\qquad z = \biggl( - \frac{\tau}{\tau_{1}}\biggr).
\label{EX7}
\end{equation}
It is therefore possible to work directly either with $c(z)$ or with $(- q \eta)$ depending on the 
convenience. 

\subsection{Scales of the problem}
\begin{figure}[!ht]
\centering
\includegraphics[height= 8cm]{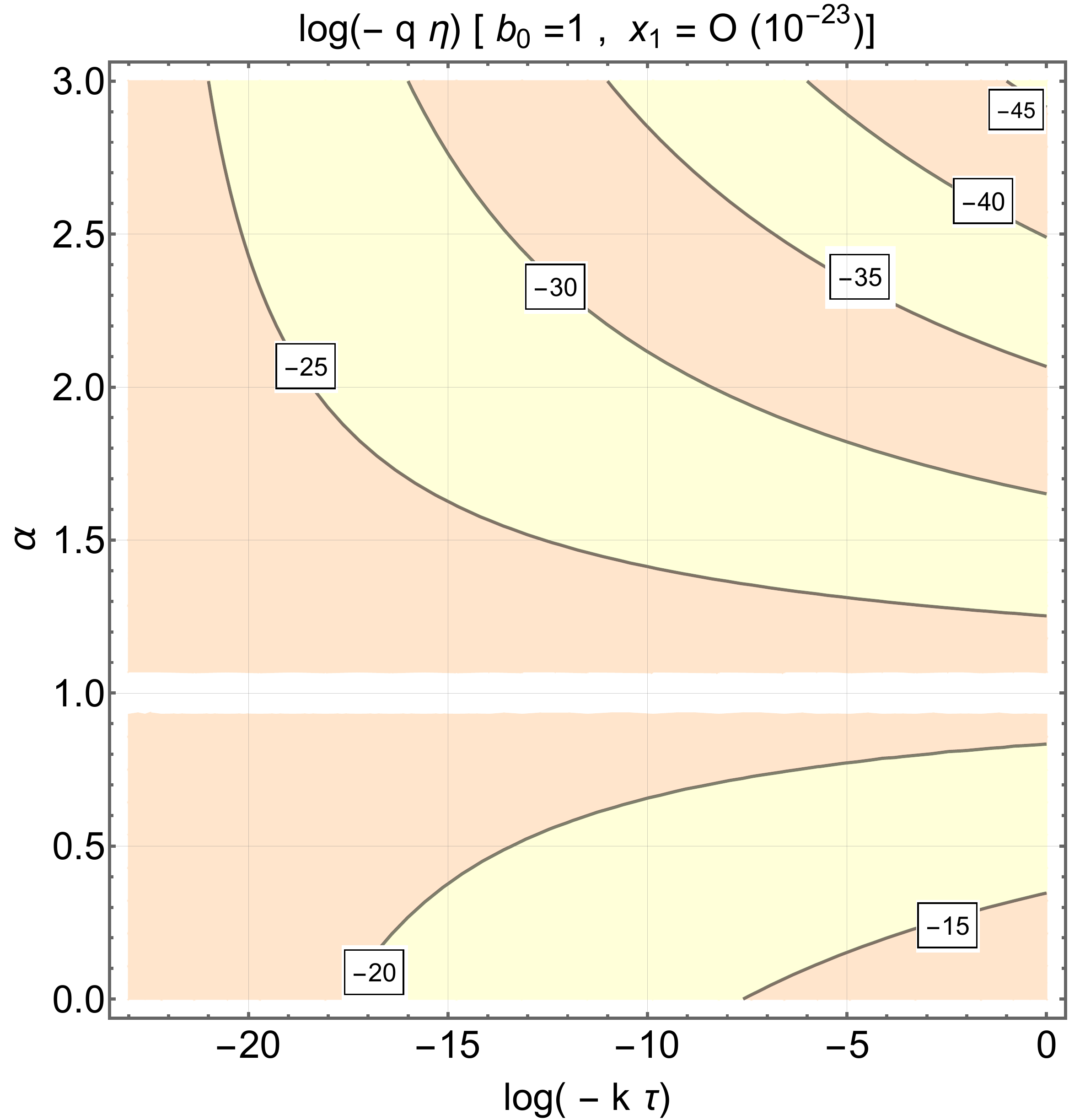}
\includegraphics[height= 8cm]{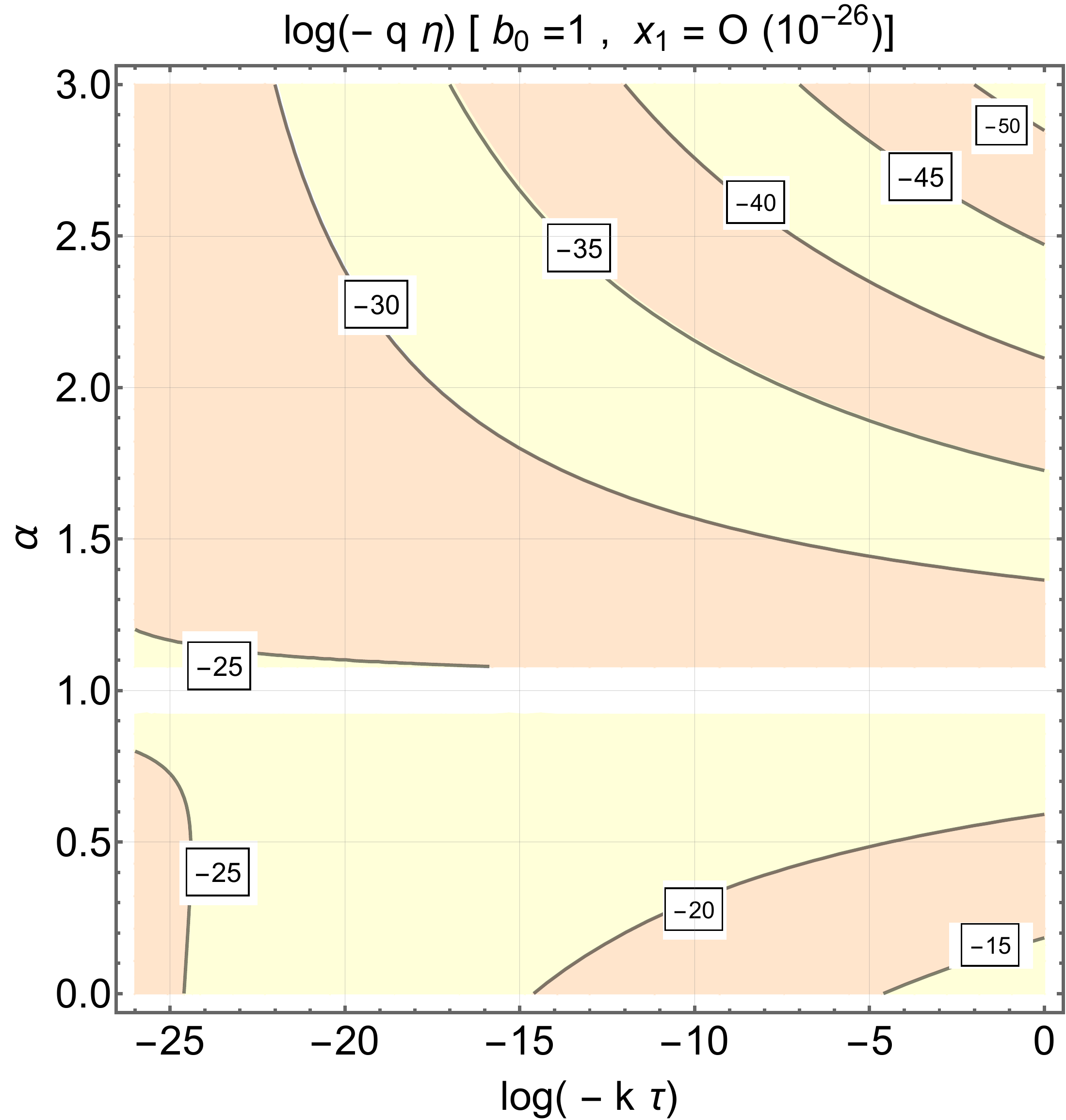}
\caption[a]{The relation given in Eq. (\ref{EX7}) is graphically illustrated for $x_1 ={\mathcal O}(10^{-23})$  (right plot) and for $ x_{1} = {\mathcal O}(10^{-26})$ (left plot). The 
labels on the various contours denote the common logarithm of $(-q \eta)$ while on the horizontal axis we report the common logarithm 
of $(-k \tau)$. When $k = {\mathcal O}(\mathrm{Mpc}^{-1})$ (or smaller) the regions where 
$(-q \eta) \ll 1$ coincide with the wavelengths that are larger than the Hubble radius (i.e.  $(-k \tau) \ll 1$). We recall, in this respect, that the connection between $x_{1}$ and $k$ follows from Eq. (\ref{EX10}). This means that if we want to compute the power spectra 
for typical wavelengths larger than the Hubble radius we can safely take the limit  $(-q \eta) \ll 1$ in the solutions 
of Eq. (\ref{EX3}). Two singular cases must be separately treated, namely $\alpha \to 1$ and $\alpha \to 0$. The details of this 
discussion can be found in appendix \ref{APPB}. }
\label{GGG1}      
\end{figure}

\begin{figure}[!ht]
\centering
\includegraphics[height= 8cm]{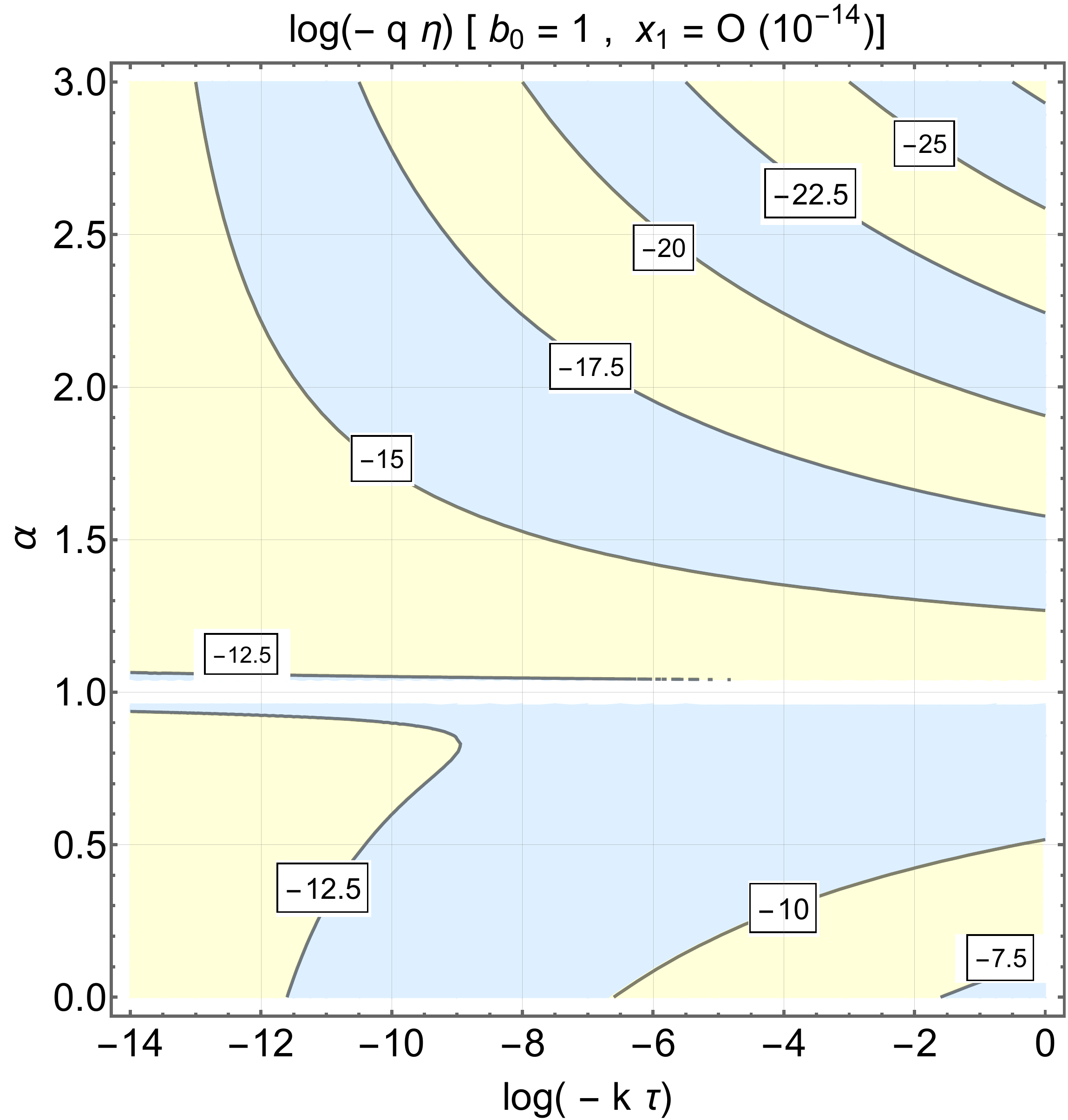}
\includegraphics[height= 8cm]{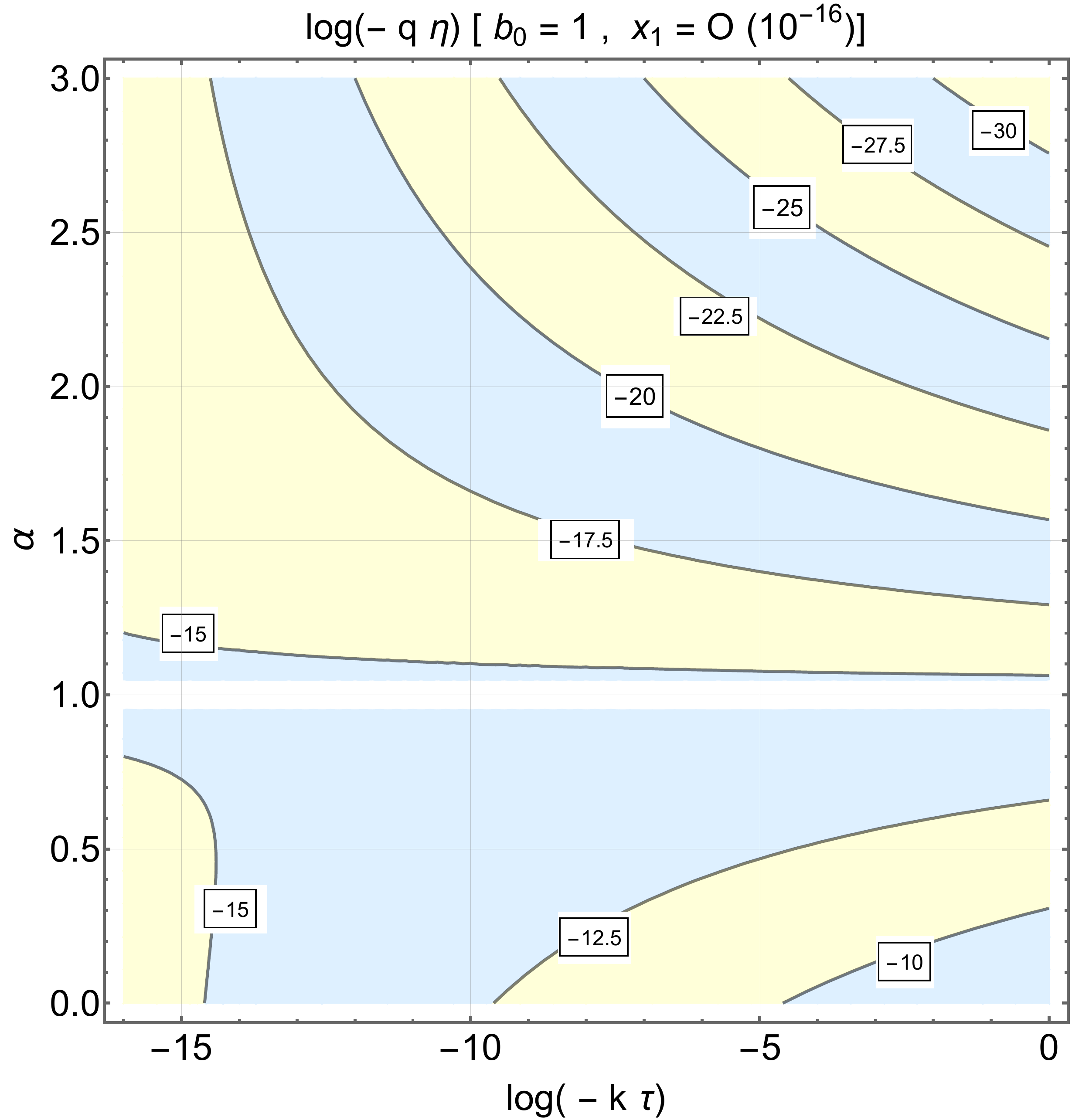}
\caption[a]{We graphically illustrate the relation (\ref{EX7}) for $x_1 ={\mathcal O}(10^{-14})$  (left plot) and for $ x_{1} = {\mathcal O}(10^{-16})$ (right plot). As in Fig. 
\ref{GGG1} the labels on the various contours denote the common logarithm of $(-q \eta)$ while on the horizontal axis we illustrated the common logarithm 
of $(-k \tau)$. The results of this figure show that when $k$ is much larger than  ${\mathcal O}(\mathrm{Mpc}^{-1})$ we still have that $(-q \eta) \ll 1$ provided $(-k \tau) \ll 1$. Even if the ranges of $(-q \eta)$ and $(- k \tau)$ are different,
 if we want to compute the power spectra for typical wavelengths larger than the Hubble radius we can take the limit  $(-q \eta) \ll 1$ in the solutions of Eq. (\ref{EX3}). We remind again that the explicit connection bewteen $x_{1}$ and $k$ can be found in Eq. (\ref{EX10}). }
\label{GGG2}      
\end{figure}

For the typical scales of the problem the condition $(- q \eta) \ll 1$ is always verified.
The magnetogenesis requirements involve typical wavenumbers
${\mathcal O}(\mathrm{Mpc}^{-1})$ so that the corresponding wavelengths reenter the effective horizon prior 
to equality:
\begin{eqnarray}
\frac{\tau_{k}}{\tau_{eq}} &=& \sqrt{2} \biggl(\frac{H_{0}}{k} \biggr) \biggl(\frac{\Omega_{M0}}{\sqrt{\Omega_{R0}}}\biggr) 
\nonumber\\
&=& 1.06 \times 10^{-2} \biggl(\frac{h_{0}^2 \Omega_{M0}}{0.1386} \biggr) 
\biggl(\frac{h_{0}^2 \Omega_{R 0}}{4.15\times 10^{-5}}\biggr)^{-1/2} \, \biggl( \frac{k}{\mathrm{Mpc}^{-1}}\biggr)^{-1}, 
\label{PH1}
\end{eqnarray}
where $\tau_{k}=1/k$ denotes the reentry time of a generic wavelength and $\tau_{eq}$ is the time of matter-radiation equality.  As long as $k = {\mathcal O}(\mathrm{Mpc}^{-1})$ the relation between $(- q\eta)$ and $(-k \tau)$ is illustrated in  Fig. \ref{GGG1} where the contours actually correspond to the common logarithm of $(-q \eta)$ when $(-k \tau)$ and $\alpha$ vary in their respective physical ranges. 
In Figs. \ref{GGG1} and \ref{GGG2} we illustrated different values of $x_{1}$. The rationale for these 
values can be understood by looking at Eqs. (\ref{EX10}) and (\ref{PH1}). In short the idea is the following.
\begin{itemize}
\item{} Let us start from the 
scales ${\mathcal O}(\mathrm{Mpc}^{-1})$ which are the ones relevant for magnetogenesis; 
according to Eq. (\ref{EX10}) we see that for $k = \mathrm{Mpc}^{-1}$ that $x_{1} = {\mathcal O}(10^{-23})$.
From Fig. \ref{GGG1} we see that $(-q \eta) < {\mathcal O}(10^{-20})$ this means that the solutions 
of the auxiliary equations (i.e. Eqs. (\ref{EX5})--(\ref{EX6})) can always be evaluated in the limit of 
small arguments i.e. for $(- q \eta) \ll 1$. 
\item{} The same conclusion holds when $x_{1} \ll 10^{-23}$: in the right plot of Fig. \ref{GGG1} 
we took a smaller value, i.e.  $x_{1} = {\mathcal O}(10^{-26})$. Also in this case
the results of Fig. \ref{GGG1} show that $(-q \eta)\ll 1$.
\item{} Finally the condition $(- q \eta) \ll 1$ is also verified for $x_{1} \gg 10^{-23}$: in Fig. \ref{GGG2} 
we illustrated the cases $x_{1} = {\mathcal O}(10^{-14})$ and ${\mathcal O}(10^{-16})$ and we can clearly 
see that $( - q \eta) \ll 1$ 
\end{itemize} 
 As expected on the basis of the general arguments given above, for $\alpha \to 0$ the relation between $(-q \eta)$ and $(- k\tau)$ is singular; the same is true 
when $\alpha \to 1$; this is why both cases shall be separately treated. If the typical wavelength is increased (i.e. for smaller $k$) the smallness of $(- q \eta)$ persists as it can be deduced from the right plot in Fig. \ref{GGG1} where $x_{1} = {\mathcal O}(10^{-26})$. The values of $b_{0}$ are not crucial and it can be directly checked that whenever $b_{0}$ increases from $1$ to $10^{6}$ the patterns of the relation illustrated in Fig. \ref{GGG1} are very similar.  

For the present purposes also the scales associated with the electroweak physics will be particularly important. 
In particular, as we shall see, the scales $k = {\mathcal O}(k_{ew})$ will reenter prior to symmetry 
breaking while the magnetogenesis scales will reenter after the electroweak phase trsnistion
Depending on the various parameters the bunch of wavenumbers corresponding to the electroweak scale will be ${\mathcal O}(10^{9}) \,\, \mathrm{Mpc}^{-1}$  
\begin{eqnarray}
k_{ew} &=& \biggl(\frac{8 \pi^3\, N_{eff} \, \Omega_{R0}}{45} \biggr)^{1/4} \, \sqrt{\frac{H_{0}}{M_{P}}} \, \, T_{ew}
\nonumber\\
&=& 2.6 \times 10^{9} \, \biggl(\frac{N_{eff}}{106.75}\biggr)^{1/4} \, \biggl(\frac{h_{0}^2 \Omega_{R0}}{4.15\times 10^{-5}}\biggr)^{1/4} \, \biggl(\frac{T_{ew}}{100\, \mathrm{GeV}}\biggr) \, \, \mathrm{Mpc}^{-1}.
\label{PH1aa}
\end{eqnarray}
The $\mathrm{Mpc}^{-1}$ units are not ideal but give an idea of the hiererchy of the scales. It is furthermore 
essential to bear in mind that $k_{ew}$ is the (comoving) electroweak wavenumber and not simply the 
Hubble rate at the electroweak time.
According to Eq. (\ref{PH1}) the scales corresponding to $k_{ew}$ will reenter the effective horizon when
$\tau_{k} = {\mathcal O}(10^{-11}) \tau_{eq}$, i.e. much earlier than the magnetogenesis wavelengths. From Eq. (\ref{EX10}) the value of $x_{1}$ corresponding to $k_{ew}$  
is therefore ${\mathcal O}(10^{-14})$. In Fig. \ref{GGG2} the plot at the left illustrated the relation between 
$(- q \eta)$ and $(- k \tau)$ for $x_{1} = {\mathcal O}(10^{-14})$; in the plot at the right we assumed 
a slightly smaller wavenumber with $x_{1} = {\mathcal O}(10^{-16})$. 

\subsection{Comparison with the WKB results}
The problem we are now facing is to compare the approximate results of the WKB 
approximation with the explicit examples following from the exact solutions 
of Eq. (\ref{EX3}). This analysis is actually quite lengthy but essential: the details of this comparison can be found in appendix \ref{APPB} and here we shall focus on the final results.  Let us start with the case of increasing 
gauge coupling; in this case the approximate results for the mode functions 
have been deduced in Eqs. (\ref{WKB22})--(\ref{WKB23}). The hypermagnetic and the hyperelectric power spectra have been 
instead computed in Eqs. (\ref{WKB24})--(\ref{WKB25}); the corresponding gyrotropic contributions are reported in Eqs. (\ref{WKB26a})--(\ref{WKB26b}). The solutions of the auxiliary equation (\ref{EX3}) can be classified from the values of $\alpha$ and $\gamma$. In particular the values of $\alpha$ and $b_{0}$ control the pseudoscalar coupling while the value of $\gamma$ controls the scalar coupling; the explicit expressions of the pump fields have been given in Eq. (\ref{EX2a}). 

The strategy followed in the comparison (see appendix \ref{APPB}) has been to solve the auxiliary equation (\ref{EX3})  and to compute the exact form of the mode functions. If we are interested in the spectra for typical wavelengths 
larger than the Hubble radius (i.e. $(-k\tau)\ll1$),  Figs. \ref{GGG1} and \ref{GGG2}  show that the solutions of the auxiliary equations must be  evaluated in their small argument limit (i.e. $(- q \eta) \ll 1$). If this limit is 
taken consistently the spectra can be explicitly computed and finally compared with the WKB results.
The essence of the comparison can be summarized as follows. 
\begin{itemize}
\item{} In the case $\alpha>0$ WKB results and the approach based on the auxiliary equation (\ref{EX3}) give coincident 
results. The difference between the two strategies is that 
the results based on Eq. (\ref{EX3}) capture with greater accuracy the numerical prefactor 
which is however not essential to estimate the power spectra at later time (see also, in this resepct, the results of section \ref{sec5}).
\item{} The explicit analysis of \ref{APPB} assumes
that the gauge coupling increases (see Eq. (\ref{EqP13a}) and discussion therein).
The WKB estimates obtained in the case of decreasing gauge coupling (see Eqs. (\ref{EqP14a}))
also match the results following from Eq. (\ref{EX3}). 
\item{} The discussion of the case $\alpha >0$ does not apply when $\alpha \to 0$ and $\alpha \to 1$. As 
discussed in connection with Figs. \ref{GGG1} and \ref{GGG2} in these two cases the relation 
between $(-q\eta)$ and $(-k\tau)$ gets singular. In these two separate situations the explicit solutions 
and the power spectra have been discussed in the last part of appendix \ref{APPB}.
The general conclusion of the WKB approximation presented in section \ref{sec3} also 
holds in the limit $\alpha \to 0$ and $\alpha\to 1$. 
\item{} The same discussion carried on in the case of increasing 
gauge coupling also applies, with some differences, to the case of decreasing gauge coupling 
of Eq. (\ref{EqP14a}). To avoid lengthy digressions a swift version of this analysis has been relegated to 
appendix \ref{APPC}; the explicit discussion merely reproduces the same steps of the one already presented in appendix \ref{APPB}.
\end{itemize}
Based on the results of appendix \ref{APPB} and \ref{APPC} we therefore claim that 
the slopes of the large-scale gauge spectra are not affected 
by the strength of the pseudoscalar terms that solely determine the gyrotropic contributions. 
This conclusion is quite relevant from the phenomenological viewpoint for two 
independent reasons. If we simply look at the hypermagnetic power spectra with the aim
of addressing the magnetogenesis requirements we can expect that the role 
of the pseudoscalar interactions (associated with $\alpha$ and $b_{0}$) will be completely 
negligible. Conversely different values of $\alpha$ will be essential to deduce the gyrotropic 
contributions that determine the baryon asymmetry of the Universe.  These two complementary 
expectations will be explicitly discussed in section \ref{sec5}.

\renewcommand{\theequation}{5.\arabic{equation}}
\setcounter{equation}{0}
\section{Late-time power spectra and some phenomenology}
\label{sec5}
The pseudoscalar couplings do not affect the slopes of the large-scale hypermagnetic and hyperelectric power spectra at early times while the gyrotropic components depend
(more or less severely) on the anomalous contributions. The impact of these results on the late-time power spectra
will now be considered. For the comparison of the late-time gauge spectra with the observables we shall assume that, 
after the end of inflation,  the radiation background dominates below a typical curvature scale $H_{r}$. In the simplest situation $H_{r}$ coincides with $H_{1}$.  In this situation, according to Eq. (\ref{PH1}) the different wavelengths reenter the effective horizon at different times during the radiation-dominated stage. The first aspect to appreciate is that the hypermagnetic and the gyrotropic power spectra computed when the gauge coupling flattens out do not exactly coincide the power 
spectra outside the horizon but can be obtained from them via a specific 
unitary transformation that depends on the rate of variation of the gauge coupling after inflation.

\subsection{Comparing late-time power spectra}
If the gauge coupling $e = \sqrt{4\pi/\lambda}$ increases and then flattens out,  Eqs. (\ref{EqP13a})--(\ref{EqP13b}) imply that 
 $f_{k\, \pm}(\tau)$ and $g_{k\, \pm}(\tau)$ can be expressed for $\tau > -\tau_{1}$  in terms of the corresponding mode functions computed for  $\tau \leq - \tau_{1}$ \cite{MMM3a}:
\begin{eqnarray}
f_{k\, \pm}(\tau) &=&  A_{f\,f}^{(\pm)}(z_{1},\, z,\, \delta) \,\,\overline{f}_{k,\, \pm} +  A_{f\,g}^{(\pm)}(z_{1}, \, z,\, \delta) \,\, \frac{\overline{g}_{k,\,\pm}}{k},
\label{FFF1}\\
g_{k\, \pm}(\tau) &=&  A_{f\,f}^{(\pm)}(z_{1}, \, z,\, \delta) k\, \overline{f}_{k,\, \pm} +  A_{g\,g}^{(\pm)}(z_{1}, \, z,\, \delta) \, \overline{g}_{k,\,\pm},
\label{FFF2}
\end{eqnarray}
where $\overline{f}_{k,\,\pm}$ and $\overline{g}_{k,\,\pm}$  denote the hypermagnetic and hyperelectric mode functions mode functions at the at the end of inflation (i.e. evaluated for $\tau = -\tau_{1}$). Since in Eq. (\ref{EqP13b}) we assumed that $\overline{\lambda}$ 
is constant for $\tau > - \tau_{1}$, it follows 
that the various coefficients appearing in Eqs. (\ref{FFF1})--(\ref{FFF2}) will not be different 
for the left and right modes, e.g. $A_{f\,f}^{(+)} = A_{f\,f}^{(-)} = A_{f\,f}(z_{1}, \, z,\, \delta)$ 
and similarly for all the other coefficients whose common expressions are:
\begin{eqnarray}
A_{f\, f}(z_1,\,z,\,\delta)  &=&\frac{\pi}{2} \sqrt{z_{1}\, z} \biggl[ Y_{\sigma -1}( z_{1}) J_{\sigma}(z) - J_{\sigma-1}(z_{1}) Y_{\sigma}(z) \biggr],
\nonumber\\
A_{f\, g}(z_1,\,z,\,\delta)  &=& \frac{\pi}{2} \sqrt{z_{1}\, z} \biggl[ J_{\sigma}( z_{1}) Y_{\sigma}(z) - Y_{\sigma}(z_{1}) J_{\sigma}(z) \biggr], 
\nonumber\\
A_{g\, f}(z_1,\,z,\,\delta) &=& \frac{\pi}{2} \sqrt{z_{1}\,z} \biggl[ Y_{\sigma - 1}( z_{1}) J_{\sigma -1}(z) - J_{\sigma-1}(z_{1}) Y_{\sigma-1}(z) \biggr],
\nonumber\\
A_{g\, g}(z_1,\,z,\,\delta)  &=& \frac{\pi}{2} \sqrt{z_{1}\, z} \biggl[ J_{\sigma}( z_{1}) Y_{\sigma-1}(z) - Y_{\sigma}(z_{1}) J_{\sigma-1}(z) \biggr].
\label{FFF3}
\end{eqnarray}
In Eq. (\ref{FFF3}), as usual, $J_{\sigma}(x)$ and $Y_{\sigma}(x)$ denote the standard Bessel functions \cite{MM1,MM2}; furthermore  $z_{1}$, $z$ and $\delta$ are defined as:
\begin{equation}
z_{1} = (\delta/\gamma) k \, \tau_{1}, \qquad z = k \tau + k \tau_{1}(1 + \delta/\gamma), \qquad \sigma= \delta +1/2.
\label{FFF4}
\end{equation}
 An expression analogous to Eq. (\ref{FFF3}) can be easily derived in the case of decreasing 
gauge coupling and can be found in appendix \ref{APPC}. It can be explicitly verified that Eqs. (\ref{FFF1}) and (\ref{FFF2}) obey the Wronskian normalization. Furthermore, from Eq. (\ref{FFF4}) we have that for $\tau = -\tau_{1}$, $z(-\tau_{1}) = z_{1}$ and, consequently,
$A_{f\,f}(z_{1}, z_{1}) = A_{g\, g}(z_{1}, z_{1}) =1$  while $A_{g\,f}(z_{1}, z_{1}) = A_{f\, g}(z_{1}, z_{1}) =0$. It is finally relevant to appreciate that, in the limit $\delta \ll 1$, 
 $A_{f\,f}(z_{1}, z) = A_{g\, g}(z_{1}, z) \to \cos{k (\tau + \tau_{1})}$ and $A_{f\,g}(z_{1}, z) = - A_{g\, f}(z_{1}, z) \to \sin{k (\tau + \tau_{1})}$. 
 
 The late-time power spectra following from Eqs. (\ref{FFF1})--(\ref{FFF2}) do not coincide 
 with the early-time power spectra evaluated in the large-scale limit, as it is sometimes 
 suggested. In the case Eq. (\ref{FFF3}) 
 the obtained expressions get simpler if we observe that 
 \begin{eqnarray}
&& \bigl| \,A_{f\,f}(z_{1}, \, z,\, \delta) \overline{f}_{k,\, \pm}\,\bigr|^2 \ll  \biggl| A_{f\,g}(z_{1}, \, z,\, \delta) \, \frac{\overline{g}_{k,\,\pm}}{k}\biggr|^2,
 \label{FFF5}\\
&& \bigl| \,A_{g\,f}(z_{1}, \, z,\, \delta) k\, \overline{f}_{k,\, \pm}\,\bigr|^2 \ll \biggl| \,A_{g\,g}(z_{1}, \, z,\, \delta) \, \overline{g}_{k,\,\pm}\,\biggr|^2. 
\label{FFF6}
 \end{eqnarray}
 Thanks to Eqs. (\ref{FFF5})--(\ref{FFF6}) all the late-time comoving spectra easily follow. In view of the applications the following three relevant results 
will be mentioned:
  \begin{eqnarray}
 P_{B}(k,\tau) &=& a_{1}^4 H_{1}^4 Q(\alpha, \gamma, \delta) \biggl(\frac{k}{a_{1} H_{1}}\biggr)^{4 - 2 \gamma - 2 \delta} \, 
 F_{B}(k \tau, \delta),
 \label{FFF7}\\ 
 P_{E}(k,\tau) &=& a_{1}^4 H_{1}^4 Q(\alpha, \gamma, \delta) \biggl(\frac{k}{a_{1} H_{1}}\biggr)^{4 - 2 \gamma - 2 \delta} \, 
 F_{E}(k \tau, \delta),
 \label{FFF8}\\
 P^{(G)}_{B}(k,\tau) &=& a_{1}^4 H_{1}^4 Q^{(G)}(\alpha, \gamma, \delta, b_{0}) \biggl(\frac{k}{a_{1} H_{1}}\biggr)^{4 - \alpha -2 \gamma - 2 \delta} \, F_{B}(k \tau, \delta).
\label{FFF9}
\end{eqnarray}
Since we want to be able to take smoothly the limit where the post-inflationary 
gauge coupling is completely frozen (i.e. $\delta \to 0$), 
the explicit expressions of $F_{B}(k\, \tau, \delta)$ and $F_{E}(k\, \tau, \delta)$ will be evaluated in the regime $ 0\leq \delta< 1/2$:
\begin{equation}
F_{B}(k\, \tau, \delta) =  |k\, \tau| J^2_{\delta+1/2}(k\, \tau),\qquad F_{E}(k\, \tau, \delta) =  |k\, \tau| J^2_{\delta-1/2}(k\, \tau).
\label{FFF10}
\end{equation}
Equations (\ref{FFF7}) and (\ref{FFF8}) are consistent with the main findings of this analysis namely the fact that for any $\alpha \geq 0$ the slopes of the gauge power spectra do not depend upon $\alpha$ which instead appears in the spectral slope of Eq. (\ref{FFF9}). Similarly $Q(\alpha, \gamma, \delta)$ and $Q^{(G)}(\alpha, \gamma, \delta, b_{0})$ do depend on $\alpha$ but not on $k$. Provided $\alpha \neq 0$ and $\alpha \neq 1$ we have, in particular, 
\begin{eqnarray}
Q_{B}(\alpha, \gamma, \delta) &=& \frac{2^{ 2 \delta -4}(\gamma^2 +1)}{\pi^2} \biggl(\frac{\delta}{\gamma}\biggr)^{-2 \delta}  \, \Gamma^2(\delta +1/2), 
\nonumber\\
Q_{B}^{(G)}(\alpha, \gamma, \delta, b_{0}) &=& b_{0} \frac{2^{ 2 \delta -4}\{ \gamma [ 3 + 2 \gamma (\gamma -1)]\}}{\pi^2} \biggl(\frac{\gamma}{\delta}\biggr)^{-2 \delta}  \, \Gamma^2(\delta +1/2), \qquad \alpha > 0.
\label{FFF11}
\end{eqnarray}
In the case $\alpha=1$ and $\alpha =0$ the expressions (\ref{FFF11}) are slightly different and follow from the results obtained in section \ref{sec4}.
The late-time power spectra of Eqs. (\ref{FFF7}), (\ref{FFF8}) and (\ref{FFF9}) have been obtained in the case of increasing gauge coupling. From the results of appendix \ref{APPC} the relevant expressions valid in the case of decreasing gauge coupling easily follow, if needed. If the gauge coupling decreases the evolution is likely to start in a non-perturbative regime.
For this reason we shall consider this case as purely academic as recently pointed out in a related context \cite{MMM3a}.

In what follows we shall consider the situation where, for  $T> T_{ew}$, the electroweak symmetry is restored. Around $T\simeq T_{ew}$ the  ordinary magnetic fields are proportional to the hypermagnetic fields through the cosine of the Weinberg's angle $\theta_{W}$, i.e. $\cos{\theta_{W}} \, \vec{B}$.  To illustrate the gauge spectra for different values of $\alpha$ we shall consider the simplest scenario where the modes reentering above the electroweak temperature will affect the baryon asymmetry of the Universe (BAU). Conversely the magnetic power spectra obtained from the modes reentering for $T< T_{ew}$ will be compared with the magnetogenesis requirements. 
\subsubsection{Magnetogenesis considerations}
 While the modes inside the Hubble radius at the 
electroweak time reentered right after inflation,  the magnetogenesis  wavelengths crossed the effective horizon much later but always prior to matter-radiation equality (see Eq. (\ref{PH1}) and discussion therein). For $\tau > \tau_{k}$ the conductivity dominates and while the electric fields are suppressed by the finite value of the conductivity, the magnetic fields are not dissipated at least for typical scales smaller than the magnetic diffusivity scale. The mode functions 
for $\tau \geq \tau_{k}$ will be suppressed with respect to their values at $\tau_{k}$:
\begin{equation}
f_{k}(\tau) = f_{k}(\tau_{k}) e^{- k^2/k_{d}^2}, \qquad g_{k}(\tau) = (k/\sigma_{em}) g_{k}(\tau_{k}) e^{- k^2/k_{d}^2},\qquad 
k_{d}^{-2} = \int_{\tau_{k}}^{\tau} \,\,d \, z/\sigma_{em}(z),
\label{PH2}
\end{equation}
where $\sigma_{em}$ is the standard conductivity of the plasma  and $k_{d}$ denotes the magnetic diffusivity 
momentum. The ratio $(k/k_{d})^2$ appearing in Eq. (\ref{PH2}) is actually extremely small in the 
phenomenologically interesting situation since around $\tau = \tau_{\mathrm{eq}}$ (and for $k = {\mathcal O}(\mathrm{Mpc}^{-1})$) the ratio $(k/k_{d})^2 = {\mathcal O}(10^{-26})$.

While so far we just considered comoving fields, what matters for the 
magnetogenesis requirements are instead the physical power spectra prior to the gravitational collapse of the protogalaxy. Recalling Eq. (\ref{PXphys}) the physical power spectrum is:
\begin{equation}
P_{B}^{(phys)}(k,\tau) = \frac{P_{B}(k,\tau)}{a^4 \lambda}\, \cos^2{\theta_{W}} , \qquad \Rightarrow \quad P_{B}^{(phys)}(k,\tau_{*}) = \frac{P_{B}(k,\tau_{k})}{a_{k}^4 \lambda_{k}}\biggl(\frac{a_{k}}{a_{*}}\biggr)^4\, \cos^2{\theta_{W}}.
\label{PH4}
\end{equation} 
The first expression of Eq. (\ref{PH4}) is just the definition of the physical power spectrum obtained by evaluating, after symmetry breaking, the two-point function of Eqs. (\ref{WKB10})--(\ref{WKB11}) as a function of the physical fields; $P_{B}^{(phys)}(k,\tau_{*})$ is instead the physical power spectrum computed at a reference time  $\tau_{*} > \tau_{k}$ under the further assumption that the mode functions follow from Eq. (\ref{PH2}) for $\tau > \tau_{k}$.  In a conservative perspective  the magnetogenesis requirements 
 roughly demand that the magnetic fields at the time of the gravitational collapse of the protogalaxy should be approximately larger than a (minimal) power spectrum which can be estimated between ${\mathcal O}(10^{-32})\,\mathrm{nG}^2$ and ${\mathcal O}(10^{-22})\, \mathrm{nG}^2$. 
 The least demanding requirement  
 \begin{equation}
 \sqrt{P^{(phys)}_{B}(k,\tau_{*})} > 10^{-16} \, \, \mathrm{nG},
 \label{RQ1}
 \end{equation}
 should then be complemented with the stricter limit:
 \begin{equation} 
 \sqrt{P^{(phys)}_{B}(k,\tau_{*})} > 10^{-11} \, \, \mathrm{nG}.
 \label{RQ2}
 \end{equation}
The value $10^{-16} \mathrm{nG}$ follows by assuming 
that, after compressional amplification, every rotation of the galaxy increases the initial magnetic field of one $e$-fold. According to some this requirement is not completely realistic since it takes more than one $e$-fold  to increase the value of the magnetic field by one order of magnitude and this is the rationale for the most demanding condition associated with $10^{-11}$ nG. 

\subsubsection{Baryogenesis considerations}
The considerations associated with the BAU involve typical $k$-modes in the range $a_{ew} \, H_{ew} \leq k < k_{\sigma}$ where $k_{\sigma}$ denotes the diffusivity scale associated with electroweak conductivity. While the contribution of the hypermagnetic gyrotropy determines the baryon to entropy ratio $\eta_{B} = n_{B}/\varsigma$, 
 the hyperelectric gyrotropy is washed out inside the Hubble radius.  Denoting by $N_{eff}$  the effective number of relativistic degrees of freedom at the electroweak epoch, the expression of the BAU \cite{FIVEa0,FIVEa1,MMM3b} is:
\begin{equation}
\eta_{B}(\vec{x},\tau) = \frac{n_{B}}{\varsigma} = \frac{ 3 g^{\prime\, 2} n_{f}}{32\,\pi^2 \, H} \biggl(\frac{T}{\sigma_{c}}\biggl) \frac{{\mathcal G}^{(B)}(\vec{x}, \tau)}{{\mathcal H} \,a^4 \rho_{crit}},\qquad\qquad {\mathcal G}^{(B)}(\vec{x}, \tau) =  \vec{B} \cdot \vec{\nabla} \times \vec{B},
\label{BAU}
\end{equation}
where $\varsigma= 2 \pi^2 T^3 N_{eff}/45$ is the entropy density of the plasma, and $g^{\prime}$ (with $g'\simeq 0.3 $) is the $U(1)_{Y}$ coupling at the electroweak time and  $n_{f}$ is the number of fermionic generations. In what follows $N_{eff}$ shall be fixed to its standard model value (i.e. $N_{eff}= 106.75$).  In Eq. (\ref{BAU}) 
$\sigma_{c}$ denotes the electroweak conductivity. Equation (\ref{BAU}) holds when the rate of the slowest reactions in the plasma (associated with the right-electrons) is larger than the dilution rate caused by the hypermagnetic field itself: at the phase transition the hypermagnetic gyrotropy  is converted back into fermions since the ordinary magnetic fields does not couple to fermions. Since all quantities in Eq. (\ref{BAU})  are comoving, $\langle \,\eta_{B}(\vec{x},\tau)\,\rangle$ for $\tau = \tau_{ew}$ is 
  determined by the averaged gyrotropy:
   \begin{equation} 
 \langle \, \eta_{B}(\vec{x}, \tau_{ew})\, \rangle = \frac{3 \, n_{f}\, \alpha^{\prime\, 2 }}{ 4 \, \pi\sigma_{0}\, a_{ew}^4\, \rho_{crit} {\mathcal H}_{ew}} \int_{0}^{k_{\sigma}} \, P_{B}^{(G)}(k,\tau) \, dk,
 \label{BAU1a}
 \end{equation}
 where $\sigma_{0}$ accounts for the theoretical uncertainty associated with the determination of the chiral conductivity  of the electroweak plasma\footnote{Typical values of range $\sigma_{0}$ between $1$ and $10$. 
 For the illustrative purposes of this discussion the values of $\sigma_{0}$ are immaterial and we shall then fix 
 $\sigma_{0} =1$ 
 (see also \cite{cond1,cond2}).} according to $\sigma_{c} = \sigma_{0} T/\alpha^{\prime}$ with 
 $\alpha^{\prime} = g^{\prime\, 2}/(4 \pi)$. The upper limit of integration in Eq. (\ref{BAU1a}) coincides with  diffusivity momentum and  it is useful to express $k_{\sigma}$ in units of ${\mathcal H}_{ew} = a_{ew} H_{ew}$ where 
 $H_{ew}^{-1} = {\mathcal O}(1) \, \mathrm{cm}$ is the Hubble radius at the electroweak time:
 \begin{equation}
 \frac{k_{\sigma}}{a_{ew} \, H_{ew}} = 3.5 \times 10^{8} \, \sqrt{\frac{\sigma_{0}}{\alpha^{\prime} \, N_{eff}}}\, \biggl(\frac{T_{ew}}{100\, \mathrm{GeV}} \biggr)^{-1/2}.
 \label{ksigma}
 \end{equation}
The typical diffusion wavenumber exceeds the electroweak Hubble rate by approximately $8$ orders of magnitude. 

\subsection{The range $\alpha \geq 1$}
The slopes of the hyperelectric and hypermagnetic power spectra at early times do not depend on the strength of the anomalous interactions. The corresponding gyrotropic 
 spectra, on the contrary, depend explicitly on $\alpha$ and $b_{0}$. In what follows we shall illustrate this general aspect in terms of the late-time power spectra.
 \begin{figure}[!ht]
\centering
\includegraphics[height= 8cm]{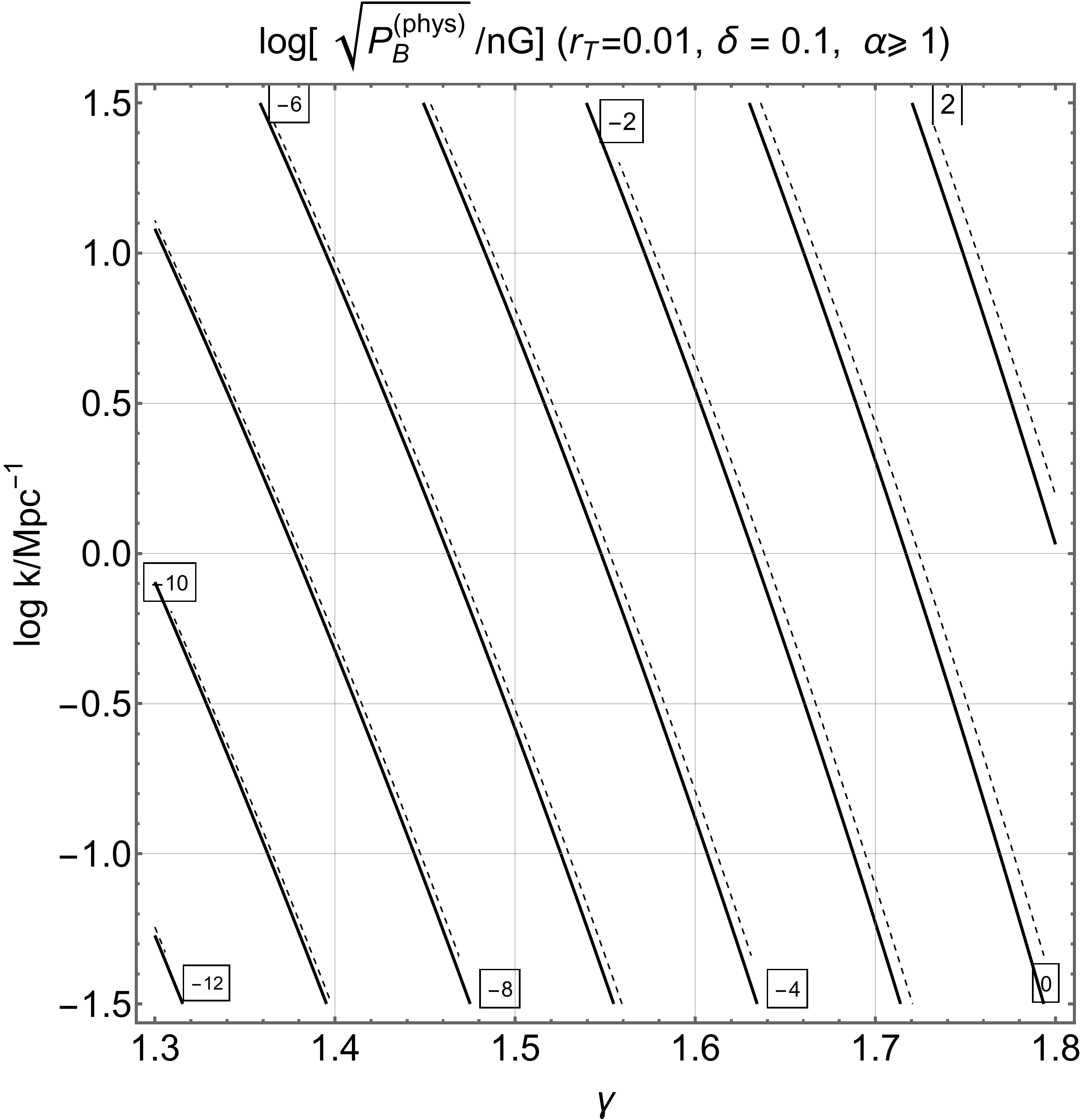}
\includegraphics[height= 8cm]{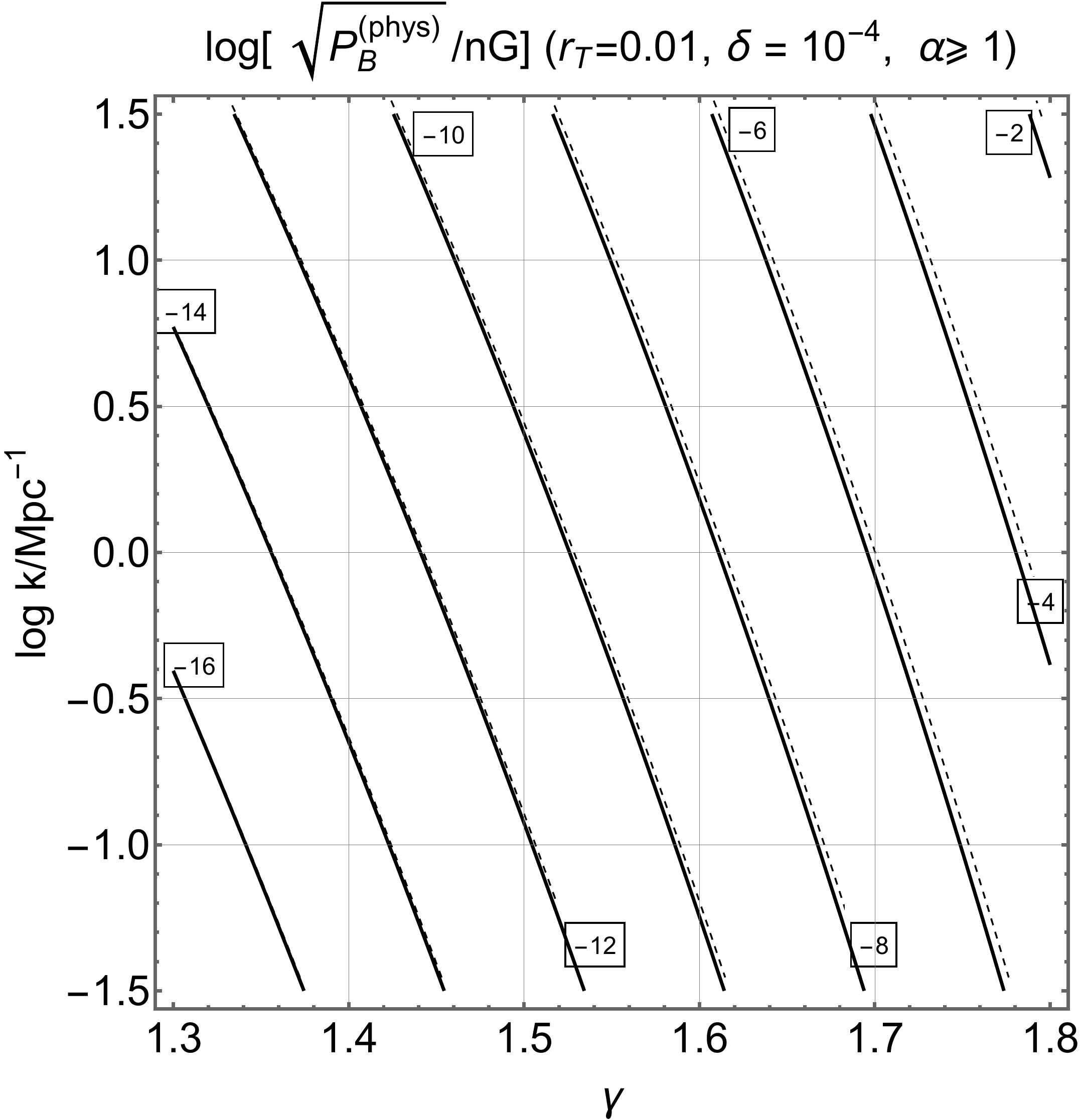}
\caption[a]{The physical spectra of the magnetic field are illustrated at late time. The dashed line denotes 
the case $\alpha =1$ while the full line corresponds to the generic case $\alpha > 1$. As explicitly indicated the left and right plots differ because of the values of $\delta$: since we are interested in the situation where the gauge coupling flattens out after inflation we will have that $\delta \ll 1$. From the comparison of the two plots of this figure the 
slopes of the power spectra are not affected when $\delta$ passes from $0.1$ to $10^{-4}$ (as long as $\delta \ll 1$). The results of both plots primarily demonstrate that different values of $\alpha$ lead to the same slopes of the magnetic power spectra at late time. We recall that $\alpha$ controls the profile of the pseudoscalar 
coupling while $\gamma$ accounts for the evolution of the scalar coupling (see Eq. (\ref{EX2a}) and discussion thereafter). The second point illustrated by both plots is that the late-time power 
spectra are phenomenologically relevant; as suggested after Eq. (\ref{PH4}) we must have $\sqrt{P^{(phys)}_{B}(k,\tau_{*})} > 10^{-11} \, \, \mathrm{nG}$ to fulfil the most demanding magnetogenesis constraints (see Eqs. (\ref{RQ1})--(\ref{RQ2}) and discussion therein). The cases $\alpha >1$ and $\alpha \to 1$ 
have been treated explicitly in appendix \ref{APPB} (see, in particular, Eqs. (\ref{cond8}) and (\ref{b06}). } 
\label{FFFG1}      
\end{figure}
In Fig. \ref{FFFG1} the various curves correspond to different values of 
 the magnetic power spectra at late times. The 
 labels appearing on the contours denote the common logarithm of  $ \sqrt{P_{B}^{(phys)}}$ expressed in nG, i.e. $\log{[\sqrt{P_{B}^{(phys)}}/\mathrm{nG}]}$. On the horizontal axis the 
values of $\gamma$ are reported while on the common 
logarithm of the comoving wavenumber $k$ is plotted in units of $\mathrm{Mpc}^{-1}$. To obtain the physical 
spectra of Fig. \ref{FFFG1} we used Eq. (\ref{FFF7}) evaluated at $\tau_{k}$ (see Eq. (\ref{PH1})) 
and then computed the physical power spectrum according to Eq. (\ref{PH4}). In Fig. \ref{FFFG1} 
the dashed and the thick lines correspond to the cases $\alpha=1$ and $\alpha > 1$, respectively.  
The slight mismatch between the thick and dashed contours 
plots does not come from the slope of the power spectra but from a minor difference in the overall amplitude which is immaterial for the present considerations. In the left plot we illustrated the case $\delta =0.1$ while in the 8cm
right plot we considered the limit $\delta \to 0$ by setting $\delta = 10^{-4}$.  In both plots of Fig. \ref{FFFG1} 
there are regions  where $ \sqrt{P_{B}^{(phys)}} > 10^{-11} $ nG and even a region where $\sqrt{P_{B}^{(phys)}}= {\mathcal O}(10^{-2})$ nG showing that magnetogenesis is possible in this case. 

The results of Fig. \ref{FFFG1} ultimately demonstrate that the values of $\alpha \geq 1$ do not modify the late-time magnetic spectra. The rationale for this result can be understood, in simpler terms, by appreciating that the slopes of the large-scale hyperelectric and hypermagnetic fields at the 
end of inflation do not depend on $\alpha$ and $b_{0}$. 
 \begin{figure}[!ht]
\centering
\includegraphics[height=8cm]{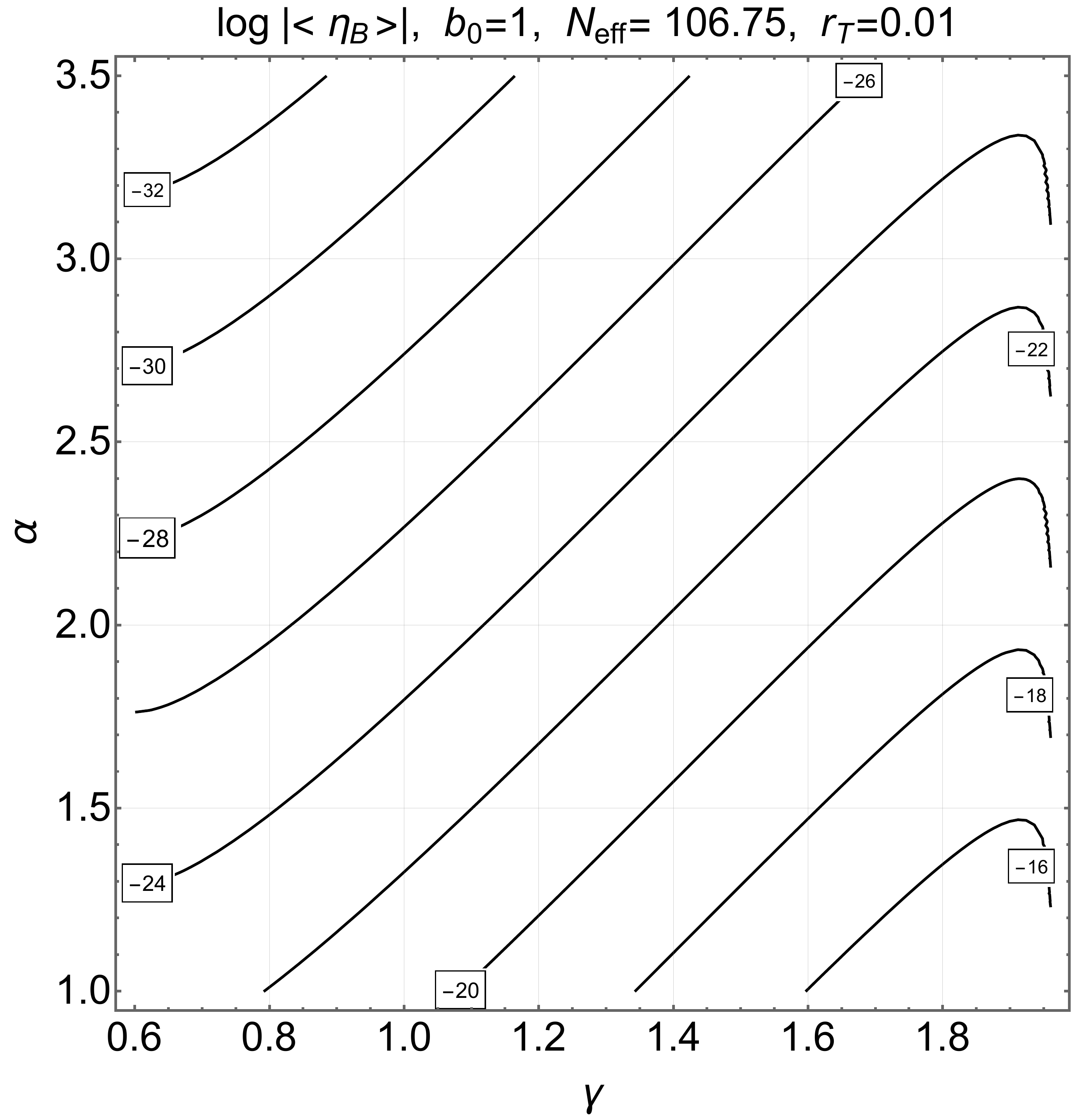}
\includegraphics[height=8cm]{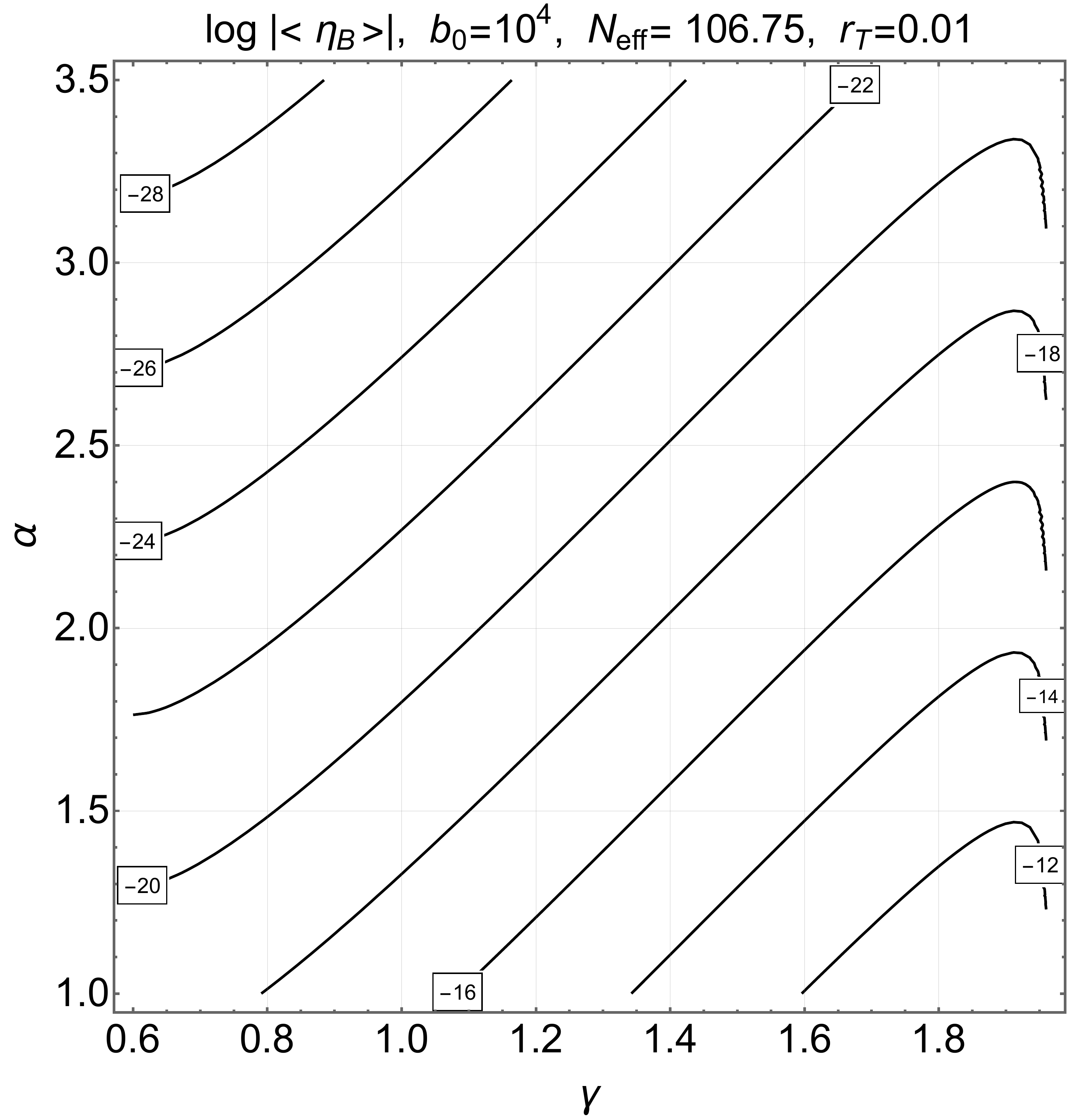}
\caption[a]{The gyrotropic contribution is  illustrated for different values of $\alpha\geq 1$ (reported on the vertical axis) 
and $\gamma$ (appearing on the horizontal axis). In both plots, for the sake 
of simplicity, we took the limit $\delta \to 0$ since, as previously established, different values of $\delta \ll 1$ are practically indistinguishable. The physical spectra of the magnetic field are illustrated at late time. The range of $\alpha$ coincides with the one of Fig. \ref{FFFG1}. We see that the values of the baryon asymmetry that are phenomenologically 
more relevant occur for large $\gamma$ and small $\alpha$ (i.e. in the bottom right corner of the right plot). 
What is more relevant for the present considerations is that different values of $\alpha$ strongly affect 
the gyrotropic spectra and the resulting values of $\eta_{B}$ (see Eq. (\ref{BAU}) and discussion thereafter).
If the two plots 
of this figure are compared with Fig. \ref{FFFG1} we are led to conclude that when $\alpha \geq 1$ that the magnetogenesis 
requirements of Eqs. (\ref{RQ1})--(\ref{RQ2}) can be easily satisfied but it is impossible to reproduce the correct value of the BAU (except 
for some corners of the parameter space characterized by extreme values of $\gamma$).}
\label{FFFG2}      
\end{figure}
The gyrotropic contributions, on the 
contrary, do depend on the values of $\alpha$ and, to a lesser extent, on $b_{0}$.  This aspect is summarized in Fig. \ref{FFFG2} where we illustrate the magnitude of the gyrotropic contributions
in the case $\alpha \geq 1$. As already mentioned, to make the comparison more physical, we directly
illustrated the baryon asymmetry $\eta_{B}$ which is proportional to the magnetic gyrotropy and it is computed from Eqs. (\ref{BAU})--(\ref{BAU1a}). The labels on the curves correspond this time to the common logarithm of $\eta_{B}$. We see that as $\alpha$ increases the associated baryon asymmetry decreases sharply. This reduction is partially compensated by an increase of $b_{0}$: while in the left plot of Fig. \ref{FFFG2} we took 
$b_{0} =1$, in the right plot $b_{0} = 10^{4}$. The rationale for this result is that, in practice, the 
magnetic gyrotropy scales linearly with $b_{0}$ while the $\alpha$ enters the gyrotropy via $x_{1}^{\alpha}$; since 
$x_{1} \ll 1$ a small increment in $\alpha$ implies a very large suppression  that cannot be compensated by $b_{0}$.  A 
 large value of $b_{0}$ can be obtained from Eq. (\ref{EX2b}) either by
increasing  $\overline{\lambda}_{2}/\lambda_{1}$ or by imposing a large hierarchy between $\tau_{1}$ and $\tau_{2}$.
Since both tunings are somehow unnatural we shall regard the case $b_{0} = {\mathcal O}(1)$ as the most plausible. Furthermore, as we shall see in a moment, large values of $b_{0}$ quickly
leads to a violation of the critical density bound. All in all when $\alpha\geq 1$ the conclusions 
can be summarized in the following manner:
\begin{itemize}
\item{} different values of $\alpha$ do not affect the late-time form of the hypermagnetic power 
spectra and of their magnetic part obtained by projecting the hypercharge field through the cosine 
of the Weinberg angle; this result also confirms, as expected from the results of section \ref{sec4}, that the pesudoscalar interactions do not help, in practice, with the magnetogenesis requirements of Eqs. (\ref{RQ1})--(\ref{RQ2});
\item{} the pseudoscalar interactions and the different values of $\alpha$ are instead crucial for the 
estimate of the gyrotropic spectra and for the calculation of the BAU;
\item{} finally if we consider the obtained results at face value we are led to conclude 
that, in the case $\alpha \geq 1$, the magnetogenesis requirements (\ref{RQ1})--(\ref{RQ2}) can be easily satisfied but 
the BAU cannot be correctly reproduced unless we choose some extreme corners of the parameter 
space.
\end{itemize}
Concerning the last point in the above list of items it is useful to stress that the typical values of of the magnetic power 
spectra appearing in Fig. \ref{FFFG1} not only satisfy the magnetogenesis requirements but can even be ${\mathcal O}(nG)$
over the typical scale of the gravitational collapse of the protogalaxy. 

\subsection{The range $0\leq \alpha < 1$}
Based on the previous trends we expect that for even smaller values 
of $\alpha$ the weight of the gyrotropic contributions will increase while the slopes of the hypermagnetic and hyperelectric power spectra will remain practically unaffected. This means, in particular, that we also expect that the BAU limits and the 
magnetogenesis requirements could be jointly satisfied.
Generally speaking this is what 
happens with one important caveat: as $\alpha \to 0$ the turning points 
will not depend, in practice, on the values of $\alpha$. 
For this reason in Fig. \ref{FFFG4} we separately discussed the case $\alpha \to 0$. 
\begin{figure}[!ht]
\centering
\includegraphics[height=8cm]{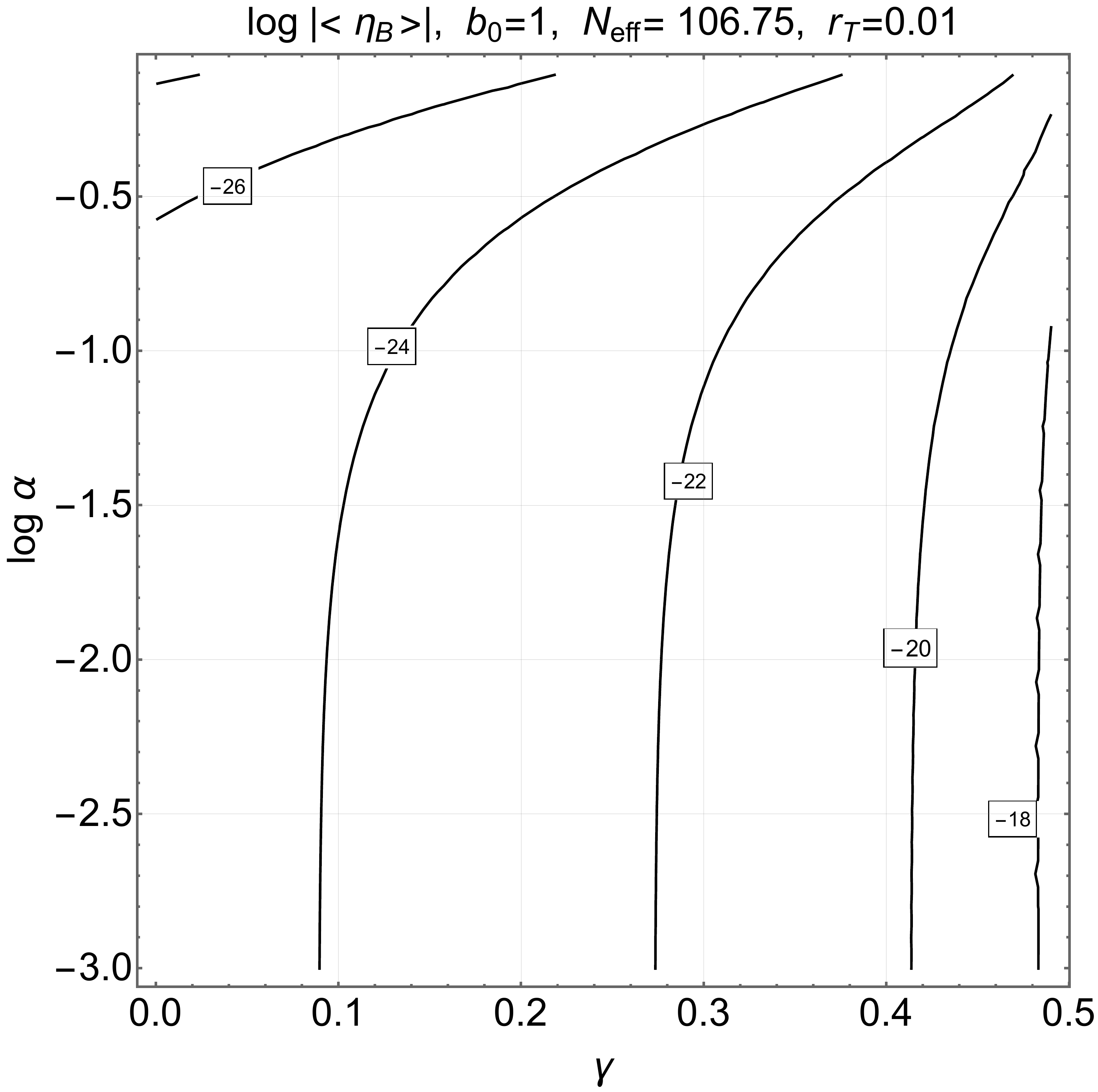}
\includegraphics[height=8cm]{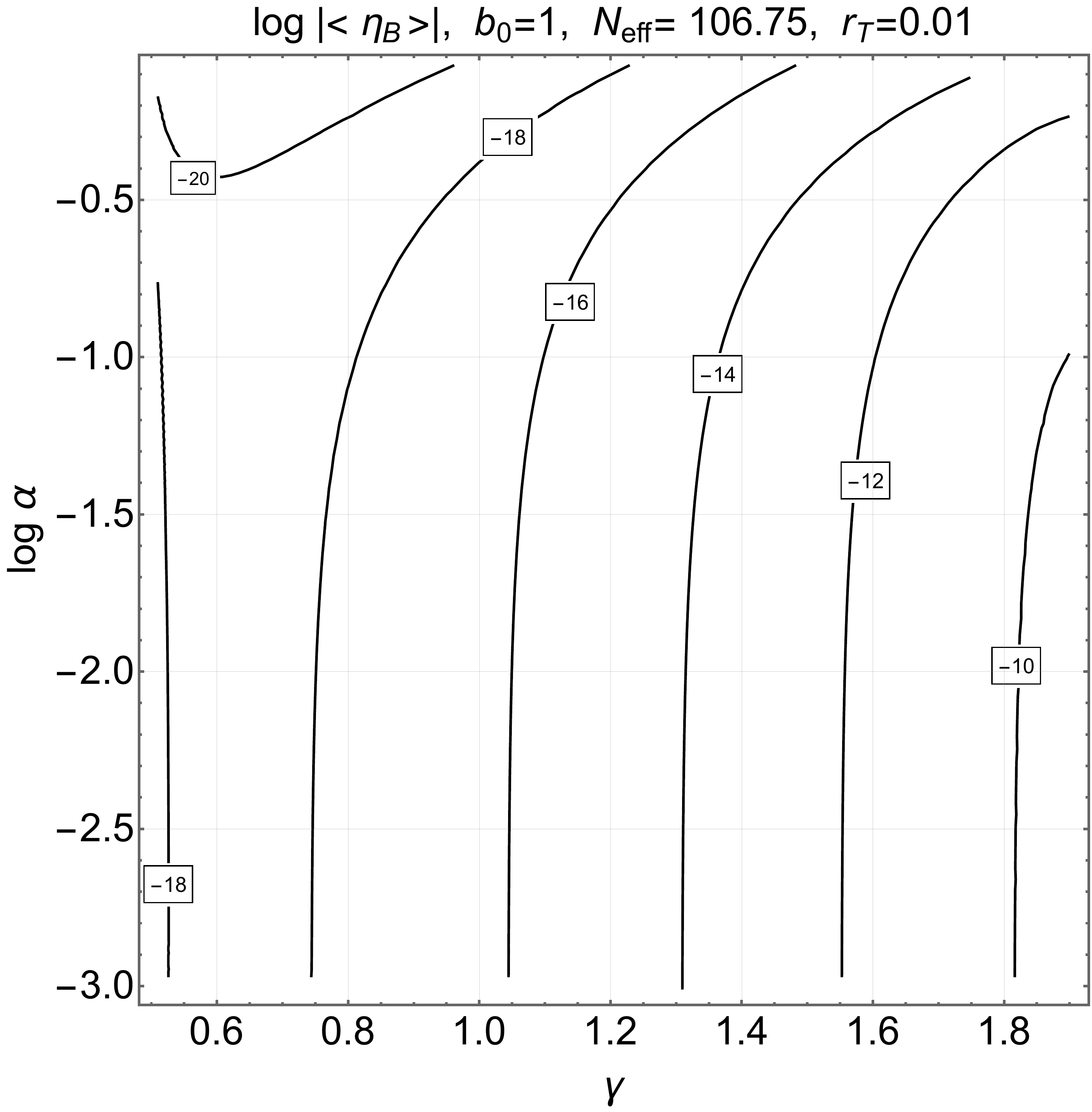}
\caption[a]{The gyrotropic contribution is  illustrated in the plane $(\gamma, \, \log{\alpha})$ and for  $0< \alpha < 1$. In this range the value of the obtained baryon asymmetry can be phenomenologically relevant. Since the 
magnetic power spectra are practically independent on $\alpha$ we conclude that for $0< \alpha < 1$ 
the requirement $ \sqrt{P_{B}^{(phys)}} > 10^{-11} $ nG is satisfied and the hypermagnetic gyrotropy is sufficiently large to seed the BAU (see also Eqs. (\ref{BAU})--(\ref{BAU1a}) and discussion therein). As we can clearly 
appreciate by comparing the two plots of this figure the preferable values of $\alpha$ and $\gamma$ are 
in the ranges $0<\alpha < 1$ and $\gamma > 1$. This conclusion excludes, by construction, the case 
$\alpha \to 0$. In this case the $\eta$-time parametrization is singular (see also appendix \ref{APPB}); when $\alpha \to 0$ the 
phenomenological implications will be separately discussed hereunder (see, in particular, Figs. \ref{FFFG4} and \ref{FFFG5}).}
\label{FFFG3}      
\end{figure}
Figure \ref{FFFG3} illustrates the different values of $\eta_{B}$ in the range $0< \alpha < 1$. 
From the comparison of Figs. \ref{FFFG2} and \ref{FFFG3} we can appreciate that as the values of $\alpha$ decrease 
below $1$ the corresponding values of the gyrotropic spectra increase sharply. In Figs.  \ref{FFFG2} and \ref{FFFG3} we considered the limit $\delta \to 0$ since larger values of $\delta$ 
only modify the actual numerical values of the gyrotropy but do not alter the main conclusions.
The difference between the two plots in Fig. \ref{FFFG3} is simply given 
by the range of $\gamma$: in the left plot $0< \gamma <1/2$ while in the right plot $\gamma >1/2$.
\begin{figure}[ht!]
\centering
\includegraphics[height=8cm]{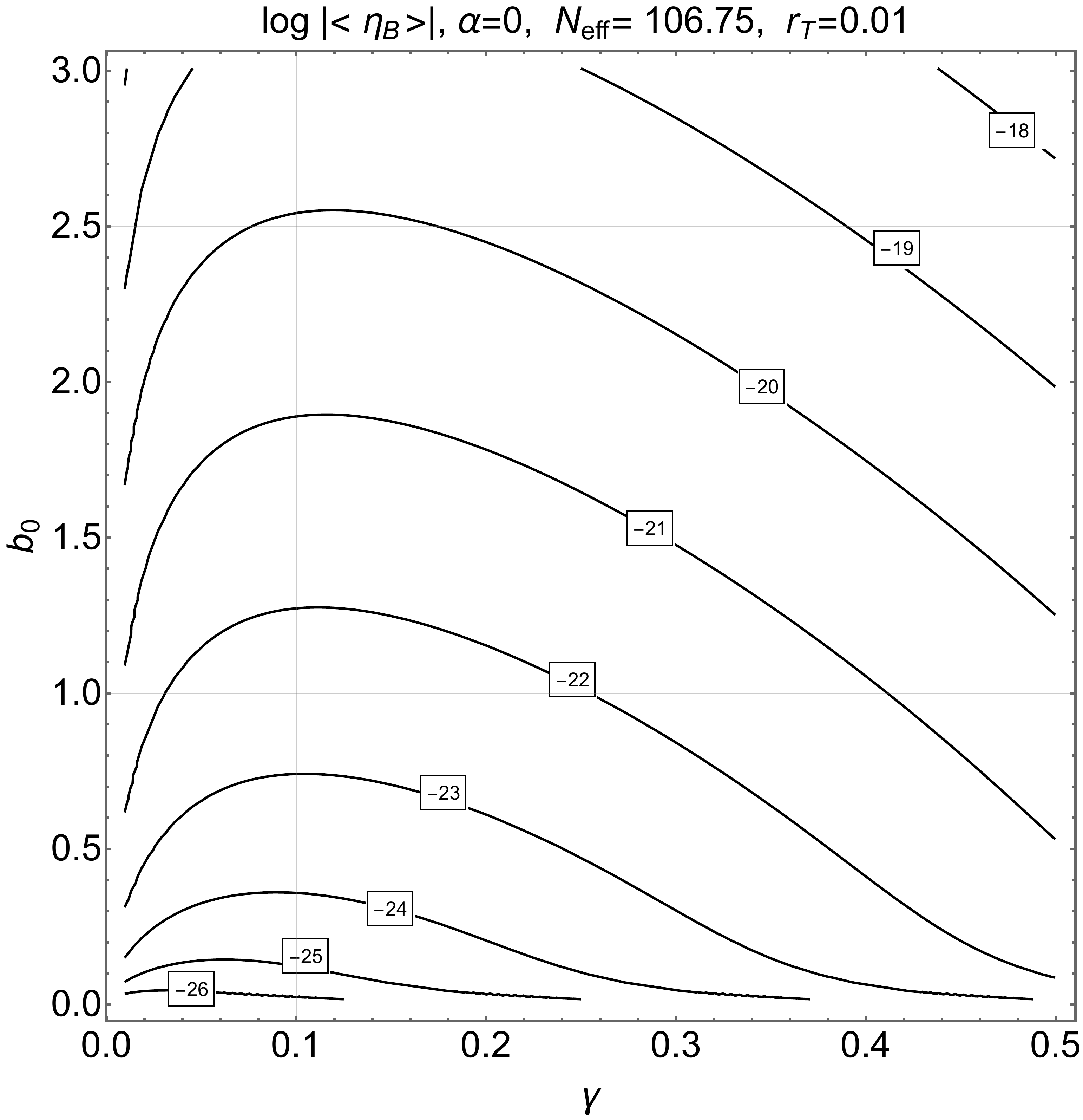}
\includegraphics[height=8cm]{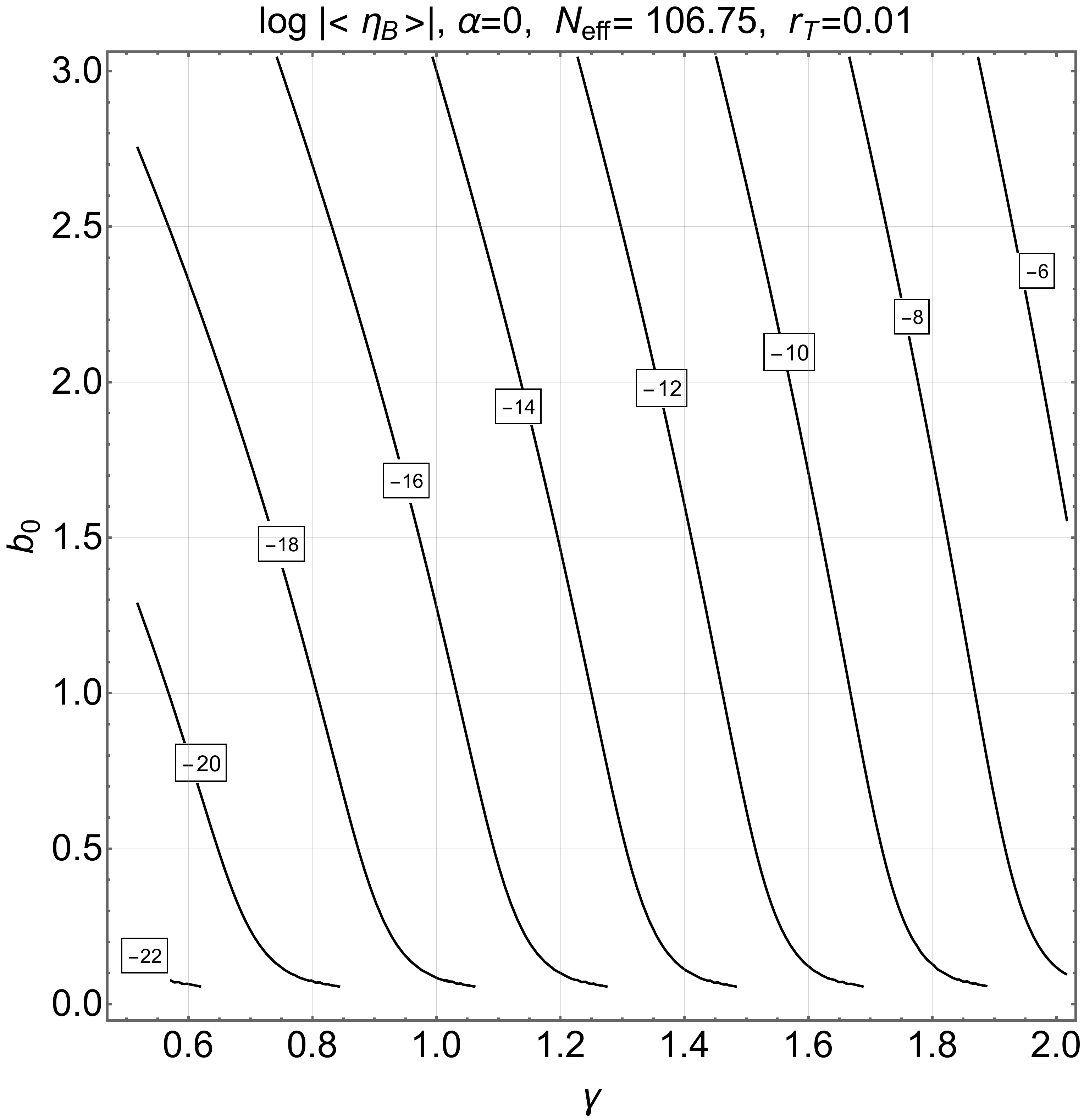}
\caption[a]{We illustrate the gyrotropic contribution in the case $\alpha \to 0$. Special care is required 
since now the gyrotropic spectra depend explicitly on $b_{0}$: this happens because, for $\alpha\to 0$ 
the solution for the mode functions involves the Whittaker's functions whose asymptotic limits involve
 $b_{0}$ (see appendix \ref{APPB}). For different values of $b_{0}$ and $\gamma$ the BAU can 
be reproduced with a preference for the range $\gamma > 1/2$. 
The obtained values of the baryon asymmetry are compatible with the ones 
already deduced in Fig. \ref{FFFG3} when $0<\alpha < 1$. The difference here is 
that the amplitude of the hypermagnetic fields is more affected than in the case $\alpha \neq 1$. This 
aspect will be more specifically illustrated in Fig. \ref{FFFG5}.}
\label{FFFG4}      
\end{figure}
As in the general case, also for $\alpha \to 0$ the 
 slopes of the hyperelectric and hypermagnetic power spectra are not modified at large-scales. However the corresponding amplitudes are comparatively more affected than in the case $\alpha > 0$. This is exactly what   happens in Fig. \ref{FFFG4} where, in the left plot, we illustrate the gyrotropic contribution and in the right plot we compute the power magnetic power spectrum. It is finally useful to discuss also the gauge power spectra in the limit $ \alpha \to 0$. To avoid repetitive remarks (and 
 the proliferation of figures) we only treat in detail the case $\alpha \to 0$ but the obtained results are also applicable also when $\alpha \ll 1$.
 In Fig. \ref{FFFG5} in the left plot we illustrate the spectral energy density 
during inflation for the maximal frequency of the spectrum. To avoid drastic departure from the isotropy $\Omega_{Y}(k,\tau)$ introduced in Eq. (\ref{OMY}) must be sufficiently small and the shaded region of the left plot in Fig. \ref{FFFG5} corresponds to the plausible requirement that $\Omega_{Y}(k_{max}, \tau) < 10^{-6}$. In the plot at the right 
the full curves correspond to $b_{0} \to 0$ while the dashed line have been computed for $b_{0} \to 2$. Since the dashed and the full lines are parallel, the slopes of the physical spectra are the same. 
Whenever $b_{0} \neq 0$ the 
amplitude of the physical power spectrum increases. This change in the 
amplitude could be however compensated by a shift in $\gamma$.
We conclude that the overall amplitude of late-time power spectra is only marginally controlled by $b_{0}$. 
\begin{figure}[ht!]
\centering
\includegraphics[height=8cm]{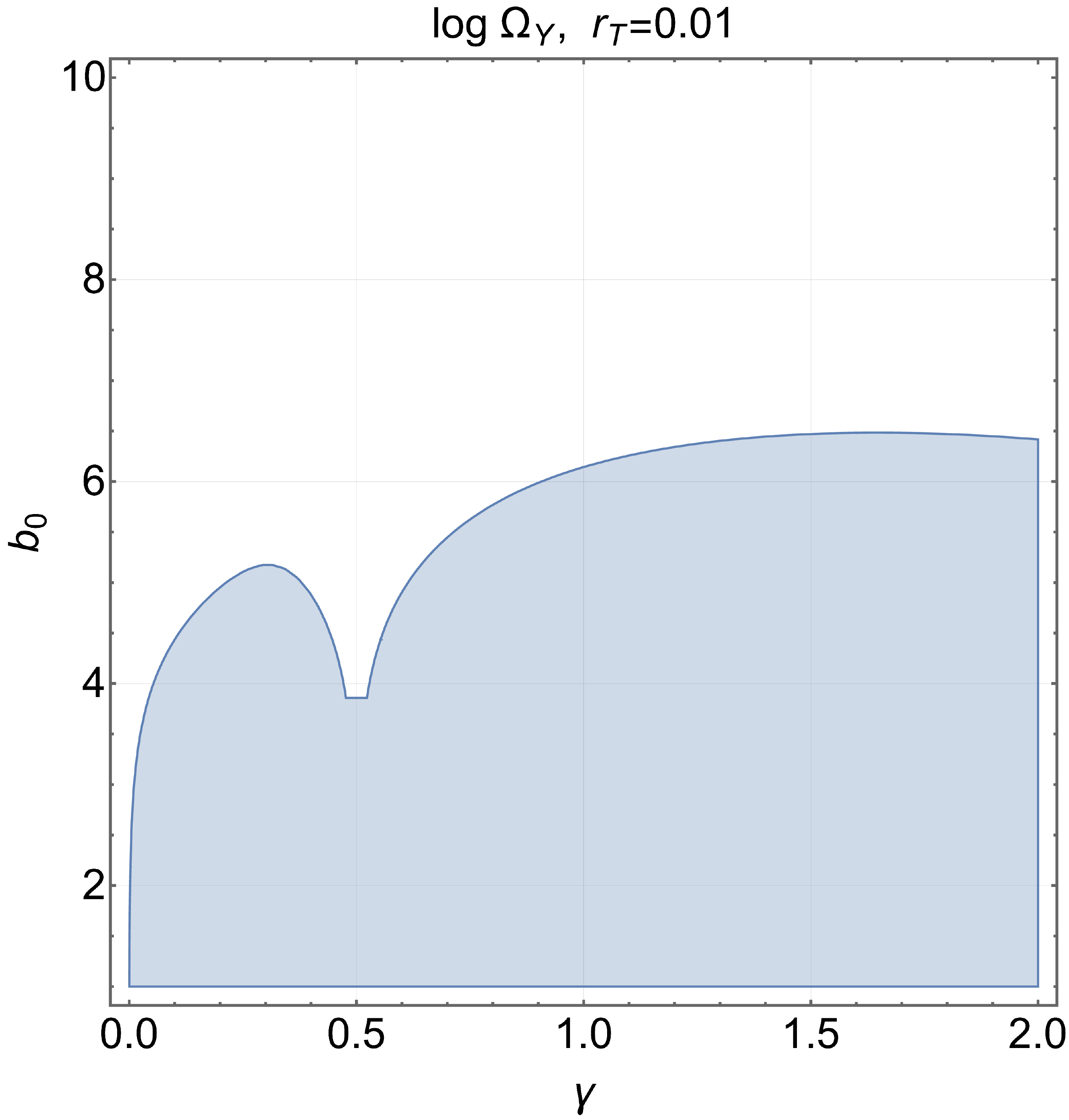}
\includegraphics[height=8cm]{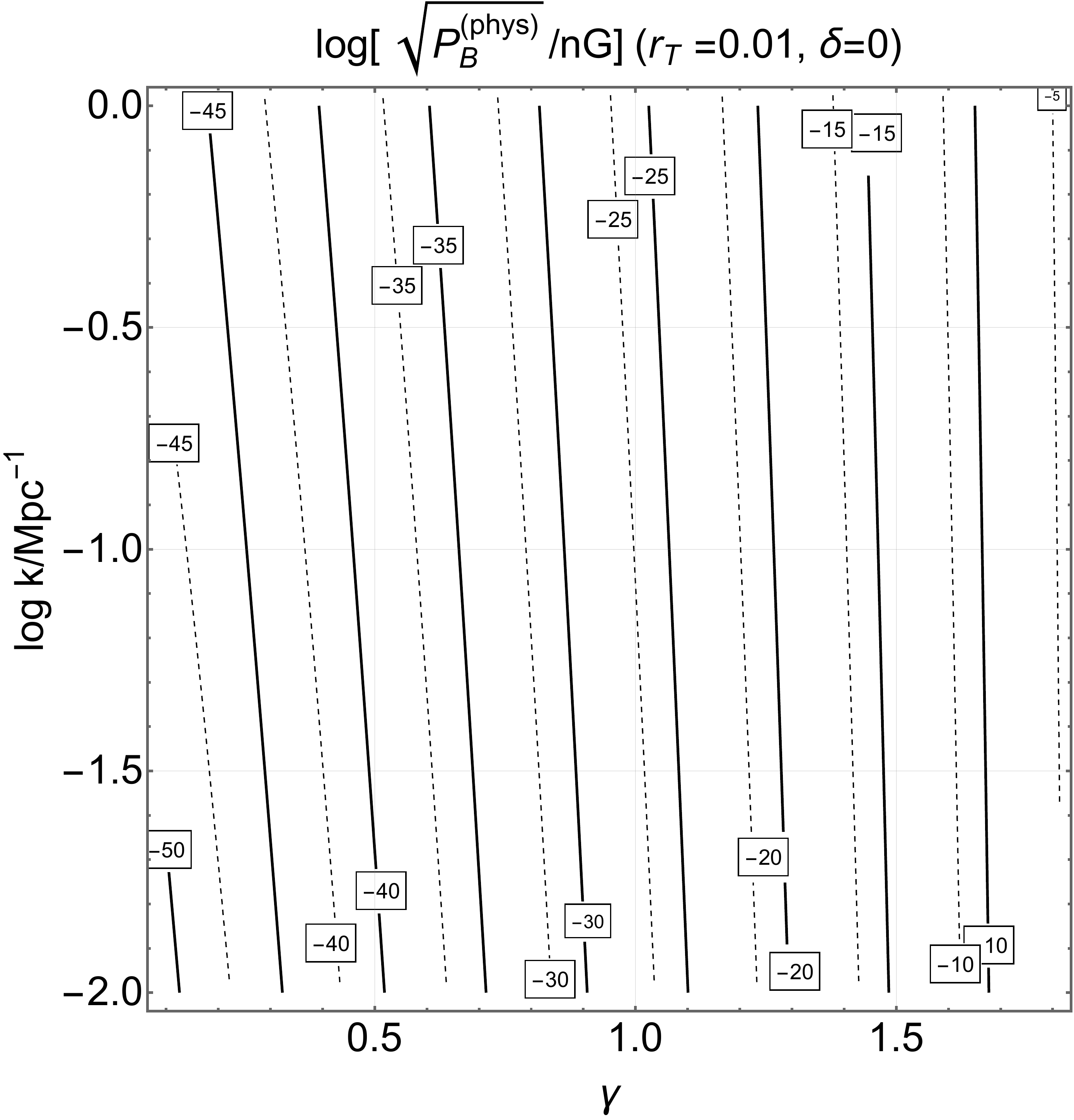}
\caption[a]{We illustrate late-time magnetic power 
spectrum in the case $\alpha \to 0$ where the amplitude of the magnetic power spectra  depends on the values of $b_{0}$ defined in Eqs. (\ref{EX2a})--(\ref{EX2b}). In the plot at the left
we illustrate the critical density bound in the plane $(b_{0}, \gamma)$: if $b_{0}$ falls outside the shaded area
the produced gauge fields are overcritical. Therefore the complementary white region is excluded. 
In the plot at the right the full line refers to $b_{0} \to 0$ while the dashed curves correspond to $b_{0} =2$. 
The results of Figs. \ref{FFFG4} and \ref{FFFG5} demonstrate that the magnetogenesis and the 
baryogenesis requirements can be met for $\alpha \to 0$ provided the valued of $b_{0}$ and $\gamma$ fall within the shaded area of the plot at the left.  In other words not all the values 
of $b_{0}$ are possible: if $b_{0}$ is larger than ${\mathcal O}(3)$ the critical density 
bound associated with the produced gauge fields is violated. This is why, in the previous figure, $b_{0}$ has been chosen to be smaller 
than about $3$. From the right plot we can also appreciate that different values of $b_{0}$ do not modify the late-time slopes of the magnetic spectrum since the dashed and full lines (corresponding to different values of $b_{0}$) are practically parallel.}
\label{FFFG5}      
\end{figure}
The phenomenological discussion of this section can be summarized as follows:
\begin{itemize}
\item{} the slopes of the late-time magnetic power spectra are  completely 
insensitive to the pseudoscalar coupling (associated with $\alpha$) but only depend 
on the scalar coupling (associated with $\gamma$); this conclusion matches 
the results discussed for $\alpha \geq 1$ and completes the analysis;
\item{} if $0\leq \alpha < 1$ the baryogenesis and the magnetogenesis requirements 
can be simultaneously satisfied so there exist some regions of the parameter space 
where the large-scale magnetic fields and the BAU are generated at once by using 
the same set of parameters; in particular it turns out that the relevant phenomenological 
region is given by $0 <\alpha < 1$ and $\gamma > 1/2$;
\item{} if $\alpha \to 0$ the power spectra can be computed exactly in terms of Whittaker's 
functions and the main features of this case coincide with what happens for $\alpha \ll 1$;
\item{} for $\alpha \to 0$ the baryogenesis and the magnetogenesis requirements 
are simultaneously satisfied provided $\gamma > 1/2$ and in the case where $b_{0} < {\mathcal O}(3)$.
\end{itemize}
This discussion presented in this section refers to the case of increasing gauge coupling. A similar discussion can be 
carried on for a decreasing gauge coupling. In this case, however, we must expect 
a strongly coupled regime at the beginning of the cosmological evolution and we therefore regard this case 
as less appealing (see also, in this respect, Ref. \cite{MMM3a}). However, with the strategy described in section \ref{sec4} 
and with the explicit results of appendix \ref{APPC} it will be possible to computed the wanted spectra.
We want to stress, in this respect, that also in the case of decreasing gauge coupling the 
hyeprmagnetic and the hyperelectric gauge spectra are, in practice, not affected by the pseudoscalar 
interactions in the same sense discussed when the gauge coupling increases and then flattens out. 
In particular it can be verified that the approximate duality symmetry connecting 
Eqs. (\ref{WKB24})--(\ref{WKB25}) and (\ref{WKB27a})--(\ref{WKB27b}) holds, in spite of the values of $\alpha$, and in the 
long wavelength limit.

\renewcommand{\theequation}{6.\arabic{equation}}
\setcounter{equation}{0}
\section{Genericness of the obtained results}
\label{sec6}
\subsection{Effective approach to inflationary scenarios}
When the dependence of the Lagrangian 
on the inflaton field is unconstrained  by symmetry principles (or by 
other aspects of the underlying theory) the effective approach suggests that corresponding
inflationary scenario is generic.  If we focus, for simplicity, on the case 
of single-field inflationary models the lowest order effective Lagrangian (corresponding to a 
portion of  Eq. (\ref{action1})) is a fair approximation to the full theory and can be written as\footnote{In Eq. (\ref{FIN1}) we considered, for simplicity, that the potential appearing in Eq. (\ref{action1}) only depend on $\varphi$ 
i.e. ${\mathcal V}(\varphi, \psi) \equiv V(\varphi)$.}:
\begin{equation}
{\mathcal L}_{inf} = \sqrt{- G} \biggl[ - \frac{ \overline{M}_{P}^2\,\, R}{ 2} + 
\frac{1}{2} G^{\alpha\beta} \partial_{\alpha} \varphi \partial_{\beta} \varphi - V(\varphi)\biggr].
\label{FIN1}
\end{equation}
In the effective approach Eq. (\ref{FIN1}) is the first term of a 
generic theory where the higher derivatives are suppressed by the negative powers of a large mass $M$ that specifies the scale of the underlying description. The leading correction to Eq. (\ref{FIN1}) consists of all possible terms containing four derivatives; following the classic analysis of Weinberg \cite{ONE0g} and barring for some minor differences the correction consists of $12$ terms:
\begin{eqnarray}
&& \Delta\,{\mathcal L}_{inf} = \sqrt{-G} \biggl[ c_{1}(\phi) \bigl(G^{\alpha\beta} \partial_{\alpha} \phi \partial_{\beta} \phi\bigr)^2 
+ c_{2}(\phi) G^{\mu\nu}\, \partial_{\mu} \phi \, \partial_{\nu} \phi\, \Box \phi + 
c_{3}(\phi) \bigl( \Box \phi \bigr)^2 
\nonumber\\
&& + c_{4}(\phi) \, R^{\mu\nu} \, \partial_{\mu} \phi \,\partial_{\nu} \phi
+ c_{5}(\phi) \, R\, G^{\mu\nu} \,\partial_{\mu} \phi \,\partial_{\nu} \phi
+ c_{6}(\phi) R \, \Box \phi + c_{7}(\phi) R^2 + c_{8}(\phi) \, R_{\mu\nu} \,R^{\mu\nu} 
\nonumber\\
&& + c_{9}(\phi) R_{\mu\alpha\nu\beta} \, R^{\mu\alpha\nu\beta} + c_{10}(\phi) C_{\mu\alpha\nu\beta} \, C^{\mu\alpha\nu\beta}
+ c_{11}(\phi)  R_{\mu\alpha\nu\beta} \, \widetilde{\,R\,}^{\mu\alpha\nu\beta} + c_{12}(\phi) C_{\mu\alpha\nu\beta} \, \widetilde{\,C\,}^{\mu\alpha\nu\beta}\biggr],
\label{FIN2}
\end{eqnarray}
where the dimensionless scalar $\phi = \varphi/M$ has been introduced for convenience. In Eq. (\ref{FIN2}) the notations are 
standard:  $R_{\mu\alpha\nu\beta}$ and $C_{\mu\alpha\nu\beta}$ denote the Riemann and  Weyl tensors while 
$ \widetilde{\,R\,}^{\mu\alpha\nu\beta}$ and $\widetilde{\,C\,}^{\mu\alpha\nu\beta}$ are the corresponding duals. Furthermore 
$\Box \phi = G^{\alpha\beta} \nabla_{\alpha} \nabla_{\beta} \phi$ and so on and so forth. 
From the parametrization of Eq. (\ref{FIN2}) it follows that the leading correction to the two-point function of the scalar mode of the geometry comes from the terms containing four-derivatives of the inflaton field while in the case of the 
tensor modes the leading corrections stem from  $C_{\mu\alpha\nu\beta} \,\widetilde{C}^{\mu\alpha\nu\beta}$ and $R_{\mu\alpha\nu\beta} \,\widetilde{R}^{\mu\alpha\nu\beta}$ which are typical of Weyl and Riemann gravity \cite{MMM2a,MMM2b}. Incidentally both terms break parity and are therefore capable of polarizing the stochastic backgrounds of the relic gravitons \cite{POLWKB} by ultimately affecting the dispersion relations of the two circular polarizations.

\subsection{Effective approach to magnetogenesis scenarios}
The  analysis leading to Eq. (\ref{FIN2}) and to the results of Ref. \cite{ONE0g} can be 
extended to include the hypercharge fields \cite{MMM3}. In full analogy with Eq. (\ref{FIN2}), rather than assuming a particular underlying description the idea is to 
include all the generally covariant terms potentially appearing with four space-time derivatives in the effective action and to weight them by inflaton-dependent couplings. The Lagrangian density associated with Eq. (\ref{action2}) will now be complemented by:
\begin{eqnarray}
&& \Delta {\mathcal L}_{gauge} = \frac{\sqrt{-G}}{16 \, \pi\, M^2} \biggl[ \lambda_{1}(\phi) \, R\, Y_{\alpha\beta}\, Y^{\alpha\beta} + \lambda_{2}(\phi) \, R_{\mu}^{\,\,\,\,\,\nu} \, Y^{\mu\alpha}\, Y_{\alpha\nu}  + \lambda_{3}(\phi) \, R_{\mu\alpha\nu\beta} \, 
Y^{\mu\alpha} \, Y^{\nu\beta} 
\nonumber\\
&& + \lambda_{4}(\phi) \, C_{\mu\alpha\nu\beta} \, Y^{\mu\alpha} \, Y^{\nu\beta} + \lambda_{5}(\phi) \Box \phi \,\, Y_{\alpha\beta} \, Y^{\alpha\beta}
+ \lambda_{6}(\phi) \partial_{\mu}\phi \partial^{\nu}\phi 
Y^{\mu\alpha} \, Y_{\nu\alpha} + \lambda_{7}(\phi) \nabla_{\mu}\nabla^{\nu} \phi\,  Y_{\nu\alpha} Y^{\mu\alpha}
\nonumber\\
&& + \overline{\lambda}_{8}(\phi) \, R\, Y_{\alpha\beta}\, \widetilde{\, Y\,}^{\alpha\beta} + \overline{\lambda}_{9}(\phi) \, R_{\mu}^{\,\,\,\,\,\nu} \, Y_{\alpha\nu}  \widetilde{\, Y\,}^{\mu\alpha}\, + \overline{\lambda}_{10}(\phi) \, R_{\mu\alpha\nu\beta} \, 
Y^{\mu\alpha} \, \widetilde{\, Y\,}^{\nu\beta} + \overline{\lambda}_{11}(\phi) \, C_{\mu\alpha\nu\beta} \, Y^{\mu\alpha} \, \widetilde{\,Y\,}^{\nu\beta} 
\nonumber\\
&& + \overline{\lambda}_{12}(\phi) \Box \phi \,\, Y_{\alpha\beta} \, \widetilde{\,Y\,}^{\alpha\beta}
+ \overline{\lambda}_{13}(\phi) \partial_{\mu}\phi \partial^{\nu}\phi 
\widetilde{\,Y\,}^{\mu\alpha} \, Y_{\nu\alpha} + \overline{\lambda}_{14}(\phi) \nabla_{\mu}\nabla^{\nu} \phi\,  Y_{\nu\alpha} \, \widetilde{\,Y\,}^{\mu\alpha}
 \biggr].
\label{FIN3}
\end{eqnarray}
Equation (\ref{FIN3}) contains $14$ distinct terms; $7$ of them do not break parity and are weighted by the couplings $\lambda_{i}(\phi)$ (with $i =1,\,\, ...\,,\,\,7$). The remaining $7$ contributions are weighted by the prefactors $\overline{\,\lambda\,}_{j}(\phi)$ (with $j =8,\,\, ...\,,\,\,14$) and contain parity-breaking terms. Equation (\ref{FIN3}) is also applicable when the various $\lambda_{i}$ and $\overline{\lambda}_{j}$ are $\psi$-dependent quantities. In the latter case the collection of the contributions with four derivatives must be considered in conjunction with the supplementary restrictions associated with the physical nature of the spectator fields. Finally if the couplings depend simultaneously on the inflaton $\phi$ and on $\psi$ further terms (containing the covariant gradients of $\psi$) will have to be included in the effective Lagrangian \cite{MMM3}. 
For the sake of illustration we shall stick to the simplest situation of the single-field inflationary scenarios.  When the $\phi$-dependent couplings disappear the first three terms have been analyzed by Drummond and Hathrell \cite{HHH1} and more recently the same terms (without the parity-breaking contributions) have been considered in Ref. \cite{HHH2} for the analysis of photon propagation in curved space-times. The Riemann coupling associated with $\overline{\lambda}_{10}(\phi)$ has been proposed in Ref. \cite{POLWKB}; this term may ultimately polarize the relic graviton background.  The effective action (\ref{FIN3}) does not include terms like 
$(Y_{\mu\nu}\, Y^{\mu\nu})^2$ (appearing, for instance, in the Euler-Heisenberg Lagrangian).  These terms should only enter the effective action if the gauge fields are a source of the background 
 and break explicitly the isotropy; in this case the gauge background affects the dispersion relations \cite{HHH3} but this is not the situation discussed here.

\subsection{Effective action during a quasi-de Sitter stage}
As in Eq. (\ref{FIN2}), the higher derivatives appearing in Eq. (\ref{FIN3}) are 
suppressed by the negative powers $M$. During a quasi-de Sitter stage Eq. (\ref{FIN3})  always leads to an asymmetry between the hypermagnetic and the hyperelectric susceptibilities. Indeed, the full gauge action obtained from the sum of Eqs. (\ref{action2}) and (\ref{FIN3}) becomes 
\begin{equation}
S_{gauge} = \int d^{3} x \int d\tau \biggl({\mathcal L}_{gauge} + \Delta {\mathcal L}_{gauge}\biggr) 
= \frac{1}{2}\int d^{3} x \int d\tau \biggl(\chi_{E}^2 \, E^2 - \chi_{B}^2 B^2 + \overline{\chi}^2 \vec{E}\cdot\vec{B} \biggr),
\label{FIN4}
\end{equation}
where $\chi_{E}^2$ and  $\chi_{B}^2$ denote the hyperelectric and the hypermagnetic susceptibilities while 
$\overline{\chi}^2$ is the strength of the anomalous couplings. In the formal limit $M \to \infty$ we
have that $ \chi_{E} = \chi_{B} \propto\sqrt{\lambda}$ and $\overline{\chi} \propto \sqrt{\overline{\lambda}}$.
The comoving fields $\vec{E}$ and $\vec{B}$ appearing in Eq. (\ref{FIN4}) are defined as $\vec{B} = a^2\,\, \chi_{B} \,\, \vec{B}^{(phys)}$ and as $\vec{E} = a^2 \,\,\chi_{E} \,\, \vec{E}^{(phys)}$; in the limit $\chi_{E} \to \chi_{B}$ the two previous rescalings exactly coincide with the ones already discussed in section \ref{sec3}.
The explicit expressions of the hyperelectric and the hypermagnetic susceptibilities can be computed in general 
terms however, for the present ends, it is sufficient to consider the case of a quasi-de Sitter stage: 
of expansion 
 \begin{eqnarray}
 && \chi_{E}^2  = \frac{\lambda}{4 \pi} \biggl( 1 + \frac{H^2 }{M^2} d_{E}^{(1)} - \epsilon \frac{H^2 }{M^2} d_{E}^{(2)} - \epsilon \frac{H^2}{M^2} d_{E}^{(3)} +  \sqrt{\epsilon} \frac{H^2 M_{P}}{M^3} d_{E}^{(4)} + \sqrt{\epsilon}\,\, \eta \, \frac{H^2 M_{P}}{M^3} \,d_{E}^{(5)}\biggr),
\label{FIN5}\\
&& \chi_{B}^2 = \frac{\lambda}{4 \pi} \biggl( 1 + \frac{H^2 }{M^2} d_{B}^{(1)} - \epsilon \frac{H^2}{M^2} d_{B}^{(2)} - \sqrt{\epsilon}\, \frac{H^2 M_{P}}{M^3} d_{B}^{(3)} +\sqrt{\epsilon} \,\, \eta\, \frac{H^2 M_{P}}{M^3} \,\, d_{B}^{(4)}\biggr),
\label{FIN6}\\
&& \overline{\chi}^2 = \frac{\overline{\lambda}}{4\pi}\biggl( 1 + \frac{H^2}{M^2}  \overline{d}^{(1)} - \epsilon \frac{H^2 }{M^2} \overline{d}^{(2)} - \epsilon \frac{H^2 M_{P}^{2}}{M^{4}}  \overline{d}^{(3)}
+ \sqrt{\epsilon}\, \frac{H^2 M_{P}\,}{M^{3}}\, \overline{d}^{(4)} +  \sqrt{\epsilon} \,\,\eta \frac{M_{P} H^2 }{M^{3}}\overline{d}^{(5)}  \biggr),
\label{FIN7}
\end{eqnarray}
where $\epsilon = - \dot{H}/H^2$ and $ \eta = \ddot{\phi}/(H\, \dot{\phi})$ are the relevant show-roll parameters\footnote{The slow-roll parameter defining $\eta$ should not be confused 
with the $\eta$-time defined in section \ref{sec4}; there is 
no possible misunderstanding since the two quantities are never 
used in the same context.} (see, for instance,  \cite{ONE0h,THIRTEENa}).
The coefficients of Eqs. (\ref{FIN5}), (\ref{FIN6}) and (\ref{FIN7}) can be accurately 
computed in terms of $\lambda_{i}(\phi)$ and $\overline{\lambda}_{j}(\phi)$ appearing in Eq. (\ref{FIN3}) \cite{MMM3}. However  the naturalness of the couplings and the absence of fine-tunings implies that all the $\lambda_{i}(\phi)$ are all of the order of $\lambda(\phi)$ and similarly 
for the $\overline{\lambda}_{j}(\phi)$ which should all be ${\mathcal O}(\overline{\lambda})$. In this situation the leading contribution to the gauge power spectra is given by the leading-order action. The same conclusion follows if $\lambda_{i}(\phi) \ll \lambda(\phi)$ and $\overline{\lambda}_{i}(\phi) \ll \overline{\lambda}(\phi)$. In the opposite situation $\lambda_{i}(\phi) \gg \lambda(\phi)$ and $\overline{\lambda}_{i}(\phi) \gg \overline{\lambda}(\phi)$ the hyperelectric and the hypermagnetic susceptibilities may evolve at different rates.  Barring for this possibility that implies an explicit fine-tuning, the results of the previous sections hold provided the higher-order corrections are generically subleading and this will be 
the last step of this discussion.

\subsection{Generic corrections during a quasi-de Sitter stage}
If none of the couplings $\lambda_{i}(\phi)$ and $\overline{\lambda}_{j}(\phi)$ are 
fine-tuned to be artificially much larger than all the others, the first possibility suggested by Eqs. (\ref{FIN5}), (\ref{FIN6}) and (\ref{FIN7}) is that $\epsilon$ is smaller than $1$ but not too small.  In this case the change of $\dot{\phi}$ during a Hubble time $H^{-1}$ follows from the background evolution and, in this limit,  $M \simeq \sqrt{2 \epsilon} \,\overline{M}_{P}$. This means that, for generic theories of inflation (i.e. when $\varphi$ is  not constrained by symmetry principles) $M$ cannot be much smaller than $\sqrt{2 \epsilon} \,\,\overline{M}_{P}$, otherwise $\dot{\phi}/H$ would diverge. 
If $M= \sqrt{2 \epsilon} \,\,\overline{M}_{P}$ then $H/M$ will be slightly larger than $H/\overline{M}_{P}$. 
In the case of conventional inflationary scenarios we will have that $\overline{M}_{P}^2 H^2/M_{P}^4 = \epsilon {\mathcal A}_{{\mathcal R}}/8$ where 
${\mathcal A}_{{\mathcal R}} = 2.41\times 10^{-9}$ is the amplitude of the curvature inhomogeneities assigned at the pivot scale $k_{p} = 0.002\, \mathrm{Mpc}^{-1}$.
If we keep track of the various factors the leading contributions to $\chi_{E}^2$, $\chi_{B}^2$ and $\overline{\chi}^2$ are all ${\mathcal O}({\mathcal A_{0}}/\epsilon)$ where ${\mathcal A}_{0} = 8 \pi^3 {\mathcal A}_{{\mathcal R}} \simeq 6 \times 10^{-7}$. 
 
The situation described in the previous paragraph is not the one compatible with the current 
phenomenological estimates of $r_{T}$ ranging between $r_{T} < 0.07$ \cite{RT0} and $r_{T}< 0.01$ \cite{RT1,RT2}. Since the consistency relations 
stipulate that $\epsilon \simeq r_{T}/16$, we have to acknowledge that  $\epsilon< 10^{-3} $ which is not the situation discussed in the previous paragraph. We should then require, in the present context,  that 
$M \gg \sqrt{2\epsilon} \, \overline{M}_{P}$ implying $ M\simeq \overline{M}_{P}$ and $\epsilon \ll 1$. This means that the leading contributions appearing in Eqs. (\ref{FIN5}), (\ref{FIN6}) and (\ref{FIN7}) will be associated with $d_{E}^{(1)}$, $d_{B}^{(1)}$ and $\overline{d}^{(1)}$. 

All in all if  $\lambda_{i}$ and $\overline{\lambda}_{j}$ are not fine-tuned  
the leading-order expressions of the susceptibilities, as established above, is obtained 
by setting $M \sim \overline{M}_{P}$ and $\epsilon\ll 1$:
\begin{equation}
\chi_{X} = \sqrt{\frac{\lambda}{4\pi} }\,\,\sqrt{ 1 + \alpha_{X} \biggl(\frac{H}{M_{P}}\biggr)^2}, \qquad 
\overline{\chi}= \sqrt{\frac{\overline{\lambda}}{4\pi} }\,\,\sqrt{ 1 + \overline{\alpha} \biggl(\frac{H}{M_{P}}\biggr)^2}, 
\label{FIN8}
\end{equation}
where $X = E,\,\, B$ so that $\alpha_{X}$ and $\overline{\alpha}$ do not depend on $\phi$. It should now 
be clear that Eqs. (\ref{FIN4}) and (\ref{FIN8}) lead exactly to the same conclusions of the leading-order 
contribution of Eq. (\ref{action2}). This conclusion can also be reached in rigorous terms 
by noting that the explicit form of the gauge action can be extended 
to the case (\ref{FIN4}); this extension follows by adopting a new time parametrization and a consequent redefinition of the susceptibilities, namely:
\begin{equation}
\tau\to s = s(\tau), \qquad  d\tau= n(s)\, ds,\qquad n^2 = \chi_{E}^2/\chi_{B}^2, \qquad \chi = \sqrt{\chi_{E}\, \chi_{B}}.
\label{FIN9}
\end{equation}
The $s$-time parametrization is vaguely analogue to the $\eta$-time and this is why we always used an overdot. It should be stressed, however, 
that the auxiliary equation written in the $\eta$-time (see Eq. (\ref{EX3}) and discussion therein) applies when the gauge coupling is not asymmetric. In terms of $\chi$ and $n$ the comoving  fields are now given by 
$\vec{B} = \vec{\nabla} \times \vec{{\mathcal Y}}/\sqrt{n}$ and by $\vec{E} = - \, (\chi/\sqrt{n})\,\partial_{s} (\vec{{\mathcal Y}}/\chi)$  where $\vec{{\mathcal Y}}$ is, as usual, the comoving vector potential.  If the these expressions are inserted into Eq. (\ref{FIN4}) the full action takes the following simple form:
\begin{equation}
S_{gauge}= \frac{1}{2} \int d^3 x\, \int \, ds \biggl[ \dot{{\mathcal Y}}_{a}^2 + \biggl(\frac{\dot{\chi}}{\chi}\biggr)^2 {\mathcal Y}_{a}^2  - 2  \biggl(\frac{\dot{\chi}}{\chi}\biggr) 
{\mathcal Y}_{a} \, \dot{\mathcal Y}_{a} - \partial_{i} {\mathcal Y}_{a} \partial^{i} {\mathcal Y}_{a} - {\mathcal C}(s) {\mathcal Y}_{a} \partial_{b} {\mathcal Y}_{m} \, \epsilon^{a \, b\, m} \biggr],
\label{FIN10}
\end{equation}
where ${\mathcal C}(s) = \partial_{s} \overline{\chi}^2/ \chi^2$
where the overdots now denote a derivation with respect to the new time coordinate $s$ and should not be confused with the derivation with respect to the $\eta$-time. 
From the action (\ref{FIN10}) it is possible to solve the dynamics also in the case 
when the gauge couplings are asymmetric. In our case, however, it is sufficient to note 
that from Eq. (\ref{FIN8}) 
\begin{equation}
\chi_{E} = \chi_{B} \biggl[ 1 + {\mathcal O}(10^{-10})\biggr], \qquad n = 1 +  {\mathcal O}(10^{-10}).
\label{FIN11}
\end{equation}
In  Eq. (\ref{FIN10}), to leading order, $s \to \tau$, $\chi \to \sqrt{\lambda}$ and $\overline{\chi} \to \sqrt{\overline{\lambda}}$. The resulting expression for the gauge action will then be
\begin{equation}
S_{gauge}= \frac{1}{2} \int d^3 x\, \int \, d\tau \biggl[ {\mathcal Y}_{a}^{\,\prime\,\,2} + {\mathcal F}^2  {\mathcal Y}_{a}^2  - 2  {\mathcal F} {\mathcal Y}_{a} \,{\mathcal Y}^{\,\prime}_{a} - \partial_{i} {\mathcal Y}_{a} \partial^{i} {\mathcal Y}_{a} - \frac{\overline{\lambda}^{\prime}}{\lambda} {\mathcal Y}_{a} \partial_{b} {\mathcal Y}_{m} \, \epsilon^{a \, b\, m} \biggr].
\label{FIN12}
\end{equation}
From Eq. (\ref{FIN12})  the canonical momenta can be deduced as $\pi_{a} = {\mathcal Y}_{a}^{\prime} - {\mathcal F}\, {\mathcal Y}_{a}$; the canonical Hamiltonian associated with Eq. (\ref{FIN12}) turns out to be exactly the one already discussed in Eq. (\ref{WKB2}).  
 
The results of this investigations are, overall, as generic as the conventional models of inflation where the  dependence of the Lagrangian on the inflaton field is practically unconstrained by symmetry. This means that there are classes of models where this conclusion does not immediately follow, at least in principle. One possibility, as already mentioned, is that some of the 
 couplings $\lambda_{i}(\phi)$ and $\overline{\lambda}_{i}(\phi)$ are artificially tuned to be very large. 
From the viewpoint of the underlying inflationary model it could also happen that the inflaton has some particular symmetry (like a shift symmetry $\varphi \to \varphi + \mathrm{const}$); this possibility reminds of the relativistic theory of Van der Waals (or Casimir-Polder) interactions  \cite{SYM1,SYM2} and leads to a specific class of magnetogenesis scenarios \cite{SYM3}.
Another non-generic possibility implies that the rate of inflaton roll defined by $\eta$ remains constant (and possibly much larger than $1$), as it happens in certain fast-roll scenarios \cite{SYM4,SYM5,SYM6}. In all these cases $\chi_{E}$ and $\chi_{B}$ may have asymmetric evolutions and the general results reported here cannot be applied. 
 
\newpage
\renewcommand{\theequation}{7.\arabic{equation}}
\setcounter{equation}{0}
\section{Final remarks}
\label{sec7}
The slopes of the large-scale hypermagnetic and hyperelectric power spectra amplified by the variation of the gauge coupling from their quantum mechanical fluctuations are insensitive to the relative strength of the parity-breaking terms. The pseudoscalar contributions to the effective action control instead the slopes and amplitudes of the gyrotropic spectra. After 
proposing a strategy for the approximate estimate of the gauge power spectra we analyzed a number of explicit examples and found that they all corroborate the general results. The form of the gauge  spectra for a generic variation of the pseudoscalar interaction term $\overline{\lambda}$  has been discussed in connection 
with the dynamics of the gauge coupling which is related, within the present notations, to the inverse of $\lambda$. The scaling of $(\overline{\lambda}^{\, \prime}/\lambda)$ (where the prime denotes the conformal time derivative) ultimately determines the properties of the corresponding power spectra. If $(\overline{\lambda}^{\, \prime}/\lambda)$ decreases as $\tau^{-1- \alpha}$ two different physical regimes emerge. When $\alpha>1$ the hypermagnetic and hyperelectric power spectra are practically unaltered by the presence of the pseudoscalar terms. This is true both for the early-time and for the late-time power spectra. If $0 < \alpha <1$ the slopes of the hypermagnetic and hyperelectric power spectra are still not affected but the overall amplitude gets modified depending on the value of $\alpha$. Two particular limits must be separately treated and they correspond to the boundaries of the two regions (i.e. $\alpha\to 1$ and $\alpha\to 0$). 

The production of the Chern-Simons condensates has been investigated under the assumption that the gauge coupling smoothly evolves during a quasi-de Sitter phase and then flattens out in the radiation epoch by always remaining perturbative. 
In all physical limits the gauge power spectra have been also illustrated at late times with the purpose of discussing the phenomenological impact of the various regions of the parameter space. By focussing on the case of increasing gauge coupling (which we regard as the most plausible) we showed that the magnetogenesis requirements are satisfied in spite of the pseudoscalar couplings. In practice only the region $0 \leq \alpha \ll 1$ is relevant for the generation of the baryon asymmetry. In the case of decreasing gauge couplings the results are quantitatively different but the general logic remains the same. All in all 
we can summarize the phenomenological implications of this analysis in the following manner:
\begin{itemize}
\item{} if $\alpha \geq 1$ the magnetogenesis requirements can be reproduced but the BAU is not generated 
except for a corner of the parameter space where the rate of variation of the gauge coupling 
is close to the critical density limit;
\item{} if $0 \leq \alpha < 1$ we have instead that the baryogenesis and the magnetogenesis requirements 
are simultaneously satisfied so that the large-scale magnetic fields and the BAU can be 
seeded by the same mechanism and in the same region of the paraneter space;
\item{} in the most promising region of the parameter space $\eta_{B} = {\mathcal O}(10^{-10})$ (or slightly larger)
while the magnetic power spectra associated with the modes reentering after symmetry breaking may even be of the order of a few hundredths of a nG over typical length scales comparable with the Mpc prior to the collapse of the protogalaxy.  
\end{itemize}
Concerning the above statements we first remark that the regions where $\eta_{B}$ is a bit larger than $10^{-10}$ should not be excluded 
since various processes can independently reduce the BAU generated in this way. The 
second comment is that, in this analysis, we mainly focussed on the case of increasing gauge coupling. For decreasing gauge couplings 
only few results have been reported just to avoid a repetitive analysis. In spite of that when the gauge coupling decreases 
the main theoretical result still holds: the slopes of the large-scale hypermagnetic and hyperelectric power spectra are insensitive to the relative strength of the parity-breaking terms that are instead essential to compute the gyrotropies and the values of 
the Chern-Simons condensates.

We finally demonstrated that the proposed approaches and the obtained results hold generically for the whole class of inflationary models where the inflaton is not constrained by any underlying symmetry. This question has been addressed in the framework of the effective field theory description of the inflationary scenarios which can be extended to include the contributions of the hypercharge field. Rather than assuming a particular underlying description, all the generally covariant terms potentially appearing with four space-time derivatives in the effective action have been included and weighted by inflaton-dependent couplings. During a quasi-de Sitter stage the corrections are immaterial in the case of generic inflationary models but may become relevant in some non-generic scenarios where either the inflaton has some extra symmetry or the higher-order terms are potentially dominant. In this sense the present findings both simplify and generalize the effective description of the gauge fields during inflation and in the subsequent stages of the expansion.

\section*{Acknowledgements}

It is a pleasure to thank T. Basaglia, A. Gentil-Beccot and S. Rohr of the CERN scientific 
information service for their kind help.

\newpage 

\begin{appendix}

\renewcommand{\theequation}{A.\arabic{equation}}
\setcounter{equation}{0}
\section{Explicit forms of the mode functions in the general case}
\label{APPA}
Since for $\tau < -\tau_{\pm}$ the Wronskian normalization  must be enforced (see Eq. (\ref{WKB6}) and discussion thereafter),
the explicit form of the mode functions follows from Eqs. (\ref{EX5}) and (\ref{EX6}).
In particular, the hypermagnetic mode functions $f_{k,\,\pm}(\tau)$ are: 
\begin{eqnarray}
f_{k,\,+}(\tau) &=& \frac{e^{i k \tau_{+}}}{\sqrt{2 k} \, \Delta_{+}} \sqrt{\frac{z}{z_{+}}} \biggl\{ \biggl[ I_{\nu}(z)\, \biggl(\frac{ d K_{\nu}}{d z}\biggr)_{+} - K_{\nu}(z) \, \, \biggl(\frac{ d I_{\nu}}{d z}\biggr)_{+}\biggr] 
\nonumber\\
&+& \biggl(\frac{1}{2} - i\biggr) \, x_{1} \,\biggl[I_{\nu}(z) \, K_{\nu}(z_{+}) - I_{\nu}(z_{+}) \, K_{\nu}(z) \biggr]
\biggr\},
\label{APPA1}\\
f_{k,\,-}(\tau) &=& \frac{e^{i k \tau_{-}}}{\sqrt{2 k} \, \Delta_{-}} \sqrt{\frac{z}{z_{-}}} \biggl\{ \biggl[ J_{\nu}(z)\, \biggl(\frac{ d Y_{\nu}}{d z}\biggr)_{-} - Y_{\nu}(z) \, \, \biggl(\frac{ d J_{\nu}}{d z}\biggr)_{-}\biggr] 
\nonumber\\
&+& \biggl(\frac{1}{2} - i\biggr) \, x_{1} \,\biggl[J_{\nu}(z) \, Y_{\nu}(z_{-}) - J_{\nu}(z_{-}) \, Y_{\nu}(z) \biggr]
\biggr\}.
\label{APPA2}
\end{eqnarray}
If $S_{\nu}(z)$ represents any of the four different Bessel functions appearing above, 
the concise notation employed in Eqs. (\ref{APPA1})--(\ref{APPA2})  corresponds to:
\begin{equation}
 S_{\nu}( z) = S_{\nu}[c(z)], \qquad S_{\nu}( z_{\pm}) = S_{\nu}[c_{\pm}] = S_{\nu}[c(z_{\pm})], \qquad
 \biggl(\frac{d S_{\nu}}{dz}\biggr)_{\pm} =  \biggl(\frac{d S_{\nu}}{dz}\biggr)_{z= z_{\pm}}.
 \label{APPA2a}
\end{equation} 
Recalling  Eqs. (\ref{EX3}) and (\ref{EX7}) the definitions of $\nu$ and $c(z)$ are:
\begin{equation}
c(z) = \frac{2 \sqrt{ b_{0} \, x_{1}}}{|1- \alpha|} \, z^{(1-\alpha)/2} = \frac{2 \sqrt{ q \, x_{1}}}{|1- \alpha|} \, \biggl(- \frac{\tau}{\tau_{1}}\biggr)^{(1-\alpha)/2}, \qquad \nu = \biggl| \frac{ 2 \gamma - 1}{1 - \alpha} \biggr|,
\label{APPA2b}
\end{equation}
where $x_{1} = k \tau_{1} \ll 1$. In full analogy with Eq. (\ref{APPA2b}) the explicit expression of $c(z_{\pm})$ is given by 
\begin{equation}
c(z_{\pm}) = \frac{2 \, \sqrt{b_{0}\, x_{1}}}{|1 - \alpha|} \, z_{\pm}^{(1-\alpha)/2} = \frac{2 \sqrt{b_{0}}}{|1 - \alpha|} x_{1}^{\alpha/2}[1 + \epsilon_{\pm}(k)]^{(1-\alpha)/2}.
\label{APPA2c}
\end{equation}  (see also Figs. \ref{GGG1} and \ref{GGG2} and discussion therein)
Since  $x_{1}$ ranges  between ${\mathcal O}(10^{-23})$ (or smaller)  
and  ${\mathcal O}(10^{-14})$ (see Eq. (\ref{EX10}) and discussion therein)
 there three complementary cases where Eqs. (\ref{APPA2b})--(\ref{APPA2c}) can be analyzed:
\begin{itemize}
\item{} when $x_{1} \ll 1$ we have that $|c(z_{\pm})| \ll 1$ ( both for $0 < \alpha < 1$ and for $\alpha >1$);
\item{} if $\alpha \to 1$ Eqs. (\ref{APPA2b})--(\ref{APPA2c}) are formally divergent ;
\item{} finally for $\alpha \to 0$ we have that $c(z_{\pm}) = 2 \sqrt{b_{0}}[ 1 + \epsilon_{\pm}]$.
\end{itemize}
All in all the cases $\alpha =0$ and $\alpha =1$ are not singular but they must be separately treated 
as we showed in the bulk of the paper.
In Eqs. (\ref{APPA1})--(\ref{APPA2}) we also introduced $\Delta_{\pm}$ which are
the Wronskians of the corresponding solutions:
\begin{eqnarray}
\Delta_{+} &=& I_{\nu}(z_{+}) \,\biggl(\frac{ d K_{\nu}}{d z}\biggr)_{+} - K_{\nu}(z_{+}) \biggl(\frac{ d I_{\nu}}{d z}\biggr)_{+} = - \frac{(1-\alpha)}{2 z_{+}} = - \, \frac{(1- \alpha) x_{1}}{2},
\label{APPA3}\\
\Delta_{-}&=& J_{\nu}(z_{-}) \,\biggl(\frac{ d Y_{\nu}}{d z}\biggr)_{-} - Y_{\nu}(z_{-}) \biggl(\frac{ d J_{\nu}}{d z}\biggr)_{-} =  \frac{(1-\alpha)}{\pi\, z_{-}} =\, \frac{(1- \alpha) x_{1}}{\pi},
\label{APPA4}
\end{eqnarray}
where $z_{\pm} = (-\tau_{\pm}/\tau_{1}) = (1+ \epsilon_{\pm})/x_{1}$. 
Finally, recalling the notations of Eqs. (\ref{APPA1})--(\ref{APPA2}) and (\ref{APPA3})--(\ref{APPA4}) the 
hyperelectric mode function  $g_{k,\,\pm}(\tau)$ are:
\begin{eqnarray}
\frac{g_{k,\, +}}{k} &=& - \frac{e^{i k\tau_{+}}}{x_{1} \, \Delta_{+} \, \sqrt{2 k}} \, \sqrt{\frac{z}{z_{+}}} \biggl\{ 
\frac{x_{1}}{z} \biggl(\frac{1}{2} -\gamma\biggr) \biggl(\frac{1}{2} -i\biggr) \biggl[ K_{\nu}(z_{+}) \, I_{\nu}(z) - K_{\nu}(z) \, I_{\nu}(z_{+}) \biggr] 
\nonumber\\
&+& \biggl[ \biggl(\frac{d K_{\nu}}{dz} \biggr)_{+} \biggl(\frac{d I_{\nu}}{dz} \biggr) - \biggl(\frac{d I_{\nu}}{dz} \biggr)_{+} 
\biggl(\frac{d K_{\nu}}{dz} \biggr)\biggr]
+ \biggl(\frac{1}{2} -i\biggr) x_{1} \biggl[ K_{\nu}(z_{+}) \biggl(\frac{d I_{\nu}}{dz} \biggr) - I_{\nu}(z_{+}) \biggl(\frac{d K_{\nu}}{dz} \biggr) \biggr]
\nonumber\\
&+& \frac{1}{z} \biggl(\frac{1}{2} -\gamma\biggr)\biggl[ I_{\nu}(z) \biggl(\frac{d K_{\nu}}{dz} \biggr)_{+} - K_{\nu}(z) \biggl(\frac{d I_{\nu}}{dz} \biggr)_{+} \biggr] \biggr\},
\label{APPA5}\\
\frac{g_{k,\, -}}{k} &=& - \frac{e^{i k\tau_{-}}}{x_{1} \, \Delta_{-} \, \sqrt{2 k}} \, \sqrt{\frac{z}{z_{-}}} \biggl\{ 
\frac{x_1}{z} \biggl(\frac{1}{2} -\gamma\biggr) \biggl(\frac{1}{2} -i\biggr) \biggl[ Y_{\nu}(z_{-}) \, J_{\nu}(z) - Y_{\nu}(z) \, J_{\nu}(z_{-}) \biggr] 
\nonumber\\
&+& \biggl[ \biggl(\frac{d Y_{\nu}}{dz} \biggr)_{-} \biggl(\frac{d J_{\nu}}{dz} \biggr) - \biggl(\frac{d J_{\nu}}{dz} \biggr)_{-} 
\biggl(\frac{d Y_{\nu}}{dz} \biggr)\biggr] + \biggl(\frac{1}{2} -i\biggr) x_{1} \biggl[ Y_{\nu}(z_{-}) \biggl(\frac{d J_{\nu}}{dz} \biggr) - J_{\nu}(z_{-}) \biggl(\frac{d Y_{\nu}}{dz} \biggr) \biggr]
\nonumber\\
&+& \frac{1}{z} \biggl(\frac{1}{2} -\gamma\biggr)\biggl[ J_{\nu}(z) \biggl(\frac{d Y_{\nu}}{dz} \biggr)_{-} - Y_{\nu}(z) \biggl(\frac{d J_{\nu}}{dz} \biggr)_{-} \biggr] \biggr\}.
\label{APPA6}
\end{eqnarray}
It is is useful to mention that in the examples discussed here we always 
considered the case $\alpha \geq 0$. For the sake of completeness we want also to discuss the case $-1<\alpha<0$ where the equation defining the turning points can be written as:
\begin{equation}
(- k \tau)^2 \simeq \pm (- k \tau)^{1 - \alpha} \, b_{0} \, x_{1}^{\alpha} + \gamma(\gamma -1).
\label{NN1}
\end{equation}
If $\alpha$ is negative, the term $x_{1}^{\alpha}$ in Eq. (\ref{NN1}) 
will be, in principle, extremely large (recall, in this respect, that according to Eq. (\ref{EX10}) 
$x_{1} = {\mathcal O}(10^{-20})$). The second term at the right-hand side of (\ref{NN1})  can then be 
neglected and the solution would be 
\begin{equation}
- k \tau_{\pm} \simeq b_{0}^{1/(1+\alpha)} \, x_{1}^{\alpha/(\alpha+1)}\, e^{i \sigma_{\pm}},
\label{NN2}
\end{equation}
where $\sigma_{+} =0$ and $\sigma_{-} = \pi/(1+\alpha)$. It is inappropriate to talk about turning points in this case since we are in the situation where $c(z)$ and $c_{\pm}$ are both very large and the magnetic spectra are exponentially divergent. The mode functions (\ref{APPA1})--(\ref{APPA2}) and (\ref{APPA5})--(\ref{APPA6}) must be evaluated for $c(z) \gg 1$ and $c_{\pm} \gg1$. Up to irrelevant 
numerical factors the magnetic power spectrum can be written, in this case, as
\begin{equation}
P_{B}(k,\tau) \simeq a^4\, H^4 (- k \,\tau)^{(9 +\alpha)/2} \, x_{1}^{-\alpha/2} b_{0}^{-1/2} e^{ c_{+}(b_{0}, x_{1})}, \qquad c_{+}(b_{0}, x_{1}) = \frac{2}{|1 - \alpha|} b_{0}^{1/(\alpha +1)}\, x_{1}^{\alpha/[2 (\alpha +1)]}.
\label{NN3}
\end{equation}
From Eq. (\ref{NN3}) we observe that the exponential divergence in the hypermagnetic power spectra 
has a counterpart in the electric case. To avoid that the critical density bound during inflation is
strongly violated we must therefore require that the power spectra are not affected by the exponential 
increase. From Eq. (\ref{NN3}) this happens provided $\alpha < -0.01$ assuming $x_{1} = {\mathcal O}(10^{-23})$, as required by Eq. (\ref{EX10}). But this means, according to Eq. (\ref{NN3})  that the power spectrum at the end of inflation will be of the order of $H_{1}^4 x_{1}^{9/2}$ which is irrelevant for any observational purpose. 

\newpage
\renewcommand{\theequation}{B.\arabic{equation}}
\setcounter{equation}{0}
\section{Comparing the WKB results and the exact solutions}
\label{APPB}
In section \ref{sec4} we showed, in general terms, that the auxiliary equation (i.e. Eq. (\ref{WKB16})) 
can be solved explicitly by introducing a new time 
parametrization (i.e. the $\eta$-time). In terms of the $\eta$-time Eq. (\ref{WKB16}) 
assumes a more friendly aspect (see e.g. Eqs. (\ref{EX1}) and (\ref{EX3})); the resulting 
equation can then be solved for different values of $\alpha$ and compared with the WKB 
expectation. In what follows, for the sake of accuracy, we shall analyze the whole range of $\alpha$ by separating the 
generic case (i.e. $\alpha > 0$) from the particular cases $\alpha=0$ and $\alpha=1$. 
We remind that the parametrization of the pump fields $\overline{\lambda}$ and $\lambda$ 
is the one introduced in Eqs. (\ref{EX2a}) and (\ref{EX2b}); in particular 
$\alpha$ and $b_{0}$ control the strength of the pseudoscalar term while 
$\gamma$ is associated with the evolution of the scalar  contribution.
\subsection{The generic case $\alpha>0$}
\label{APPB44}
Recalling the relation between $\eta$-parametrization and conformal time 
(see, for instance, Eq. (\ref{EX4})) we have from Eq. (\ref{EX13}) that at the turning points $\tau = \tau_{\pm}$:
\begin{equation}
  - q \eta_{\pm}= \frac{2 \sqrt{b_{0}}}{|1 - \alpha|} x_{1}^{\alpha/2} ( 1 + \epsilon_{\pm})^{(1-\alpha)/2}\ll 1.
 \label{cond2b}
 \end{equation}
Since the critical values of $\alpha$ are avoided by requiring $\alpha \neq 1$ and $\alpha\neq 0$, Eq. (\ref{cond2b}) holds separately for the ranges $0< \alpha<1$ and for $ \alpha >1$.
The mode functions (\ref{EX5}) and (\ref{EX6}) can therefore be matched 
with their plane-wave limit. As already mentioned these results  can be founds in Eqs. (\ref{APPA1})--(\ref{APPA2}) and (\ref{APPA3})--(\ref{APPA4}). To compute the power 
spectra it is necessary to study the hypermagnetic and hyperelectric mode functions 
for typical wavelengths larger than the effective horizon. This will be the purpose 
of the present appendix. The obtained results will be compared with the WKB 
results valid in the same limit.

\subsubsection{Explicit expression of the mode functions in the long-wavelength limit}
\label{APPB441}
In the small argument limit Eqs. (\ref{APPA1})--(\ref{APPA2})  become:
 \begin{eqnarray}
 f_{k,\pm}(\tau) &=& - \frac{1}{\sqrt{2 k}} \biggl[ A_{k,\, \pm} ( - k \tau)^{1/2 + (1-\alpha) \nu/2} - B_{k,\,\pm} (- k\tau)^{1/2- (1-\alpha)\nu/2} \biggr],
\label{cond3a}\\
 A_{k,\, \pm} &=& \frac{(1+ \epsilon_{\pm})^{-1/2 -(1-\alpha)\nu/2}}{2 \nu}\biggl[ \frac{2}{(1 - \alpha)}\biggl( \frac{1}{2} - i\biggr) - \frac{\nu}{1 + \epsilon_{\pm}}\biggr],
 \label{cond3b}\\
 B_{k,\, \pm} &=& \frac{(1+ \epsilon_{\pm})^{-1/2 +(1-\alpha)\nu/2}}{2 \nu}\biggl[ \frac{2}{(1 - \alpha)}\biggl( \frac{1}{2} - i\biggr) + \frac{\nu}{1 + \epsilon_{\pm}}\biggr].
 \label{cond3c}
 \end{eqnarray}
Equations (\ref{cond3a}) and (\ref{cond3b})--(\ref{cond3c}) hold provided $\nu \neq 0$; as $\nu\to 0$ the small argument limit of the corresponding Bessel functions  involves a logarithmic correction \cite{MM1,MM2} which is also present in the absence of anomalous contributions (i.e. for $\overline{\lambda} \to 0$). For the sake 
of conciseness this discussion will be omitted but the final result will be  the same since the limit 
of Eqs. (\ref{APPA1})--(\ref{APPA2}) for $\nu\to 0$ will reproduce that WKB results in the same limit (i.e. $\gamma\to 1/2$). The hyperelectric mode functions of Eqs. (\ref{APPA3})--(\ref{APPA4}) 
are obtained with the same strategy:
 \begin{eqnarray}
 g_{k,\,\pm}(\tau) &=& \sqrt{\frac{k}{2}}  \biggl[ C_{k,\, \pm}\, (-k\tau)^{-1/2 + (1-\alpha)\nu/2} + D_{k,\, \pm}\, (-k\tau)^{-1/2 - (1-\alpha)\nu/2} \biggr],
 \label{cond4a}\\
 C_{k,\, \pm} &=&  \frac{(1+ \epsilon_{\pm})^{1/2 -(1-\alpha)\nu/2}}{2 \nu} \biggl[ \biggl( \frac{1}{2} - i\biggr) \biggl(\frac{1 - 2\gamma}{1 -\alpha} + \nu\biggr) 
 - \frac{\nu( 1-\alpha)}{2 ( 1 + \epsilon_{\pm})} \biggl( \frac{1 - 2\gamma}{1 -\alpha} + \nu\biggr) \biggr],
 \label{cond4b}\\
D_{k,\, \pm} &=&   \frac{(1+ \epsilon_{\pm})^{1/2 +(1-\alpha)\nu/2}}{2 \nu} \biggl[ \biggl( \frac{1}{2} - i\biggr) \biggl(\frac{2\gamma-1}{1 -\alpha} + \nu\biggr) 
 - \frac{\nu( 1-\alpha)}{2 ( 1 + \epsilon_{\pm})} \biggl( \frac{1 - 2\gamma}{1 -\alpha} - \nu\biggr) \biggr].
\label{cond4c}
\end{eqnarray}

\subsubsection{Hypermagnetic power spectra in the long-wavelength limit }
\label{APPB442}
 If Eqs. (\ref{cond3a}) and (\ref{cond3b})--(\ref{cond3c}) are inserted into Eq. (\ref{WKB12}) the hypermagnetic power spectrum turns out to be:
 \begin{eqnarray}
 P_{B}(k,\tau) &=& \frac{ a^4 \, H^4}{8 \pi^2} \biggl\{ \biggl| A_{k,\,+} \,\, (- k \tau)^{5/2 + (1-\alpha)\nu/2} - B_{k,\,+} \,\, (- k \tau)^{5/2 - (1-\alpha)\nu/2} \biggr|^2 
 \nonumber\\
 &+& \biggl| A_{k,\,-} \,\, (- k \tau)^{5/2 + (1-\alpha)\nu/2} - B_{k,\,-} \,\, (- k \tau)^{5/2 - (1-\alpha)\nu/2} \biggr|^2 \biggr\}
 \label{cond5}
 \end{eqnarray}
Equation (\ref{cond5}) implies that the power spectrum outside the horizon has always the same slope for $| k \tau | \ll 1$. Indeed, depending on the interval of $\alpha$, the combination $(1-\alpha) \nu$ takes two opposite expressions: in the range $0< \alpha < 1$ (thanks to the absolute values entering the definition of $\nu$) we have that 
 $ (1-\alpha) \nu = |2 \gamma -1|$; for the same reason when $\alpha >1$ we rather have $(1-\alpha) \nu = - |2 \gamma -1|$. In summary we have:
\begin{eqnarray} 
0< \alpha < 1 \qquad &\Rightarrow& \qquad (1-\alpha) \nu = (1- \alpha) \biggl| \frac{2 \gamma -1}{1 - \alpha}\biggr| = |2 \gamma -1|,
\label{cond6a}\\
\alpha > 1 \qquad &\Rightarrow& \qquad (1-\alpha) \nu = (1- \alpha) \biggl| \frac{2 \gamma -1}{1 - \alpha}\biggr| = - |2 \gamma -1|.
\label{cond6b}
\end{eqnarray}
For $0< \alpha < 1$ the contributions 
proportional to $B_{k,\pm}$ dominate in Eq. (\ref{cond5}) and therefore we obtain:
\begin{equation}
P_{B}(k,\tau) = \frac{a^4 H^4}{8 \pi^2} \biggl[ \bigl|B_{k,\,+}\bigr|^2 + \bigl|B_{k,\,-}\bigr|^2\biggr] ( - k\tau)^{ 5 - |2 \gamma -1|},\qquad\qquad 0< \alpha < 1.
\label{cond7a}
\end{equation}
Conversely in the range $\alpha > 1$ the contributions 
proportional to $A_{k,\pm}$ dominate in Eq. (\ref{cond5}) and the hypermagnetic power spectrum becomes:
\begin{equation}
P_{B}(k,\tau) = \frac{a^4 H^4}{8 \pi^2} \biggl[ \bigl|A_{k,\,+}\bigr|^2 + \bigl|A_{k,\,-}\bigr|^2\biggr] ( - k\tau)^{ 5 - |2 \gamma -1|}, \qquad\qquad \alpha > 1.
\label{cond7b}
\end{equation}
Equations (\ref{cond7a})--(\ref{cond7b}) imply that  the WKB result of Eq. (\ref{WKB24}) is recovered up to 
an overall amplitude. Even though the prefactors are  
immaterial for the comparison\footnote{The reason is that the WKB result reproduces the exact amplitude up to overall 
constant terms ${\mathcal O}(1)$.} it is amusing to remark that, to lowest order in $|\epsilon_{\pm}| < 1$,  the amplitudes in Eqs. (\ref{cond7a})--(\ref{cond7b}) coincide. Indeed,
using the shorthand notation $ \mu = |\gamma-1/2|$ the prefactors of Eqs. (\ref{cond7a})--(\ref{cond7b}) give exactly the same result:
\begin{equation}
P_{B}(k,\tau) =  \frac{a^4 H^4}{8 \pi^2}\biggl[ \frac{4 \mu  (\mu +1)+5}{16 \mu ^2}+\frac{\left(8 \mu ^3-4 \mu ^2+2 \mu -5\right)
   (\epsilon_{+} + \epsilon_{-}) }{16 \mu ^2} + {\mathcal O}\left(\epsilon_{+}^2\right) +  {\mathcal O}\left(\epsilon_{-}^2\right)\biggr] ( - k\tau)^{ 5 - 2\mu }.
\label{cond8}
\end{equation}
As in the case of Eq. (\ref{cond3a}), also Eq. (\ref{cond8}) holds provided $\mu \neq 0$; once more if $\mu \to 0$ 
there will be logarithmic divergences coming from the small argument limit of the Bessel functions appearing in Eqs. (\ref{APPA1})--(\ref{APPA2}) that must be separately treated. This is actually not surprising since 
$\nu \neq 0$ implies $\mu\neq 0$ (provided, as we are assuming in this portion of the discussion, $\alpha \neq 1$).
All in all we then conclude that, within the accuracy of the approximation, Eq. (\ref{cond8}) coincides with the result of Eq. (\ref{WKB24}). 
The same analysis leading to the general result (\ref{cond8}) can be repeated in the case of the  gyrotropic spectra by inserting Eqs. (\ref{cond3a}) and (\ref{cond3b})--(\ref{cond3c}) into Eq. (\ref{WKB13}):
\begin{eqnarray}
0< \alpha < 1 \qquad \Rightarrow \qquad P^{(G)}_{B}(k,\tau) &=& \frac{a^4 H^4}{8 \pi^2} \biggl[ \bigl|B_{k,\,+}\bigr|^2 - \bigl|B_{k,\,-}\bigr|^2\biggr] ( - k\tau)^{ 5 - 2\mu},
\nonumber\\
&=& \frac{a^4 H^4}{8 \pi^2} \biggl[\frac{\left(8 \mu ^3-4 \mu ^2+2 \mu -5\right)
   (\epsilon_{+} - \epsilon_{-}) }{16 \mu ^2}\biggr] ( - k\tau)^{ 5 - 2\mu},
\label{cond9a}\\
\alpha > 1 \qquad  \Rightarrow \qquad P^{(G)}_{B}(k,\tau) &=& \frac{a^4 H^4}{8 \pi^2} \biggl[ \bigl|A_{k,\,+}\bigr|^2 - \bigl|A_{k,\,-}\bigr|^2\biggr] ( - k\tau)^{ 5 -2\mu}.
\nonumber\\
&=& \frac{a^4 H^4}{8 \pi^2} \biggl[\frac{\left(8 \mu ^3-4 \mu ^2+2 \mu -5\right)
   (\epsilon_{+} - \epsilon_{-}) }{16 \mu ^2}\biggr] ( - k\tau)^{ 5 - 2\mu}.
\label{cond9b}
\end{eqnarray}
Again Eqs. (\ref{cond9a})--(\ref{cond9b}) coincide, up to numerical factors, 
with the spectra already deduced in Eq. (\ref{WKB26a}). 

\subsubsection{Hyperelectric power spectra in the long-wavelength limit }
\label{APPB443}
 Since the coefficients of Eqs. (\ref{cond4b})--(\ref{cond4c}) do not only contain $\nu$ but also the explicit value of $\gamma$ (see Eq. (\ref{WKB14}) and comment thereafter), in the derivation of the hyperelectric power spectra we must distinguish the different ranges of $\gamma$. The reason of this difference is that the hyperelectric mode functions do not simply coincide with the derivative of $f_{k,\pm}(\tau)$ but they are shifted by ${\mathcal F} f_{k,\pm}(\tau)$. Inserting then Eq. (\ref{cond4a})  into 
Eq. (\ref{WKB12}) and assuming $0< \alpha < 1$ the hyperelectric power spectra turn out to be:
\begin{eqnarray}
P_{E}(k,\tau) &=& \frac{a^4 H^4}{8 \pi^2} \biggl[ \bigl| D_{k,\, +}\bigr|^2 + \bigl| D_{k,\, -}\bigr|^2\biggr] ( - k \tau)^{4 - 2 \gamma}, \qquad 0<\alpha< 1, \qquad \gamma > 1/2,
\nonumber\\
&=&\frac{a^4 H^4}{8 \pi^2} \biggl[  \frac{(4 \mu  (\mu +1)+5) (2 \gamma +2 \mu -1)^2}{64 \mu ^2}
 \nonumber\\
 &+& \frac{(2 \mu +1) \left(4 \mu ^2+5\right) (2 \gamma +2 \mu -1)^2}{64 \mu
   ^2} (\epsilon_{+} + \epsilon_{-}) +
 {\mathcal O}\left(\epsilon_{+} ^2\right) + {\mathcal O}\left(\epsilon_{-} ^2\right)\biggr]
   ( - k \tau)^{4 - 2 \gamma}
\label{cond10a}\\
P_{E}(k,\tau) &=& \frac{a^4 H^4}{8 \pi^2} \biggl[ \bigl| C_{k,\, +}\bigr|^2 + \bigl| C_{k,\, -}\bigr|^2\biggr] ( - k \tau)^{4 - 2 \gamma}, \qquad 0<\alpha< 1, \qquad 0< \gamma < 1/2,
\nonumber\\
&=&  \frac{a^4 H^4}{8 \pi^2} \biggl[ \frac{(4 (\mu -1) \mu +5) (-2 \gamma +2 \mu +1)^2}{64 \mu ^2}
\nonumber\\
&-& \frac{ \left[(2 \mu -1) \left(4 \mu ^2+5\right) (-2 \gamma +2 \mu +1)^2\right]}{64 \mu ^2} (\epsilon_{+} + \epsilon_{-}) + {\mathcal O}\left(\epsilon_{+} ^2\right) + {\mathcal O}(\epsilon_{-}^2) \biggr] ( - k \tau)^{4 - 2 \gamma}.
\label{cond10b}
\end{eqnarray}
As in the hypermagnetic case Eqs. (\ref{cond10a})--(\ref{cond10b}) ultimately coincide\footnote{In fact, when $\gamma >1/2$ (as in Eq. (\ref{cond10a})) 
we have that $\mu = \gamma -1/2$; conversely if $\gamma < 1/2$   (as in Eq. (\ref{cond10b})) we must 
have that $ \mu = 1/2 -\gamma$.}:
\begin{equation}
P_{E}(k,\tau) = \frac{a^4 H^4}{8 \pi^2} \biggl\{\left(\gamma ^2+1\right)+\gamma  [2 (\gamma -1) \gamma +3](\epsilon_{+} + \epsilon_{-})  +{\mathcal O}\left(\epsilon_{+}^2\right)+{\mathcal O}\left(\epsilon_{-}^2\right)\biggr\}
(- k \tau)^{4 - 2 \gamma}, \qquad 0< \alpha< 1,
\label{cond11}
\end{equation}
and this result holds in spite of the interval of $\gamma$. Finally the gyrotropic spectrum associated with Eq. (\ref{cond11}) is:
\begin{equation}
P^{(G)}_{E}(k,\tau) = \frac{a^4 H^4}{8 \pi^2} \biggl[\gamma  (2 (\gamma -1) \gamma +3) (\epsilon_{+} + \epsilon_{-}) 
+{\mathcal O}\left(\epsilon_{+}^2\right)-{\mathcal O}\left(\epsilon_{-}^2\right)\biggr] (- k \tau)^{4 - 2 \gamma}, \qquad 0< \alpha< 1.
\label{cond12}
\end{equation}
Equations (\ref{cond11})--(\ref{cond12}) hold in the interval $0 <\alpha <1$. Since the logic has been 
already illustrated we now simply mention that when $\alpha > 1$ 
the roles of $C_{k,\,\pm}$ and $D_{k,\,\pm}$ appearing in Eqs. (\ref{cond4b})--(\ref{cond4c}) are exchanged.
This means $C_{k,\,\pm}$ determines the spectra for $\gamma >1/2$ while $D_{k,\,\pm}$ will be relevant  
for the range $0<\gamma <1/2$. Thus, also for $\alpha>1$ the power spectra coincide exactly with Eqs. (\ref{cond11})--(\ref{cond12}).  Indeed, as already observed in Eqs. (\ref{cond6a})--(\ref{cond6b}) 
this happens  since $ 2 \mu = (1 -\alpha)\nu$ where, as usual, $\nu = |2\gamma -1|/|1-\alpha|$. If $ \alpha > 1$ we have $2 \mu = - |2\gamma -1$ and this ultimately implies that Eqs. (\ref{cond11}) and (\ref{cond12}) are verified.

\subsection{The particular case $\alpha = 0$}
\label{APPB45}
\subsubsection{Exact form of the mode functions}
When $\overline{\lambda}$ and $\lambda$ are proportional we have that $\overline{\lambda}^{\prime}/\lambda= 
b_{0}/\tau$ and $\alpha \to 0$ in  Eqs. (\ref{EX2a})--(\ref{EX2b}). In this situation  the explicit expression of Eq. (\ref{WKB14}) is:
\begin{eqnarray}
f_{k,\, \pm}^{\prime\prime} + \biggl[ k^2 \pm \frac{ k b_{0}}{\tau} - \frac{\gamma (\gamma-1)}{\tau^2} \biggr] f_{k, \pm} =0, \qquad\qquad
g_{k,\,\pm} = f_{k,\,\pm}^{\prime} - \frac{\gamma}{\tau} f_{k,\,\pm}.
\label{a02}
\end{eqnarray}
Equation (\ref{EX7}) implies that,  in the limit, $(- q \eta) = 2 \sqrt{b_{0}} (- k\tau)$ which is smaller than $1$ even for moderate values of $b_{0}$. Since $| q \eta_{\pm}| = 2 b_{0}$ there are in fact 
two possibilities: $(i)$ if $b_{0} \leq {\mathcal O}(1)$ we are, in practice, in the same situation of $\alpha >0$;
$(ii)$ conversely if $b_{0} > {\mathcal O}(1)$  at the tuning points $|q \eta_{\pm}|  > {\mathcal O}(1)$ while, depending on the specific value of $b_{0}$ $(- q \eta)$ can either be smaller or larger than $1$. This 
means that, for
 $\alpha \to 0$ and $b_{0} \gg 1$, the Bessel functions should not be expanded in the limit of small arguments but rather in the large argument limit. 
 
 We could consider, again, the limits of Eqs. (\ref{APPA1})--(\ref{APPA2}) and (\ref{APPA5})--(\ref{APPA6}) in the various regions of the parameter space. However to avoid a repetitive (and lengthy) discussion it is preferable  to solve directly Eq. (\ref{a02}) in terms of Whittaker's functions \cite{MM1,MM2}. 
After an appropriate rotation in the complex plane  Eq. (\ref{a02}) becomes:
\begin{eqnarray} 
\frac{d^2 f_{k,\, \pm}}{d y^2 } + \biggl[ - \frac{1}{4} \mp \frac{i \, b_{0}}{2 y} - \frac{\gamma ( \gamma -1)}{y^2} \biggr] f_{k,\, \pm}=0,
\label{a03}
\end{eqnarray}
where  $y =  2 i k \tau$. The solution of Eq. (\ref{a03}) with the correct boundary 
conditions is
\begin{eqnarray}
 f_{k\, \pm}(y) &=& \frac{e^{-i \pi/4 \pm \, \pi \,b_{0}/4}}{\sqrt{2 k}} \, W_{\mp i\, b_{0}/2,\, \mu}(y), \qquad\qquad \mu = |\gamma -1/2|,
\label{a04}\\
g_{k\, \pm}(y) &=& i \, \sqrt{2\, k}\, e^{-i \pi/4 \pm \pi b_{0}/4} \, \biggl[ \frac{y \pm (i b_{0} \mp 2 \gamma)}{2 y} W_{\mp i b_{0}/2, \, \mu}(y) 
- \frac{W_{1 \mp i b_{0}/2,\, \mu}(y)}{y} \biggr],
\label{a05} 
\end{eqnarray} 
where  $W_{\zeta, \mu}(x)$ is the Whittaker's function with generic argument $x$ and indices $(\zeta, \, \mu)$.

\subsubsection{Gauge spectra and comparison with the WKB result}
After inserting Eq. (\ref{a04}) into Eqs. (\ref{WKB12})--(\ref{WKB13})  the hypermagnetic power spectra in the 
 large-scale limit during inflation (i.e.  $|k \tau| \ll 1$) become: 
\begin{eqnarray}
P_{B}(k,\tau) &=& a^4 H^4 \, {\mathcal C}_{B}(\gamma, b_{0}) \, (- k \tau)^{5 - 2 |\gamma -1/2|},
\label{a06}\\
P_{B}^{(G)}(k,\tau) &=& a^4 H^4 \, {\mathcal C}^{(G)}_{B}(\gamma, b_{0})  \, (- k \tau)^{5 - 2 |\gamma -1/2|}.
\label{a07}
\end{eqnarray}
where ${\mathcal C}_{B}(\gamma,b_{0})$  and ${\mathcal C}^{(G)}_{B}(\gamma, b_{0})$ are two $k$-independent prefactors
\begin{eqnarray}
&& {\mathcal C}_{B}(\gamma, b_{0}) = \frac{2^{2\mu -4}\, \Gamma^2(\mu)}{\pi^3}  \biggl[ \frac{e^{- \pi b_0/2} \,\Gamma^2(\mu +1/2)}{\bigl|\Gamma(1/2 - i b_0/2 + \mu)\bigr|^2} + \frac{e^{\pi b_{0}/2} \,\Gamma^2(\mu +1/2)}{\bigl|\Gamma(1/2 + i b_{0}/2 + \mu)\bigr|^2} \biggr],
 \nonumber\\
&&  {\mathcal C}^{(G)}_{B}(\gamma, b_{0}) = \frac{2^{2 \mu -4}\, \Gamma^2(\mu)}{\pi^3} \biggl[ \frac{e^{- \pi b_0/2} \,\Gamma^2(\mu +1/2)}{\bigl|\Gamma(1/2 - i b_0/2 + \mu)\bigr|^2} - \frac{e^{\pi b_{0}/2} \,\Gamma^2(\mu +1/2)}{\bigl|\Gamma(1/2 + i b_{0}/2 + \mu)\bigr|^2} \biggr],
\label{a07a}
\end{eqnarray}
where, as in Eq. (\ref{a04}) we use $\mu = |\gamma -1/2|$. From Eqs. (\ref{a05}) and (\ref{WKB12})--(\ref{WKB13}) the hyperelectric power spectra evaluated in the large-scale limit are:
\begin{eqnarray}
P_{E}(k,\tau) &=& a^4 H^4 \, {\mathcal C}_{E}(\gamma,\,b_{0})(- k\tau)^{4 - 2 \gamma},
\label{a08}\\
P^{(G)}_{E}(k,\tau) &=& a^4 H^4 \,{\mathcal C}^{(G)}_{E}(\gamma,b_{0}) \, (- k\tau)^{4 - 2 \gamma}.
\label{a09}
\end{eqnarray}
As in the case of Eqs. (\ref{a06})--(\ref{a07}), the $k$-independent prefactors ${\mathcal C}_{E}(\gamma,\,b_{0})$ and ${\mathcal C}^{(G)}_{E}(\gamma,\,b_{0})$ turn out to be:
\begin{eqnarray}
{\mathcal C}_{E}(\gamma,\,b_{0}) &=& \frac{2^{2\gamma -3}\, \Gamma^2(\gamma+1/2)}{\pi^3}  \biggl[ \frac{e^{- \pi b_{0}/2} \,\Gamma^2(\gamma)}{\bigl|\Gamma(\gamma - i b_{0}/2 )\bigr|^2} + \frac{e^{\pi b_{0}/2} \,\Gamma^2(\gamma)}{\bigl|\Gamma(\gamma + i b_{0}/2)\bigr|^2} \biggr],
\nonumber\\
{\mathcal C}^{(G)}_{E}(\gamma,\,b_{0}) &=& \frac{2^{2\gamma -3}\, \Gamma^2(\gamma+1/2)}{\pi^3}  \biggl[ \frac{e^{- \pi b_{0}/2} \,\Gamma^2(\gamma)}{\bigl|\Gamma(\gamma - i b_{0}/2 )\bigr|^2} - \frac{e^{\pi b_{0}/2} \,\Gamma^2(\gamma)}{\bigl|\Gamma(\gamma + i b_{0}/2)\bigr|^2} \biggr].
\label{a010}
\end{eqnarray}
Before concluding this part of the analysis we make two remarks.
While it is true that for $\alpha \to 0$ the scalar and the pseudoscalar couplings are proportional, Eq. (\ref{a02}) may also arise when $\lambda$ does not depend on $\tau$ and $\overline{\lambda}
\propto \ln{(- \tau/\tau_{1})}$. This case is however automatically included in the results of Eqs. (\ref{a06})--(\ref{a07}) and (\ref{a08})--(\ref{a09}) by demanding that  $\gamma \to 0$.  The second remark is that 
the same structure of the turning points occurs when $0< \alpha \ll 1$ and this case is briefly analyzed 
at the and of appendix \ref{APPA}.

\subsection{The particular case $\alpha = 1$}
\label{APPB46}
\subsubsection{Exact form of the mode functions}
In Eqs. (\ref{EX3})-(\ref{EX4}) the limit $\alpha \to 1$ implies $\nu \to \infty$ and $\eta \to 
\mathrm{constant}$, respectively. The singularity of $\nu$ just signals 
that the rescaling to the $\eta$-time is immaterial since 
$\overline{\lambda}^{\prime}/\lambda = - b_{0}\tau_{1}/\tau^2$ and 
$\sqrt{\lambda}^{\prime\prime}/\sqrt{\lambda}$ scale in the same way as a function of $\tau$. It is 
therefore simpler to go back to Eq. (\ref{WKB14}) 
in the case $\alpha =1$:
\begin{equation}
f_{k,\, \pm}^{\prime\prime} + \biggl[ k^2 \mp \frac{ x_{1}\, b_{0}}{\tau^2} - \frac{\gamma (\gamma-1)}{\tau^2} \biggr] f_{k, \pm} =0.
\label{b01}
\end{equation}
The solution of Eq. (\ref{b01}) is:
\begin{equation}
f_{k\, \pm}(\tau) = \frac{{\mathcal N}_{\pm}}{\sqrt{ 2 k}} \, \sqrt{- k\tau} \, H_{\mu_{\pm}}^{(1)}(-k \tau), 
\label{b02}
\end{equation} 
where  $H_{\mu_{\pm}}^{(1)}(-k \tau)$ denote the standard Hankel's functions \cite{MM1,MM2} and:
\begin{equation}
{\mathcal N}_{\pm} = \sqrt{\frac{\pi}{2}} e^{i \pi ( 2 \mu_{\pm} +1)},\qquad \mu_{\pm}^2 = (\gamma - 1/2)^2 \pm x_{1} \, b_{0}.
\label{b03}
\end{equation}
As previously noted in the case generic case $\alpha > 0$, when $\gamma \to 1/2$ the Bessel index is 
$\mu_{\pm} = \pm b_{0} x_{1}$. Since $b_{0} x_{1}$ falls between $10^{-26}$ and $10^{-14}$ we have, in practice, that $\mu_{\pm} \to 0$ when $\gamma \to 1/2$. In this case a logarithmic enhancement of the Hankel functions is expected large scales. 

\subsubsection{Gauge spectra and comparison with the WKB result}
Inserting Eqs. (\ref{b01}) into Eqs. (\ref{WKB12})--(\ref{WKB13}) the hypermagnetic power spectra are:
\begin{eqnarray}
P_{B}(k,\tau) &=& \frac{H^4 a^4}{16\pi^3} \biggl[2 ^{\mu_{+}} \Gamma^2(\mu_{+}) (- k \tau)^{5 - 2 \mu_{+}} + 
2^{2 \mu_{-}} \, \Gamma^2(\mu_{-}) (- k \tau)^{5 - 2 \mu_{-}}\biggr],
\label{b06}\\
P_{B}^{(G)}(k,\tau) &=& \frac{H^4 a^4}{16\pi^3} \biggl[
2^{2 \mu_{-}} \, \Gamma^2(\mu_{-}) (- k \tau)^{5 - 2 \mu_{-}}\, -\, 2 ^{\mu_{+}} \Gamma^2(\mu_{+}) (- k \tau)^{5 - 2 \mu_{+}} \biggr].
\label{b07}
\end{eqnarray}
Equations (\ref{b06}) and (\ref{b07}) can always be expanded in the limit $b_{0} x_{1} \ll1 $ and the final result is
\begin{eqnarray}
P_{B}(k,\tau) &=& \overline{P}_{B}(k,\tau) \biggl[1 + {\mathcal O}(x_{1}^2 \, b_{0}^2)\biggr],
\label{b08}\\
P_{B}^{(G)}(k,\tau) &=& x_{1} b_{0} \,\,\overline{P}_{B}(k,\tau)  \frac{[\ln{(-k\tau/2)}  - \psi(|\gamma -1/2|)]}{|2 \gamma -1|}\, \biggl[1 + {\mathcal O}(x_{1}^2 \, b_{0}^2)\biggr],
\label{b09}
\end{eqnarray}
where $\psi(x)$ denotes here the Digamma function and 
\begin{equation}
 \overline{P}_{B}(k,\tau)  = \frac{H^4 a^4}{\pi^3}\, 2^{2 |\gamma -1/2| -3}\, \Gamma^2(|\gamma -1/2|)\, (- k \tau)^{ 5 - 2 |\gamma -1/2|}.
\label{b010}
\end{equation}
Again Eqs. (\ref{b08})--(\ref{b09}) and (\ref{b010}) are consistent with all the previous estimates and 
corroborate the WKB results.  

Before computing the hyperelectric power spectra we note that 
the mode functions $g_{k,\,\pm}$ obtained from Eq. (\ref{b01}) take two different forms depending upon the value of $\gamma$; more 
specifically:
\begin{eqnarray}
g_{k,\, \pm}(\tau) &=& \sqrt{\frac{k}{2}} {\mathcal N}_{\pm} \biggl\{\biggl[ \mu_{\pm} + (\gamma -1/2)\biggr] \frac{H_{\mu_{\pm}}^{(1)}(- k\tau)}{\sqrt{-k \tau}}
- \sqrt{- k\tau} \, H_{\mu_{\pm}-1}^{(1)}(-k\tau) \biggr\}, ~~0 < \gamma <1/2,
\label{b04}\\
g_{k,\, \pm}(\tau) &=& \sqrt{\frac{k}{2}} {\mathcal N}_{\pm} \biggl\{\biggl[ (\gamma -1/2) - \mu_{\pm}\biggr] \frac{H_{\mu_{\pm}}^{(1)}(- k\tau)}{\sqrt{- k\tau}}
+\sqrt{- k\tau} \, H_{\mu_{\pm}+1}^{(1)}(-k\tau) \biggr\}, ~~\gamma >1/2.
\label{b05}
\end{eqnarray}
When $0 < \gamma <1/2$ we have that  $H_{\mu_{\pm}-1}^{(1)}(-k\tau)= H_{-\gamma-1/2}^{(1)}(-k\tau)$;  
since, in general,  $H_{-\gamma-1/2}^{(1)}(-k\tau) = e^{i\pi(\gamma+1/2)}H_{\gamma1/2}^{(1)}(-k\tau)$ \cite{MM1,MM2} the overall 
result for the hyperelectric spectrum is the the same for Eqs. (\ref{b04}) and (\ref{b05}):
\begin{eqnarray}
P_{E}(k,\tau) &=&   \overline{P}_{E}(k,\tau)\biggl[1 + {\mathcal O}(x_{1}^2 \, b_{0}^2)\biggr],
\label{b011}\\
P_{E}^{(G)}(k,\tau) &=& x_{1} b_{0} \,\,\overline{P}_{E}(k,\tau)  \frac{[\ln{(-k\tau/2)}  - \psi(|\gamma -1/2|)]}{|2 \gamma -1|}\, \biggl[1 + {\mathcal O}(x_{1}^2 \, b_{0}^2)\biggr],
\end{eqnarray}
where, in analogy with Eq. (\ref{b010}) we introduced  $\overline{P}_{E}(k,\tau)$ which is defined as 
\begin{equation}
\overline{P}_{E}(k,\tau)= \frac{H^4 a^4}{\pi^3}\, 2^{2 \gamma-2} \Gamma^2(\gamma +1/2)\, (- k \tau)^{ 4 - 2 \gamma}.
\label{b012aa}
\end{equation}
To simplify the forthcoming phenomenological considerations it is practical to remark that the the amplitudes of the power spectra appearing in Eqs. (\ref{cond8})--(\ref{cond11}) (and obtained in the generic case $\alpha >0$) coincide, in practice, with the prefactors of Eqs. (\ref{b010})--(\ref{b011}). For instance, by comparing Eqs. (\ref{cond8}) and (\ref{b010}) it is possible to verify that 
\begin{equation}
\frac{4 \mu  (\mu +1)+5}{128 \pi^2  \mu ^2} \sim  \frac{2^{|2 \gamma -1| -3}\, \Gamma^2(|\gamma -1/2|)}{\pi^3},
\label{b012}
\end{equation}
where, according to the previous results, $\mu = |\gamma -1/2|$.
Since the approximate equality of Eq. (\ref{b012}) holds within an order of magnitude we can argue that there are indeed only two complementary phenomenological ranges namely the region $\alpha \geq 1$ and the case $0 \leq \alpha <1$.

\newpage
\renewcommand{\theequation}{C.\arabic{equation}}
\setcounter{equation}{0}
\section{Decreasing gauge coupling: particular cases}
\label{APPC}
For the sake of completeness in this appendix we are collecting some of the results that are relevant when the gauge coupling decreases during inflation and then flattens out later on. 
\subsection{The particular case $\alpha=0$}
In the limit $\alpha \to 0$ Eq. (\ref{a03}) will keep the same form with slightly different parameters; more specifically we will have 
\begin{equation}
b_{0} \to - \widetilde{\,b\,}_{0} = (2 \widetilde{\gamma}/\tau_{1}) (\tau_{1}/\tau_{2})^{2 \widetilde{\gamma}} (\overline{\lambda}_{2}/\lambda_{1}), \qquad \qquad \mu \to \widetilde{\mu} = (\widetilde{\gamma} +1/2).
\label{NNdef}
\end{equation}
In terms of the parameters defined in Eq. (\ref{NNdef}) the analogs of Eqs. (\ref{a04}) and (\ref{a05}) are therefore given by: 
\begin{eqnarray}
 f_{k,\, \pm}(y) &=& \frac{e^{-i \pi/4 \mp \, \pi \,\widetilde{b}_{0}/4}}{\sqrt{2 k}} \, W_{\pm i\, \widetilde{b}_{0}/2,\, \widetilde{\mu}}(y), \qquad\qquad \widetilde{\mu} = \widetilde{\gamma}+1/2,
\label{a04a}\\
g_{k,\, \pm}(y) &=& i \, \sqrt{2\,k}\, e^{-i \pi/4 \mp \pi \widetilde{b}_{0}/4} \, \biggl[ \frac{y + ( 2 \widetilde{\gamma} \mp i \widetilde{b}_{0})}{2 y} W_{\pm i \widetilde{b}_{0}/2, \, \widetilde{\mu}}(y) 
- \frac{W_{1 \pm i \widetilde{b}_{0}/2,\, \widetilde{\mu}}(y)}{y} \biggr].
\label{a05a} 
\end{eqnarray} 
From Eqs. (\ref{a04a}) and (\ref{a05a}) we can estimate the hypermagnetic power 
spectra: 
\begin{eqnarray}
\widetilde{\,P\,}_{B}(k,\tau)&=&  a^4 \, H^4 \widetilde{{\mathcal C}}_{B}(\widetilde{\gamma}, \widetilde{b}_{0}) (- \, k\, \tau)^{4 - 2 \widetilde{\,\gamma\,}} ,
\label{a06a}\\
\widetilde{\,P\,}^{(G)}_{B}(k,\tau)&=&  a^4 \, H^4 \widetilde{{\mathcal C}}^{(G)}_{B}(\widetilde{\gamma}, \widetilde{b}_{0}) (- \, k\, \tau)^{4 - 2 \widetilde{\,\gamma\,}} ,
\label{a08a}
\end{eqnarray}
where $\widetilde{{\mathcal C}}_{B}(\widetilde{\gamma}, \widetilde{b}_{0}) $ and $\widetilde{{\mathcal C}}^{(G)}_{B}(\widetilde{\gamma}, \widetilde{b}_{0})$ are defined as:
\begin{eqnarray}
\widetilde{{\mathcal C}}_{B}(\widetilde{\gamma}, \widetilde{b}_{0}) &=& \frac{2^{2 \widetilde{\gamma} -3}}{\pi^3} \Gamma^2(\widetilde{\gamma}+1/2) \biggl[\frac{e^{\pi \widetilde{b}_{0}/2}\,\,\Gamma^2(\widetilde{\,\gamma\,})}{\bigl| \Gamma( \widetilde{\,\gamma\,} - i \widetilde{b}_{0}/2)\bigr|^2} + \frac{e^{-\pi \widetilde{b}_{0}/2}\,\,\Gamma^2(\widetilde{\,\gamma\,})}{\bigl| \Gamma( \widetilde{\,\gamma\,} + i \widetilde{b}_{0}/2)\bigr|^2} \biggr],
\nonumber\\
\widetilde{{\mathcal C}}^{(G)}_{B}(\widetilde{\gamma}, \widetilde{b}_{0}) &=& \frac{2^{2 \widetilde{\gamma} -3}}{\pi^3} \Gamma^2(\widetilde{\gamma}+1/2)\biggl[\frac{e^{\pi \widetilde{b}_{0}/2}\,\,\Gamma^2(\widetilde{\,\gamma\,})}{\bigl| \Gamma(\widetilde{\,\gamma\,} - i \widetilde{b}_{0}/2)\bigr|^2} - \frac{e^{-\pi \widetilde{b}_{0}/2}\,\,\Gamma^2(\widetilde{\,\gamma\,})}{\bigl| \Gamma(\widetilde{\,\gamma\,}+ i \widetilde{b}_{0}/2)\bigr|^2} \biggr].
\label{a09a}
\end{eqnarray}
Finally in the hyperelectric case the gauge power spectra are:
\begin{eqnarray}
\widetilde{\,P\,}_{E}(k,\tau) &=& a^4 H^4 \,\biggl[\widetilde{{\mathcal C}}^{(0)}_{E}(\widetilde{\gamma}) \, (- k \tau)^{5 - | 2 \widetilde{\gamma} -1|} + \widetilde{{\mathcal C}}^{(1)}_{E}(\widetilde{\gamma}, \widetilde{b}_{0}) (- k \tau)^{4 - 2 \widetilde{\gamma}} \biggr],
\label{a010a}\\
\widetilde{\,P\,}_{E}^{(G)}(k,\tau) &=& a^4 H^4 \, \biggl[\widetilde{{\mathcal C}}^{(G)}_{E}(\widetilde{\gamma}, \widetilde{b}_{0}) (- k \tau)^{4 - 2 \widetilde{\gamma}} \biggr],
\label{a011a}
\end{eqnarray}
where the various coefficients are now given by:
\begin{eqnarray}
\widetilde{{\mathcal C}}^{(0)}_{E}(\widetilde{\gamma})  &=& \frac{2^{|2\widetilde{\gamma}-1| - 3}}{\pi^3} 
\Gamma^2(\widetilde{\gamma}+1/2),
\nonumber\\
\widetilde{{\mathcal C}}^{(1)}_{E}(\widetilde{\gamma}, \widetilde{b}_{0})  &=& \widetilde{b}_{0}^2 
\frac{2^{2\widetilde{\gamma} - 3}}{\pi^3} 
\Gamma^2(\widetilde{\gamma}+1/2)\biggl[ \frac{e^{\pi \widetilde{b}_{0}/2}\,\Gamma^2( \widetilde{\,\gamma\,})}{\bigl| \Gamma(\widetilde{\,\gamma\,} + i \widetilde{b}_{0}/2)\bigr|^2} + \frac{e^{-\pi \widetilde{b}_{0}/2}\,\Gamma^2(\widetilde{\,\gamma\,})}{\bigl| \Gamma(\widetilde{\,\gamma\,} - i \widetilde{b}_{0}/2)\bigr|^2 }\biggr], 
\nonumber\\
\widetilde{{\mathcal C}}^{(G)}_{E}(\widetilde{\gamma}, \widetilde{b}_{0})  &=& \widetilde{b}_{0}^2 
\frac{2^{2\widetilde{\gamma} - 3}}{\pi^3} 
\Gamma^2(\widetilde{\gamma}+1/2)\biggl[ \frac{e^{\pi \widetilde{b}_{0}/2}\,\Gamma^2( \widetilde{\,\gamma\,})}{\bigl| \Gamma(\widetilde{\,\gamma\,} + i \widetilde{b}_{0}/2)\bigr|^2} - \frac{e^{-\pi \widetilde{b}_{0}/2}\,\Gamma^2(\widetilde{\,\gamma\,})}{\bigl| \Gamma(\widetilde{\,\gamma\,} - i \widetilde{b}_{0}/2)\bigr|^2 }\biggr].
\label{a012a}
\end{eqnarray}

\subsection{The particular case $\alpha =1$}
In the case $\alpha =1$ the basic evolution of the mode function appearing 
in Eq. (\ref{b01}) gets modified as
\begin{equation}
f_{k,\, \pm}^{\prime\prime} + \biggl[ k^2 \mp \frac{ x_{1}\, \widetilde{\,b\,}_{0}}{\tau^2} - \frac{\widetilde{\mu}_{\pm}^2 -1/4 }{\tau^2} \biggr] f_{k, \pm} =0, \qquad \qquad \widetilde{\mu}_{\pm}^2 = (\widetilde{\gamma} +1/2)^2 \pm x_{1} \widetilde{\,b\,}_{0}.
\label{b01a}
\end{equation}
The same steps of Eqs. (\ref{b02}) and (\ref{b03}) can be repeated since Eq. (\ref{b01a}) 
has an exact solution of the same kind of the one previously discussed. As before 
the crucial observation is that, in spite of the values of $b_{0}$, $x_{1}$ is 
so small that the obtained results can always be expanded for $x_{1} \ll 1$.
Large-scale power spectra will be unaffected by the anomalous contribution. For instance the hypermagnetic power 
spectra will be given by
\begin{eqnarray}
\widetilde{\,P\,}_{B}(k,\tau) &=& \frac{H^4 a^4}{16\pi^3} \biggl[2 ^{\widetilde{\mu}_{+}} \Gamma^2(\widetilde{\mu}_{+}) (- k \tau)^{5 - 2 \widetilde{\mu}_{+}} + 
2^{2 \widetilde{\mu}_{-}} \, \Gamma^2(\widetilde{\mu}_{-}) (- k \tau)^{5 - 2 \widetilde{\mu}_{-}}\biggr],
\label{b06a}\\
\widetilde{\,P\,}_{B}^{(G)}(k,\tau) &=& \frac{H^4 a^4}{16\pi^3} \biggl[2 ^{\widetilde{\mu}_{+}} \Gamma^2(\widetilde{\mu}_{+}) (- k \tau)^{5 - 2 \widetilde{\mu}_{+}} - 
2^{2 \widetilde{\mu}_{-}} \, \Gamma^2(\widetilde{\mu}_{-}) (- k \tau)^{5 - 2 \widetilde{\mu}_{-}}\biggr].
\label{b07a}
\end{eqnarray}
With the same technique we can obtain the explicit expressions of $\widetilde{\,P\,}_{E}(k,\tau)$
and $\widetilde{\,P\,}^{(G)}_{E}(k,\tau)$. If $\widetilde{\,P\,}_{B}(k,\tau)$ and $\widetilde{\,P\,}_{E}(k,\tau) $ 
are expanded in the limit  $| k\tau | \ll1 $ the resulting expressions are: 
\begin{eqnarray}
\widetilde{\,P\,}_{B}(k,\tau) &=&  \frac{H^4 a^4}{\pi^3}\, 2^{2 \widetilde{\gamma}  -2}\, \Gamma^2(\widetilde{\gamma} +1/2)\, (- k \tau)^{4 - 2 \widetilde{\gamma}}\biggl[1 + {\mathcal O}(x_{1}^2 \, \widetilde{\,b\,}_{0}^2)\biggr],
\label{b08a}\\
\widetilde{\,P\,}_{E}(k,\tau) &=& \frac{H^4 a^4}{\pi^3}\, 2^{|2\widetilde{\gamma} -1| -3}\, \Gamma^2(|\widetilde{\gamma} -1/2|)\, (- k \tau)^{ 5 - |2 \widetilde{\gamma} -1|} \biggl[1 + {\mathcal O}(x_{1}^2 \, \widetilde{\,b\,}_{0}^2)\biggr].
\label{b09a}
\end{eqnarray}
and similarly for the gyrotropic contribution that is proportional to $x_{1} \widetilde{\,b\,}_{0}$.

\subsection{Late-time gauge spectra}
In the case of decreasing gauge coupling Eq. (\ref{FFF3}) has basically the same form 
but the Bessel indices and the related arguments are clearly different. For the present 
purposes the relevant results are:
\begin{eqnarray}
\widetilde{\,A\,}_{f\, f}(z_{1}, z) &=& \frac{\pi}{2} \sqrt{z_{1}\,z} \biggl[ Y_{\widetilde{\,\sigma\,} -1}( z_{1}) J_{\widetilde{\,\sigma\,}}(z) - J_{\widetilde{\,\sigma\,}-1}(z_{1}) Y_{\widetilde{\,\sigma\,}}(z) \biggr],
\nonumber\\
\widetilde{\,A\,}_{f\, g}(z_{1}, z) &=& \frac{\pi}{2} \sqrt{z_{1}\,z} \biggl[ J_{\widetilde{\,\sigma\,}}(z_{1}) Y_{\widetilde{\,\sigma\,}}(z) - Y_{\widetilde{\,\sigma\,}}(z_{1}) J_{\widetilde{\,\sigma\,}}(z) \biggr],
\nonumber\\
\widetilde{\,A\,}_{g\, f}(z_{1}, z) &=& \frac{\pi}{2} \sqrt{z_{1}\,z} \biggl[ Y_{\widetilde{\,\sigma\,} - 1}(z_{1}) J_{\widetilde{\,\sigma\,} -1}(z) - J_{\widetilde{\,\sigma\,}-1}(z_{1}) Y_{\widetilde{\,\sigma\,}-1}(z) \biggr],
\nonumber\\
\widetilde{\,A\,}_{g\, g}(z_{1}, z)&=& \frac{\pi}{2} \sqrt{z_{1}\,z} \biggl[ J_{\widetilde{\,\sigma\,}}(z_{1}) Y_{\widetilde{\,\sigma\,}-1}(z) - Y_{\widetilde{\,\sigma\,}}(z_{1}) J_{\widetilde{\,\sigma\,}-1}(z) \biggr],
\label{APPB4}
\end{eqnarray}
where now $z_{1}$, $z$ and $\widetilde{\,\sigma\,}$ are defined as
\begin{equation}
z_{1} = (\widetilde{\,\delta\,}/\widetilde{\,\gamma\,}) k \tau_{1}, \qquad z= \tau + \tau_{1}[ 1 + (\widetilde{\,\delta\,}/\widetilde{\,\gamma\,})], \qquad  \widetilde{\,\sigma\,}  = |\widetilde{\,\delta\,} -1/2|.
\label{APPB5}
\end{equation}
\end{appendix}
Equations (\ref{APPB4})--(\ref{APPB5}) have been deduced by assuming that the gauge coupling 
first decreases and then flattens out after inflation. This happens, for instance, in the case of the profile of Eqs. (\ref{EqP14a})--(\ref{EqP14b}). Equations (\ref{APPB4})--(\ref{APPB5})  (which are the analogs of  Eqs. (\ref{FFF3})--(\ref{FFF4})) apply in the situation where the gauge coupling increases and the flattens out after inflation.


\newpage

\begin{thebibliography}{99}

\itemsep -5.5pt

\bibitem{ONE0a} P.~J.~E.~Peebles and J.~T.~Yu, Astrophys. J. {\bf 162}, 815 (1970).

\bibitem{ONE0b} D.~N.~Spergel \textit{et al.} [WMAP],  Astrophys. J. Suppl. {\bf 148}, 175 (2003).

\bibitem{ONE0c} D.~N.~Spergel \textit{et al.} [WMAP], Astrophys. J. Suppl. {\bf 170}, 377 (2007).

\bibitem{ONE0d} G.~Hinshaw \textit{et al.} [WMAP], Astrophys. J. Suppl. {\bf 208}, 19 (2013).

\bibitem{ONE0f}  P.~A.~R.~Ade \textit{et al.} [Planck],  Astron. Astrophys. {\bf 571}, A1 (2014).

\bibitem{ONE0e}  N.~Aghanim \textit{et al.} [Planck],  Astron. Astrophys. {\bf 641}, A6 (2020).

\bibitem{ONE0g} S. Weinberg, Phys. Rev. D {\bf 77}, 123541 (2008).

\bibitem{ONE0h}  S. Weinberg, {\it Cosmology} (Oxford University Press, Oxford, UK, 2008).

\bibitem{ONE0m} L.~Parker,  Phys.\ Rev.\ Lett.\  {\bf 21},  562 (1968).

\bibitem{TWOa}  S. Carroll, G. Field and R. Jackiw, Phys. Rev. D {\bf 41},  1231 (1990).

\bibitem{TWOb}  W. D. Garretson, G. Field and S. Carroll,  Phys. Rev. D {\bf 46}, 5346 (1992).

\bibitem{TWOc} G. Field and S. Carroll Phys. Rev. D, {\bf 62}, 103008 (2000).

\bibitem{ONE0n} R. D. Peccei and H. R. Quinn, Phys. Rev. Lett. {\bf 38}, 1440  (1977).

\bibitem{ONE0o}  R. D. Peccei and H. R. Quinn,  Phys. Rev. D {\bf 16}, 1791 (1977).

\bibitem{ONE0p} J. Kim, Phys. Rep. {\bf 150}, 1 (1987).

\bibitem{FOUR} B. Ratra,  Astrophys.\, J.\, Lett.  {\bf 391}, L1 (1992).

\bibitem{FIVE} M.~Gasperini, M.~Giovannini, and G.~Veneziano, Phys. Rev. Lett. {\bf 75}, 3796 (1995).

\bibitem{SIX} M.~Giovannini, Phys. Rev. D {\bf 56}, 631 (1997).

\bibitem{SEVEN} M. Giovannini, Phys.\ Rev.\ D {\bf 61}, 063004 (2000).

\bibitem{EIGHT} M.~Giovannini,  Phys.\ Rev.\  D {\bf 64}, 061301 (2001).

\bibitem{NINE} M. Giovannini, Phys.\ Lett.\  B {\bf 659}, 661 (2008).

\bibitem{TEN} K.~Bamba and J.~Yokoyama, Phys. Rev. D {\bf 69}, 043507 (2004).

\bibitem{ELEVEN} L.~Campanelli and M.~Giannotti, Phys. Rev. D {\bf 72}, 123001 (2005).

\bibitem{TWELVEa} L.~Campanelli and M.~Giannotti, Phys. Rev. Lett. {\bf 96}, 161302 (2006).

\bibitem{TWELVEb} K.~Bamba JCAP {\bf 10}, 015 (2007).

\bibitem{TWELVEc}  A.~Akhtari-Zavareh, A.~Hojjati and B.~Mirza, Prog. Theor. Phys. {\bf 117}, 803 (2007).
 
\bibitem{TWELVEd} D.~Seery, JCAP {\bf 08}, 018 (2009).

\bibitem{TWELVEdd} V.~Demozzi, V.~Mukhanov and H.~Rubinstein, JCAP {\bf 08}, 025 (2009).

\bibitem{TWELVEe} M.~Karciauskas and D.~H.~Lyth, JCAP {\bf 11}, 023 (2010). 

\bibitem{TWELVEf} I. Brown, Astrophys. J. {\bf 733}, 83 (2011).

\bibitem{TWELVEg} T.~Fujita and S.~Mukohyama,  JCAP {\bf 1210}, 034 (2012).

\bibitem{TWELVEh} R. Jain and M. Sloth, Phys. Rev. D {\bf 86}, 123528 (2012). 

\bibitem{TWELVEi}  M.~Giovannini, Phys. Rev. D {\bf 88}, 083533 (2013).

\bibitem{TWELVEl} F. Membiela, Nucl. Phys. B {\bf 885}, 196  (2014).

\bibitem{TWELVEm} R. Ferreira and J.~Ganc, JCAP {\bf 1504}, 029 (2015); 

\bibitem{TWELVEo} L.~Campanelli, Phys. Rev. D {\bf 93}, 063501 (2016).

\bibitem{TWELVEp} P.~Qian, Y.~F.~Cai, D.~A.~Easson and Z.~K.~Guo,
Phys. Rev. D {\bf 94} , 083524 (2016).

\bibitem{TWELVEq}  R.~Koley and S.~Samtani, JCAP {\bf 04}, 030 (2017).

\bibitem{TWELVEr} E.~Frion, N.~Pinto-Neto, S. Vitenti and S.~ Perez Bergliaffa,
Phys. Rev. D {\bf 101}, 103503 (2020).

\bibitem{TWELVEs} A.~Talebian, A.~Nassiri-Rad and H.~Firouzjahi,
Phys. Rev. D {\bf 102}, 103508 (2020).

\bibitem{TWELVEt} O. Sobol, A. Lysenko, E. Gorbar and S. Vilchinskii, Phys. Rev. D {\bf 102}, 123512 (2020).

\bibitem{TWELVEu} D.~Maity, S.~Pal and T.~Paul, [arXiv:2103.02411 [hep-th]].

\bibitem{TWELVEv} K.~Bamba, E.~Elizalde, S.~D.~Odintsov and T.~Paul, JCAP {\bf 04}, 009 (2021).

\bibitem{FIVEa0} M.~Giovannini,  Phys.\ Rev.\ D {\bf 61}, 063502 (2000).

\bibitem{FIVEa1}  M.~Giovannini and M.~E.~Shaposhnikov,  Phys.\ Rev.\ D {\bf 57}, 2186 (1998).

\bibitem{FIVEb1} M. Giovannini, Phys.\ Rev.\ D {\bf 61}, 063004 (2000).

\bibitem{FIVEb1a} K.~Bamba, C.~Q.~Geng and S.~H.~Ho, Phys. Lett. B  {\bf 664}, 154 (2008).

\bibitem{FIVEb2} M.~Dvornikov and V.~B.~Semikoz, JCAP {\bf 1202}, 040 (2012).

\bibitem{FIVEb3}  S.~Alexander, A.~Marciano and D.~Spergel, JCAP {\bf 1304}, 046 (2013).
  
\bibitem{FIVEb4}   N.~D.~Barrie and A.~Kobakhidze,  JHEP {\bf 1409}, 163 (2014).

\bibitem{FIVEc} M.~Giovannini, Phys. Rev. D {\bf 92},  12301 (2015).

\bibitem{SIXa} D.~E.~Kharzeev,  Prog.\ Part.\ Nucl.\ Phys.\  {\bf 75}, 133 (2014).

\bibitem{SIXb}  D. Kharzeev, L. McLerran and H. Warringa, Nucl. Phys. A {\bf 803}, 227 (2008).

\bibitem{SIXe} K.~Landsteiner, E.~Megias, L.~Melgar and F.~Pena-Benitez, JHEP {\bf 09}, 121 (2011).

\bibitem{SIXf} K.~Landsteiner, E.~Megias and F.~Pena-Benitez, Phys. Rev. D {\bf 90}, 065026 (2014).

\bibitem{SIXc} M.~Giovannini, Phys.\ Rev.\ D {\bf 88}, 063536 (2013).

\bibitem{SIXd}  M. Giovannini, Phys. Rev. D {\bf 94}, 081301 (2016).

\bibitem{POLWKB} M.~Giovannini, Phys. Rev. D {\bf 99}, 083501 (2019).

\bibitem{VP} L. H. Ford, Phys. Rev. D {\bf 31}, 704 (1985).

\bibitem{zeld2} S. I. Vainshtein, Dokl. Zh. Eksp. Teor. Fiz. {\bf 61}, 612 (1971) [Sov. Phys. JETP {\bf 34}, 327 (1971)].

\bibitem{kaza} A. P. Kazantsev, Zh. Eksp. Teor. Fiz. {\bf 53}, 1806 (1967) [Sov. Phys. JETP {\bf 26}, 1031 (1968)].

\bibitem{kraich}  R. H. Kraichnan and S. Nagarajan, Phys. Fluids {\bf 10}, 859 (1967).

\bibitem{CKN1} A.~G.~Cohen, D.~B.~Kaplan and A.~E.~Nelson,  Ann.\ Rev.\ Nucl.\ Part.\ Sci.\  {\bf 43}, 27 (1993).

\bibitem{NINEaa} S. Deser and C. Teitelboim, Phys. Rev. D 13, 1592 (1976).

\bibitem{TENaa}  S. Deser, J. Phys. A {\bf 15}, 1053 (1982).

\bibitem{MGJ} M.~Giovannini, JCAP {\bf 04}, 003 (2010).

\bibitem{MM1} M. Abramowitz and I. A. Stegun, {\it Handbook of Mathematical Functions} (Dover, New York, 1972).

\bibitem{MM2} A. Erd\'elyi, W. Magnus, F. Oberhettinger, F. G. Tricomi, {\it Higher Trascendental Functions} (Mc Graw-Hill, New York, 1953).

\bibitem{cond1} J. Ahonen and K. Enqvist, Phys. Lett. B {\bf 382}, 40 (1996).

\bibitem{cond2} J. Ahonen, Phys. Rev. D {\bf 59}, 023004 (1999).

\bibitem{MMM2a}  S.-Y. Pi and R. Jackiw, Phys. Rev. D {\bf 68}, 104012 (2003).

\bibitem{MMM2b} A. Lue, L. Wang, and M. Kamionkowski, Phys. Rev. Lett. {\bf 83}, 1506 (1999).

\bibitem{MMM3a} M.~Giovannini, Class. Quant. Grav. {\bf 38}, 135018 (2021).

\bibitem{MMM3b} M.~Giovannini, Eur. Phys. J. C {\bf 81}, 81 (2021).

\bibitem{MMM3}  M.~Giovannini,  Phys. Lett. B {\bf 819}, 136444 (2021).

\bibitem{THIRTEENa} A. Liddle, P. Parsons and J.Barrow, Phys. Rev. D {\bf 50}, 7222 (1994).

\bibitem{THIRTEENb}  A.~Liddle, Phys. Rev. D {\bf 68}, 103504 (2003).

\bibitem{HHH1} I. T. Drummond and S. J. Hathrell, Phys. Rev. D {\bf 22}, 343 (1980).

\bibitem{HHH2} T. J. Hollowood and G. M. Shore, Nucl. Phys. B {\bf 795}, 138 (2008).

\bibitem{HHH3} S. L. Adler, Ann. Phys.  {\bf 67}, 599 (1971).

\bibitem{RT0} P.~Ade {\it et al.} [BICEP2 and Keck Array Collaborations],  Phys.\ Rev.\ Lett.\  {\bf 116}, 031302 (2016).

\bibitem{RT1}  Y.~Akrami {\it et al.} [Planck Collaboration], Astron. Astrophys. {\bf 641}, A10 (2020).

\bibitem{RT2}  N.~Aghanim \textit{et al.} [Planck Collaboration], Astron. Astrophys. {\bf 641}, A6 (2020).

\bibitem{NON1} H.~Motohashi and A.~A.~Starobinsky, JCAP {\bf 11}, 025 (2019).

\bibitem{NON2} M.~Guerrero, D.~Rubiera-Garcia and D.~Saez-Chillon Gomez, Phys. Rev. D {\bf 102}, 123528 (2020).

\bibitem{NON3} A.~Mohammadi, T.~Golanbari, S.~Nasri and K.~Saaidi, Phys. Rev. D {\bf 101},  123537 (2020).

\bibitem{FOURa} M. Giovannini, Phys. Lett. B {\bf 659}, 661 (2008).

\bibitem{FOURb} M. Giovannini,  Class. Quantum Grav. {\bf 23}, 4991 (2006).

\bibitem{SYM1} G. Feinberg and J. Sucher, Phys. Rev. A {\bf 2}, 2395 (1970).

\bibitem{SYM2} G. Feinberg and J. Sucher,  Phys. Rev. D {\bf 20}, 1717 (1979).

\bibitem{SYM3} M. Giovannini,   Phys. Rev. D {\bf 92}, 043521 (2015).

\bibitem{SYM4} H.~Motohashi and A.~A.~Starobinsky, JCAP {\bf 11}, 025 (2019).

\bibitem{SYM5} M.~Guerrero, D.~Rubiera-Garcia and D.~Saez-Chillon Gomez, Phys. Rev. D {\bf 102}, 123528 (2020).

\bibitem{SYM6} A.~Mohammadi, T.~Golanbari, S.~Nasri and K.~Saaidi, Phys. Rev. D {\bf 101},  123537 (2020).


\end{thebibliography}
\end{document}